\documentclass[epj]{svjour}
\usepackage{graphicx}
\usepackage{color}
\def\mb#1{\setbox0=\hbox{$#1$}\kern-.025em\copy0\kern-\wd0
\kern-0.05em\copy0\kern-\wd0\kern-.025em\raise.0233em\box0}
\begin{document}
\title{Caloric curves fitted by polytropic distributions in the HMF model}

\author{A. Campa and P.H. Chavanis}

\institute{$^1$ Complex Systems and Theoretical Physics Unit, Health and Technology Department, Istituto Superiore di Sanit\`a, \\
and INFN Roma 1, Gruppo Collegato Sanita, 00161 Roma, Italy\\
$^2$ Laboratoire de Physique Th\'eorique (IRSAMC), CNRS and UPS, Universit\'e de Toulouse, F-31062 Toulouse, France
}

\titlerunning{Caloric curves fitted by polytropic distributions in the HMF model}

\date{To be included later }

\abstract{We perform direct numerical simulations of the Hamiltonian mean field (HMF) model starting from non-magnetized
initial conditions with a velocity distribution that is (i) gaussian, (ii) semi-elliptical, and (iii) waterbag. Below a
critical energy $E_c$, depending on the initial condition, this distribution is Vlasov dynamically unstable. The system undergoes
a process of violent relaxation and quickly reaches a quasi-stationary state (QSS). We find that the distribution function of
this QSS  can be conveniently fitted by a polytrope with index (i) $n=2$, (ii) $n=1$, and (iii) $n=1/2$. Using the values of
these indices, we are able to determine the physical caloric curve $T_{kin}(E)$ and explain the negative kinetic specific heat
region $C_{kin}=dE/dT_{kin}<0$ observed in the numerical simulations. At low energies, we find that the system takes a
``core-halo'' structure. The core corresponds to the pure polytrope discussed above but it is now surrounded by a halo of
particles. In case (iii), we recover the ``uniform'' core-halo structure previously found by
Pakter \& Levin [Phys. Rev. Lett. {\bf 106}, 200603 (2011)].
We also consider unsteady initial conditions with magnetization $M_0=1$ and isotropic waterbag distribution and report
the complex dynamics of the system creating phase space holes and dense filaments. We show that the kinetic caloric
curve is approximately constant, corresponding to a polytrope with index $n_0\simeq 3.56$ 
(we also mention the presence of an unexpected hump). Finally, we consider the
collisional evolution of an initially Vlasov stable distribution, and show that the time-evolving distribution
function $f(v,t)$ can be fitted by a sequence of polytropic distributions with a time-dependent index $n(t)$ both
in the non-magnetized and magnetized regimes. These numerical results show that polytropic distributions (also
called Tsallis distributions)
provide in many cases a good fit of the QSSs. They may even be the rule rather than the exception. 
However, in order to moderate our message, we also report a case where 
the Lynden-Bell theory (which assumes ergodicity or efficient mixing) provides an excellent prediction 
of an inhomogeneous QSS. We therefore conclude that both Lynden-Bell  and Tsallis distributions may be useful to 
describe QSSs depending on the efficiency of mixing.  }

\maketitle
%

\section{Introduction}
\label{sec_introduction}

Systems with long-range interactions are numerous in
nature. They include, for example, self-gravitating systems and
two-dimensional (2D) vortices which are systems of considerable interest \cite{houches}.
These systems may be trapped in long-lasting non-equilibrium states, called quasi-stationary states
(QSS), whose lifetime diverges with the number of particles $N$. These QSSs correspond to galaxies in astrophysics and
large-scale vortices (e.g. Jupiter's great red spot) in 2D hydrodynamics.
Therefore, in many cases of physical
interest, the system does not reach the Boltzmann distribution but
remains stuck in a non-Boltzmannian QSS. This is the case of
elliptical galaxies in stellar dynamics because the collisional relaxation
time
exceeds the age of the universe by many orders of magnitude. This is also the case in 2D geophysical and astrophysical flows
because the viscous time is generally much larger than the turnover time of a large-scale vortex.
These QSSs are known to be stable steady states of the Vlasov (or 2D Euler) equation
on a coarse-grained scale. The Vlasov equation describes
the ``collisionless'' evolution of the system before
``collisions'' (more precisely correlations, finite $N$ effects,
granularities...) drive the system towards Boltzmann's statistical equilibrium.
Since the Vlasov equation admits an infinite number of steady states,
the prediction of the QSS that is actually selected by the system is
difficult. Our understanding of these QSSs is still incomplete.

A toy model of systems with long-range interactions, called the Hamiltonian mean field (HMF)
model, has been actively studied in statistical mechanics \cite{cdr}. It consists
of $N$ particles moving on a ring and interacting via a cosine
potential\footnote{This model was first introduced in 1982 by Messer \& Spohn \cite{ms} who called it the cosine model. It was
re-introduced in the 1990s by several authors \cite{kk,ik,inagaki,pichon,ar} and considerably studied since then
(see \cite{cvb,cc} for a short historic of the HMF model). }. At statistical equilibrium, this system displays a second
order phase transition between a spatially homogeneous (non-magnetized
$M=0$) phase and a spatially inhomogeneous (magnetized $M\neq 0$)
phase. The magnetized phase appears below the critical energy $E_c=3/4$ or below the critical
temperature $T_c=1/2$. Antoni \& Ruffo \cite{ar} carried out direct numerical
simulations of the HMF model. They started from an initial condition
in which all the particles are located at $\theta=0$ (corresponding to
a magnetization $M_0=1$) with a waterbag velocity distribution. They
determined the physical caloric curve $T_{kin}(E)$ giving the average
kinetic temperature as a function of the energy. They compared their numerical results to the
theoretical caloric curve $T(E)$ corresponding to the Boltzmann equilibrium and
reported several ``anomalies''. In particular, the phase transition
takes place at an energy sensibly smaller than $E_c=3/4$ and the numerical caloric
curve $T_{kin}(E)$ presents a region of negative kinetic specific heats, unlike the theoretical caloric curve $T(E)$ corresponding
to the Boltzmann equilibrium. They
understood that these discrepancies are due to the fact that the
observed structures are out-of-equilibrium QSSs. These results were confirmed by Latora et al.
\cite{latora,lrt} who showed that the lifetime of these QSSs diverges
with $N$ and that their distribution functions are
non-Boltzmannian. They also observed many other anomalies in the region of negative kinetic specific heats such as
anomalous diffusion, L\'evy walks, aging, and dynamical correlations in
phase-space.  The observation of non-Boltzmannian QSSs was a
surprise in the community of statistical mechanics. Latora et
al. \cite{latora,lrt} proposed to interpret these QSSs in terms of Tsallis generalized
thermodynamics \cite{tsallis}. In particular, they tried to fit the QSS at $E=0.69$ by a $q$-distribution
with a power-law tail. To make the distribution normalizable, they introduced a cut-off
at large velocities. While their study definitely shows that the QSS is
non-Boltzmannian, their procedure is not very convincing and their fit is
relatively poor.

The situation changed after the conference in Les Houches in
2002 where it was indicated \cite{chavhouches} that QSSs were previously observed in stellar
dynamics and 2D turbulence. In these domains,
the QSSs are interpreted in terms of Lynden-Bell's statistical theory
of violent relaxation \cite{lb}\footnote{In 2D turbulence, this is called the
Miller-Robert-Sommeria (MRS) theory \cite{miller,rs}.}. The Lynden-Bell theory
determines the statistical
equilibrium state of the Vlasov equation, taking into account all the constraints
of the ``collisionless'' dynamics, in particular the
conservation of the Casimirs. If the system is ergodic, the QSS coincides
with the Lynden-Bell distribution (most probable state). In the
two-levels case $(f_0,0)$, the distribution
predicted by Lynden-Bell is similar to the Fermi-Dirac distribution in
quantum mechanics. When applied to the HMF model, the Lynden-Bell
theory predicts an out-of-equilibrium phase transition between a
magnetized and a non-magnetized phase, and a phenomenon of phase
re-entrance in the $(f_0,E)$ plane \cite{epjb}. The nature of these phase
transitions has been studied in detail in \cite{prl2,marseille,staniscia1,staniscia2}. In particular, there exist
a tricritical point separating first and second order phase
transitions, and a critical point (associated with the re-entrant phase) marking the onset of a second order
azeotropy \cite{staniscia2}. Direct numerical simulations \cite{staniscia1,precommun,prl1,bachelard1,bachelard2} showed
a good agreement with the Lynden-Bell prediction in certain cases\footnote{We shall 
present in  Fig. \ref{distrlynden} a new simulation  showing a perfect agreement with the Lynden-Bell theory.} but
also evidenced discrepancies in other cases. For example, in \cite{staniscia1}, the re-entrant
phase predicted from the Lynden-Bell
theory in a very small range of parameters is confirmed (which is a success of the theory), but  a secondary
re-entrant phase that is not predicted by the
Lynden-Bell theory is also observed (this secondary re-entrant phase has been
recently confirmed by another group \cite{oy} suggesting that it is not a numerical
artifact). More generally, the adequacy, or inadequacy,  of the Lynden-Bell theory
to predict the magnetization of the QSS can be read from the numerical phase diagrams plotted in \cite{staniscia1,bachelard2}.

These discrepancies can be interpreted as a
result of {\it incomplete relaxation} \cite{lb,epjb,incomplete,hb3,hb4}. Indeed,
the Lynden-Bell statistical theory is based on an assumption of
ergodicity or, at least, efficient mixing. If the system does not mix
well, the Lynden-Bell prediction fails and the system may be trapped
in a steady state of the Vlasov equation that is not the most mixed
state. This is precisely what happens in the case $M_0=1$ considered by
Antoni \& Ruffo \cite{ar} and Latora et
al. \cite{latora,lrt}. As discussed in \cite{epjb,hb3}, the failure of the Lynden-Bell prediction is
particularly clear in that case. Indeed, for an initial condition with $M_0=1$, we are in the
non-degenerate (dilute) limit of the Lynden-Bell theory. In this
limit, the Lynden-Bell distribution reduces to the Boltzmann
distribution (with a different interpretation). Therefore, as argued in \cite{hb3},
the theoretical Boltzmann caloric curve plotted by Antoni \& Ruffo \cite{ar} and Latora et
al. \cite{latora,lrt} should be interpreted as the theoretical {\it Lynden-Bell} caloric curve.
Consequently, the observed discrepancies between this theoretical caloric curve and the numerical results reveal the failure
of the Lynden-Bell prediction in that case (this is confirmed by the phase diagrams of \cite{staniscia1,bachelard2}).
Therefore, the Lynden-Bell theory does not explain everything. The limitations of the Lynden-Bell theory were
emphasized in \cite{epjb,incomplete}.

Following these observations, it has been proposed \cite{cc,epjb,cstsallis} that Tsallis $q$-distributions may provide a
good fit of the QSSs in certain cases of incomplete relaxation\footnote{This is similar to the original claim of
Latora et al. \cite{latora,lrt}, except that we consider incomplete relaxation towards the Lynden-Bell distribution
(collisionless regime), not towards the Boltzmann distribution (collisional regime). In the first case, the mixing is
due to mean field effects, while in the second case it is due to discreteness (finite $N$) effects. 
This is physically very
different \cite{incomplete}.}. At the same time, it has been
emphasized that this good agreement is not expected to be general, i.e. the Tsallis distributions are {\it not} universal
attractors. Actually, $q$-distributions correspond to what have been called {\it stellar
polytropes} in astrophysics \cite{bt}. They were introduced long ago by
Eddington \cite{eddington} as particular stationary solutions of the
Vlasov equation. They were used to construct simple self-consistent
mathematical models of galaxies.  At some time, they were found to
provide a reasonable fit of some observed star clusters, the so-called
Plummer \cite{plummer} model. Improved observation of globular
clusters and galaxies showed that the fit is not perfect and more
realistic models have been introduced since then \cite{bt}. However,
stellar polytropes are still important in astrophysics for historical reasons and for
their mathematical simplicity. Similarly, we believe that these distributions will
play a useful role in the HMF model. Of course, the relevance (or irrelevance) of $q$-distributions
can only be assessed  by the results of
numerical simulations that we now briefly review.

Campa {et al.} \cite{campa1} performed direct numerical
simulations of the HMF model starting from an initial condition with
magnetization $M_0=1$ and a waterbag velocity distribution
corresponding to an energy $E=0.69$. They obtained a non-magnetized
QSS with a velocity distribution that they called
``semi-ellipse''\footnote{This distribution function differs from the one obtained by  Latora {et al.}
\cite{lrt} for the same initial condition. However, Campa {et al.}
\cite{campa1} showed that the ordinary waterbag initial condition leads
to the presence of large sample to sample fluctuations so that many
averages are necessary. They argued that Latora {et al.}
\cite{lrt} may not have used sufficient averages, and they proposed to use {\it isotropic} waterbag
distributions to reduce the fluctuations.}. Chavanis \cite{hb3} noted
that this distribution is a particular polytropic (Tsallis)
distribution with index $n=1$.  This was the first clear evidence of a
polytropic QSS in the HMF model. Furthermore, this distribution has a
compact support which is very natural in the phenomenology of
incomplete violent relaxation. Chavanis \& Campa \cite{cc} developed a
general theory of polytropic distributions in the context of the HMF
model. In this approach, polytropic distributions are interpreted as
particular steady states of the Vlasov equation, like in astrophysics
\cite{bt}. They studied the dynamical stability problem by using
a ``thermodynamical analogy''  and evidenced a
rich variety of phase transitions depending on the polytropic index
$n$ (or $q$). They computed the physical caloric curves $T_{kin}(E)$
and found that, for $0.54<n<3.56$, these curves display a region of negative
kinetic specific heat $C_{kin}<0$ (see in particular Figure 23 of
\cite{cc}). They proposed that these results could help interpreting
the ``anomalies'' reported by Antoni \& Ruffo \cite{ar} that have
never been explained so far.

Pakter \& Levin \cite{levin} performed direct numerical simulations of the HMF model starting
from a rectangular waterbag distribution with $M_0=0.40$ and different
values of the energy. They found that the QSS has a core-halo
structure\footnote{This core-halo structure has been also observed in early numerical simulations of the
Vlasov equation in 1D and 2D gravity
\cite{hohl68,goldstein69,cuperman69,lecar71,janin71,tanekusa87,mineau,yamaguchi2008,levin2D,jw2011}.}. The
core corresponds to a completely degenerate
distribution in the Lynden-Bell theory (i.e. the ground state at
$T=0$). The halo is interpreted in terms of a parametric
resonance. This core-halo structure is not consistent with the
Lynden-Bell prediction that leads to a {\it partially degenerate} distribution without core-halo structure. In addition,
for $M_0=0.40$, the Lynden-Bell theory predicts a second order phase transition \cite{prl2} while Pakter \& Levin \cite{levin} find
a first order phase transition. We note that the homogeneous core, interpreted as a completely degenerate Lynden-Bell
distribution, is a polytrope $n=1/2$ (waterbag distribution) \cite{hmfq1}. Therefore, the first order phase transition
reported by Pakter \& Levin \cite{levin} may be connected to the first order phase transition found by Chavanis \cite{hmfq1}
for the pure waterbag
distribution (compare Figure 6 of \cite{hmfq1} to Figure 2 of \cite{levin})\footnote{We emphasize, however,
that the two results are independent since there is no core-halo state in the study of \cite{hmfq1}. We only
suggest that the nature of the phase transition (first order) is principally due to the $n=1/2$ polytropic
component (the core in \cite{levin}).}. 

Morita \& Kaneko \cite{mk} performed direct numerical simulations of
the HMF model starting from initial conditions in which the angles and the velocities of the particles have
Boltzmannian distributions with different temperatures. They found initial conditions for which the system does
not relax toward a QSS. In their simulations, the magnetization exhibits persistent
oscillations whose duration diverges with $N$. This long-lasting
periodic or quasi periodic collective motion appears through Hopf
bifurcation and is due to the presence of clumps (high density
regions) in phase space.

This brief review of numerical results in the HMF model shows that the
nature of the QSS crucially depends on the initial condition. The
purpose of this paper is to investigate other initial conditions that
have not been studied previously. We first consider non-magnetized
initial states with a velocity distribution that is (i) gaussian, (ii)
semi-elliptical, and (iii) waterbag. In each case, we determine the
physical caloric curve $T_{kin}(E)$ of the corresponding QSS and plot the distribution function
$f(\theta,v)$ as a function of the individual energy
$\epsilon=v^2/2+\Phi(\theta)$. A steady state of the Vlasov equation
is characterized by a function $f=f(\epsilon)$. In each case, we find
that the caloric curve presents a region of negative kinetic specific
heat: $C_{kin}=dE/dT_{kin}<0$.  In that region, we find that the QSS
is described by a distribution function $f=f(\epsilon)$ that can be
conveniently fitted by a pure polytrope with index (i) $n=2$, (ii)
$n=1$, and (iii) $n=1/2$. Using the values of these indices, we are
able to determine the theoretical caloric curve $T_{kin}(E)$ and explain
the negative kinetic specific heat region found in the numerical
simulations. For lower energies, we find that the system takes a
``core-halo'' structure. The core corresponds to the pure polytrope
discussed above but it is now surrounded by a halo of particles. In
case (iii), the distribution function in the core  is constant ($n=1/2$ polytrope) and
we recover the results of Pakter \& Levin \cite{levin}. We also
investigate the time evolution of the magnetization in the timescale
corresponding to the QSS. We find persistent oscillations similar to
those reported by Morita \& Kaneko \cite{mk}. In cases (i) and (ii),
they are slowly damped, and in case (iii) they are more persistent. We
interpret these oscillations in relation to the Landau damping around
a spatially inhomogeneous QSS. For the spatially homogeneous waterbag
distribution, we know that there is no Landau damping. By continuity,
the damping rate should be small in the inhomogeneous case, and this
may explain the numerical results. We also consider unsteady initial
conditions with $M_0=1$ and isotropic waterbag velocity
distribution. In that case, we find that the system forms phase space
holes and dense filaments that persist for a very long
time. Ultimately, these holes disappear and a QSS is formed. We show
that the kinetic caloric curve is approximately
constant\footnote{Actually, this curve presents an unexpected
hump with a region of negative kinetic specific heat.}, corresponding to a
polytrope with index $n_0\simeq 3.56$. This sensibly differs 
from the results reported in \cite{ar,latora}. Finally, we consider the
collisional evolution of an initially Vlasov stable distribution, and
show that the time-evolving distribution function $f(v,t)$ can be
fitted by a sequence of polytropic distributions with a time-dependent
index $n(t)$ both in the non-magnetized and magnetized regimes.

Before discussing these numerical results (Sections \ref{sec_simul}-\ref{sec_waterbagM1}), we summarize our theory
of polytropes \cite{cc}, and provide additional theoretical results that
were not given in our previous paper (Section \ref{sec_theory}). In particular, we
analyze in more detail the nature of the kinetic caloric curve
$T_{kin}(E)$ as a function of the polytropic index $n$ since these
curves are of considerable importance to interpret the
numerical simulations. Some readers may skip this theoretical part and go directly to Section \ref{sec_simul}
where the results of numerical simulations are presented.

\section{The HMF model}
\label{sec_se}

\subsection{The Hamiltonian equations}
\label{sec_he}

The HMF model can be viewed as a collection of $N$ particles of unit mass moving on a circle and
interacting via a cosine binary potential \cite{cdr}. The
dynamics of these particles is governed by the Hamilton equations
\begin{eqnarray}
\label{he1}
\frac{d\theta_{i}}{dt}=\frac{\partial H}{\partial v_{i}}, \qquad \frac{d v_{i}}{dt}=-\frac{\partial H}
{\partial \theta_{i}},\qquad \\ 
H=\frac{1}{2}\sum_{i=1}^{N}v_{i}^{2}+\frac{1}{2N}\sum_{i,j=1}^N\lbrack 1-\cos(\theta_{i}-\theta_{j})\rbrack,
\end{eqnarray}
where $\theta_{i}\in[0,2\pi[$ is the angle that particle $i$ makes with an axis
of reference $Ox$ and $v_{i}=d\theta_i/dt\in ]-\infty,+\infty[$ is its velocity. The HMF model conserves the energy $H$ and the
number of particles $N$. The order parameter is the magnetization ${\bf M}=(M_x,M_y)$ with components
\begin{eqnarray}
\label{he2}
M_x=\frac{1}{N}\sum_i\cos\theta_i,\qquad  M_y=\frac{1}{N}\sum_i\sin\theta_i.
\end{eqnarray}
The energy per particle $E=H/N$ can be written as
\begin{eqnarray}
\label{he3}
E=\frac{1}{2N}\sum_{i=1}^{N}v_{i}^{2}+\frac{1-M^2}{2},
\end{eqnarray}
where $M=(M_x^2+M_y^2)^{1/2}$ is the modulus of the magnetization. The equations of motion (\ref{he1}) can be written in
terms of the magnetization components $M_x$ and $M_y$ as
\begin{eqnarray}
\label{heeqm}
\frac{d\theta_{i}}{dt} &=& v_i,\\
\frac{d v_{i}}{dt} &=& - M_x \sin \theta_i + M_y \cos \theta_i.
\end{eqnarray}
Following the Kac prescription \cite{kac}, the interaction
strength has been rescaled by a factor $1/N$.  This is the right scaling to properly define the thermodynamic limit
$N\rightarrow +\infty$ of a system with long-range interactions such as the HMF model. In this way, the total energy of the
system $E\sim N$ is extensive. However, the energy remains fundamentally  non-additive \cite{cdr}.

The HMF model can be thought of as a set of $N$ globally coupled rotators or
$XY$-spins ($\theta_i$ represents the orientation of the $i$-th rotator and
$v_i$ is the conjugated momentum). In this respect, it is similar to the $XY$ model though the interaction is extended to all
couples of spins instead of being restricted to nearest neighbors. It is also similar to a one dimensional periodic
self-gravitating system where the potential of interaction has been truncated to the first Fourier mode.

\subsection{The Liouville equation}
\label{sec_mcd}

We introduce the $N$-body distribution function $P_N(\theta_1,v_1,...,\theta_N,v_N,t)$ giving the probability density of finding,
at time $t$, the first particle with position $\theta_1$ and velocity $v_1$, the second particle with position $\theta_2$ and
velocity $v_2$ etc. It is normalized such that $\int P_{N}\prod_{i}d\theta_{i}dv_{i}=1$. For an isolated Hamiltonian system,
such as the HMF model, the evolution of the $N$-body distribution function is governed by the Liouville equation
\begin{eqnarray}
{\partial P_{N}\over\partial t}+\sum_{i=1}^{N}\biggl (v_{i}{\partial P_{N}\over\partial\theta_{i}}+F_{i}
{\partial P_{N}\over\partial v_{i}}\biggr )=0,
\label{he4}
\end{eqnarray}
where $F_{i}=-{\partial U\over\partial\theta_{i}}=-\frac{1}{N}\sum_j \sin(\theta_i-\theta_j)$ is the force acting on
particle $i$ due to the interaction with the other particles. It can be written $F_i=-M_x \sin\theta_i+M_y \cos\theta_i$.
The Liouville equation (\ref{he4}) contains exactly the same information as the $N$-body Hamiltonian system (\ref{he1}).

\subsection{Collisionless evolution: The Vlasov equation}
\label{sec_cev}

For systems with long-range interactions, it can be shown that the mean field approximation is exact at the thermodynamic
limit $N\rightarrow +\infty$ \cite{bh}. This means that the $N$-body distribution function  is a product of $N$ one-body distributions
\begin{eqnarray}
P_{N}(\theta_1,v_1,...,\theta_N,v_N,t)=\prod_i P_1(\theta_i,v_i,t).
\label{cev1}
\end{eqnarray}
Let us introduce the one-body distribution function
$f(\theta,v,t)=\frac{1}{N}\langle \sum_{i=1}^{N}\delta(\theta-\theta_{i})\delta(v-v_{i})\rangle = P_{1}(\theta,v,t)$. It
is normalized such that
\begin{eqnarray}
I[f]\equiv \int f \, d\theta dv=1.
\label{mfa12}
\end{eqnarray}
Substituting the decomposition (\ref{cev1}) in the Liouville equation (\ref{he4}) and integrating over all variables
except $\theta_1$, $v_1$, we find that, for a fixed interval of time $[0,T]$ (any) and $N\rightarrow +\infty$, the evolution of
the distribution function $f(\theta,v,t)$ is governed by the Vlasov equation
\begin{eqnarray}
\label{cev2}
\frac{\partial f}{\partial t}+v\frac{\partial f}{\partial\theta}-\frac{\partial\Phi}{\partial\theta}\frac{\partial f}{\partial v}=0,
\end{eqnarray}
where
\begin{equation}
\Phi(\theta,t)=\int_{0}^{2\pi}\lbrack 1-\cos(\theta-\theta')\rbrack \rho(\theta',t)\, d\theta',\label{cev3}
\end{equation}
is the self-consistent potential generated by the density of particles $\rho(\theta,t)=\int f(\theta,v,t)\, dv$. The density
of particles is defined by $\rho(\theta,t)=\frac{1}{N}\langle \sum_{i=1}^{N}\delta(\theta-\theta_{i})\rangle = P_{1}(\theta,t)$
and it is normalized such that $\int\rho\, d\theta=1$. The mean force acting on a particle located at $\theta$ is
$\langle F\rangle(\theta,t)=-\partial\Phi/\partial\theta=-\int_0^{2\pi} \sin(\theta-\theta')\rho(\theta',t)\, d\theta'$.
Expanding the cosine function in equation (\ref{cev3}), we obtain
\begin{eqnarray}
\label{cev4}
\Phi(\theta,t)=1-M_x(t)\cos\theta-M_y(t) \sin\theta,
\end{eqnarray}
where
\begin{eqnarray}
\label{cev5}
M_x(t)=\int\rho(\theta,t)\cos\theta\, d\theta,
\end{eqnarray}
\begin{eqnarray}
\label{cev6}
M_y(t)=\int\rho(\theta,t)\sin\theta\, d\theta,
\end{eqnarray}
are the components of the mean magnetization ${\bf M}(t)$. The mean force acting on a particle can be written
$F(\theta,t)=-M_x(t)\sin\theta+M_y(t)\cos\theta$.

The Vlasov equation is a purely mean field equation ignoring collisions (more properly, correlations, graininess, finite
$N$-effects...) between particles. It governs the collisionless evolution of the HMF model.

The Vlasov equation admits an infinite number of stationary solutions. Any spatially homogeneous distribution
function  $f=f(v)$ is a steady state of the Vlasov equation. On the other hand, spatially inhomogeneous distributions of
the form $f=f(\epsilon)$, where $\epsilon=v^2/2+\Phi(\theta)$ is the individual energy, are also steady states of the Vlasov equation.

\subsection{The mean field energy}
\label{sec_mfa}

In the mean field approximation, the energy of the system can be written as
\begin{eqnarray}
\label{mfa3}
E[f]=\int f \frac{v^2}{2}\, d\theta dv+\frac{1}{2}\int \rho\Phi\, d\theta=E_{kin}+W,
\end{eqnarray}
where $E_{kin}$ is the kinetic energy and $W$ the potential energy. Using equations (\ref{cev4})-(\ref{cev6}), the potential
energy can be expressed in terms of the magnetization as
\begin{eqnarray}
\label{mfa4}
W=\frac{1-M^2}{2}.
\end{eqnarray}
The average kinetic temperature $T_{kin}$ is defined by
\begin{eqnarray}
E_{kin}=\frac{1}{2}T_{kin}.
\label{mfa5}
\end{eqnarray}
Therefore, the energy can be rewritten
\begin{eqnarray}
E=\frac{1}{2}T_{kin}+\frac{1-M^2}{2}.
\label{mfa7}
\end{eqnarray}
For a fixed value of the energy, the kinetic temperature directly determines the magnetization, and {\it vice versa}. The local
pressure is defined by
\begin{eqnarray}
\label{mfa8}
p(\theta,t)=\int f v^2\, dv.
\end{eqnarray}
The kinetic energy can therefore be written
\begin{eqnarray}
E_{kin}=\frac{1}{2}\int p\, d\theta.
\label{mfa9}
\end{eqnarray}
This expression will be useful in the following.

\subsection{Incomplete violent relaxation}
\label{sec_incomplete}

Starting from an unstable or unsteady initial condition, the Vlasov equation is expected to reach, on a coarse-grained
scale, a QSS. This is called weak convergence in mathematics. This QSS is a stable steady state of the Vlasov equation.
Since the Vlasov equation admits an infinity of steady states, the prediction of the QSS actually reached by the system
is difficult. A prediction can be made based on Lynden-Bell's statistical theory of violent relaxation. However, this
theory assumes that the evolution of the system is ergodic, or at least that the mixing is efficient. There are cases
where the Lynden-Bell theory gives a good prediction. However, there also exist cases where this prediction fails. In
case of incomplete relaxation, the system can be stuck in a stable steady state of the Vlasov equation that is not the
most mixed state, i.e. that differs from the Lynden-Bell distribution. In this paper, we consider a particular class
of steady states of the Vlasov equation, called polytropic distributions, that may arise in case of incomplete relaxation.
They are associated with the Tsallis ``entropy'' (in the sense given below).

\section{Reminders and complements in the theory of polytropes}
\label{sec_theory}

The theory of polytropes for the HMF model has been developed in \cite{cc}. Since this theory is rather rich
(several cases arise depending on the polytropic index) and not well-known, we provide here a summary of the main results.
In complement to our previous paper, we (i) use more conventional notations, (ii) simplify some important
formulae, (iii) treat a case that was forgotten, (iv) describe in detail the physical caloric curve $T_{kin}(E)$ that will
be needed to interpret our numerical results, and (v) specifically consider the waterbag distribution which is a particular
polytrope of index $n=1/2$.

\subsection{Polytropic distributions in phase space}
\label{sec_ps}

The Tsallis entropy is defined by
\begin{equation}
S[f]=-\frac{1}{q-1}\int (f^q-f)\, d\theta dv,   \label{tte1}
\end{equation}
with $q>0$. For $q\rightarrow 1$, it reduces to the Boltzmann entropy
\begin{equation}
S[f]=-\int f\ln f\, d\theta dv.   \label{tte2}
\end{equation}
We consider the microcanonical problem
\begin{equation}
\label{tte3}
\max_{f}\quad \lbrace S[f]\quad |\quad E[f]=E, \quad I[f]=1 \rbrace,
\end{equation}
and the canonical problem
\begin{equation}
\label{tte4}
\min_{f}\quad \lbrace F[f]=E[f]-T S[f]\quad |\quad I[f]=1\rbrace.
\end{equation}
As explained in \cite{cc,ccstab}, these optimization  problems determine particular steady states of the Vlasov equation
of the form $f=f(\epsilon)$ with $f'(\epsilon)<0$ that are dynamically stable. In this {\it dynamical} interpretation,
$S$ is a pseudo entropy, $F$ is a pseudo free energy, and $T>0$ is a pseudo thermodynamical temperature. By an abuse of
language, and to simplify the terminology, we shall omit the prefix ``pseudo'' in the sequel. However, we must keep in mind
that all the references to thermodynamics in our study are purely effective: we are dealing with dynamical stability,
not thermodynamical stability. Nevertheless, it is convenient to develop a thermodynamical analogy and use a similar
terminology\footnote{Tsallis and co-workers have tried to give a real {\it thermodynamical} interpretation to the
functional (\ref{tte1}). Here, we just consider its Vlasov dynamical stability interpretation \cite{cc,ccstab} and refer
to \cite{cstsallis,assise} for other interpretations.}.

We recall that canonical stability (criterion (\ref{tte4})) implies microcanonical stability 
(criterion (\ref{tte3})) but the converse is wrong in case of ensembles inequivalence
\cite{ellis}. Furthermore, we recall that the thermodynamic-looking optimization problems (\ref{tte3}) and (\ref{tte4})
provide just {\it sufficient} conditions of Vlasov dynamical stability. More refined dynamical stability criteria are
discussed in \cite{ccstab}.

The critical points of entropy at fixed energy and normalization are determined by the variational principle
\begin{equation}
\delta S-\beta\delta E-\alpha\delta I=0,  \label{ps1}
\end{equation}
where $\beta=1/T$ and $\alpha$  are Lagrange multipliers ($T$ is the thermodynamical temperature and $\alpha$ the chemical
potential). This yields the Tsallis $q$-distributions
\begin{equation}
f(\theta,v)=\left\lbrace \mu-\frac{(q-1)\beta}{q}\left\lbrack \frac{v^2}{2}+\Phi(\theta)\right\rbrack\right\rbrace_+^{1/(q-1)}, \label{ps2}
\end{equation}
where $\mu=\lbrack 1-(q-1)\alpha\rbrack/q$. The notation $\lbrack x\rbrack_+$  stands for $\lbrack x\rbrack_+=x$ if $x\ge 0$
and $\lbrack x\rbrack_+=0$ if $x\le 0$. These distributions are also obtained as critical points of the free energy at
fixed normalization, satisfying $\delta F+\alpha T\delta I=0$. In astrophysics, they are known
as ``polytropic distributions'' \cite{bt}. They are generally labeled by a polytropic index $n$ that is related to the
parameter $q$ by the relation
\begin{equation}
n=\frac{1}{2}+\frac{1}{q-1}.  \label{ps3}
\end{equation}
For $n=1/2$ ($q\rightarrow +\infty$), the polytropic distribution function
(\ref{ps2}) reduces  to the waterbag,
or Fermi, distribution (see Section \ref{sec_waterbag}). For $n\rightarrow +\infty$ ($q\rightarrow 1$), it reduces to the
isothermal (Maxwell-Boltzmann) distribution
\begin{equation}
f(\theta,v)=A'e^{-\beta\left \lbrack \frac{v^2}{2}+\Phi(\theta)\right \rbrack}.\label{ps3b}
\end{equation}

We need to distinguish two cases depending on the sign of $q-1$. (i) For $q>1$ ($n\ge 1/2$), the
distribution function can be written as
\begin{eqnarray}
f=A(\epsilon_{m}-\epsilon)_+^{1\over q-1},
\label{ps4}
\end{eqnarray}
where we have set $A=\lbrack\beta(q-1)/q\rbrack^{1\over q-1}$ and
$\epsilon_{m}=q\mu/\lbrack \beta(q-1)\rbrack$. Such distributions have a compact support in phase space since they vanish
for $\epsilon\ge \epsilon_m$. At a given position $\theta$, the distribution function vanishes at
$v= v_{m}(\theta)=\sqrt{2(\epsilon_{m}-\Phi(\theta))}$. For $n=1/2$, $\epsilon_m$ and $v_m(\theta)$ correspond to the Fermi
energy and to the Fermi velocity, respectively.  (ii) For $0<q<1$ ($n<-1/2$), the distribution function can be written as
\begin{eqnarray}
f=\frac{A}{(\epsilon_{0}+\epsilon)^{1\over 1-q}},
\label{ps5}
\end{eqnarray}
where we have set $A=\lbrack\beta(1-q)/q\rbrack^{1\over q-1}$ and
$\epsilon_{0}=q\mu/\lbrack\beta(1-q)\rbrack$. Such distributions are defined for all individual energies. At a given
position $\theta$, the distribution function behaves, for large velocities, as
$f\sim v^{-2/(1-q)}\sim v^{-(1-2n)}$.  In the following, we shall only consider
distribution functions for which the density and the pressure are finite. This implies $1/3<q<1$ ($n<-1$).

Polytropic distributions may arise as a result of incomplete relaxation. In that case, the system does not mix well enough
to justify the establishment of the Lynden-Bell distribution which corresponds to the  most mixed state. In general, incomplete
relaxation manifests itself by the fact that mixing takes place only in a {\it subdomain} of phase space. This generally
leads to distribution functions with a compact support (the distribution function vanishes above a certain energy
$\epsilon_m$) while the distribution function predicted by Lynden-Bell remains strictly positive for all values of the
energy (which is rather unphysical). Polytropes with index $n\ge 1/2$ have this property of confinement and this is why, in
our opinion, they play an important
role in case of incomplete relaxation\footnote{Of course, polytropic distributions are not the only distributions with a
compact support and, indeed, we will find that the QSSs may be described by more complicated distributions
(e.g. core-halo states). In other words, Tsallis distributions are not universal attractors in case of incomplete relaxation.
However, we have suggested \cite{epjb,cc} that polytropic distributions may provide a good fit of QSSs in certain cases,
and we will give numerical evidence of that claim in Sections \ref{sec_simul}-\ref{sec_waterbagM1}.}. Therefore, in the following,
we shall essentially consider
the case $n\ge 1/2$.

\subsection{Polytropic equation of state}
\label{sec_pes}

To any distribution function of the form $f=f(\epsilon)$ with $f'(\epsilon)<0$, we can associate a corresponding barotropic
gas by defining the density $\rho=\int f\, dv=\rho(\Phi)$ and the pressure $p=\int f v^2\, dv=p(\Phi)$, and eliminating
$\Phi(\theta)$ between these two expressions. This determines a barotropic equation of state $p=p(\rho)$. For the
polytropic distribution (\ref{ps2}), we get \cite{cc}:
\begin{eqnarray}
p=K\rho^{\gamma}, \qquad \gamma=1+{1\over n},
\label{pes5}
\end{eqnarray}
with
\begin{eqnarray}
K={1\over n+1}\biggl \lbrace \sqrt{2}A {\Gamma(1/2)\Gamma(1/2+n)\over \Gamma(n+1)}
\biggr \rbrace^{-{1\over n}},
\label{pes6}
\end{eqnarray}
for $n\ge 1/2$ and
\begin{eqnarray}
K=-{1\over n+1}\biggl \lbrace \sqrt{2}A {\Gamma(-n)\Gamma(1/2)\over\Gamma(1/2-n)}\biggr \rbrace^{-{1\over n}}.
\label{pes7}
\end{eqnarray}
for $n<-1$. Equation (\ref{pes5}) is the well-known polytropic equation of state. This is the reason why the
distributions (\ref{ps2}) are called polytropic distributions. The polytropic constant $K$ is sometimes called
the ``polytropic temperature''. We can show that $K$ is a monotonically increasing function of the thermodynamical
temperature $T$ \cite{cc}. For isothermal systems ($q=1$, $n=\infty$, $\gamma=1$),  we have $K=T$.

\subsection{Polytropic distributions in physical space}
\label{sec_pp}

The density is obtained by integrating Eq. (\ref{ps2}) over the velocity. We find that the density is related to the
potential $\Phi(\theta)$  by \cite{cc}:
\begin{equation}
\rho(\theta)=\left\lbrack \lambda-\frac{\gamma-1}{K\gamma}\Phi(\theta)\right
\rbrack_+^{\frac{1}{\gamma-1}},
\label{pp1}
\end{equation}
where $\lambda=\epsilon_m/(K(n+1))$ for $n\ge 1/2$ and $\lambda=\epsilon_0/(-K(n+1))$ for $n<-1$. For $\gamma=1$,
Eq. (\ref{pp1}) reduces  to the isothermal (Boltzmann) distribution
\begin{equation}
\rho(\theta)=Ae^{-\beta\Phi(\theta)},
\label{pp1b}
\end{equation}
with $A=(2\pi/\beta)^{1/2}A'$. The polytropic distribution
in physical space  $\rho=\rho(\Phi)$ given by Eq. (\ref{pp1}) has the same mathematical form as the polytropic
distribution in phase space $f=f(\epsilon)$ given by Eq. (\ref{ps2}) with  $\gamma$ playing the role of $q$ and $K$
playing the role of $T=1/\beta$. In this correspondence, $\gamma$ is related to $q$ by Eqs. (\ref{ps3}) and (\ref{pes5})
leading to $\gamma=(3q-1)/(q+1)$ and $K$ is related to $T$ by Eqs. (\ref{pes6}) and (\ref{pes7}). Polytropic
distributions (including the isothermal distribution) are apparently the only
distributions for which $f(\epsilon)$ and $\rho(\Phi)$ have the same mathematical form.

Using Eq. (\ref{pp1}), the polytropic distribution function (\ref{ps2}) can be rewritten \cite{cc}:
\begin{eqnarray}
f(\theta,v)={1\over Z}\biggl \lbrack \rho({\theta})^{1/n}-{v^{2}/2\over (n+1)K}\biggr\rbrack_+^{n-1/2},
\label{eqvp1}
\end{eqnarray}
where $Z$ is given for $n\ge 1/2$ by
\begin{eqnarray}
Z=\sqrt{2}{\Gamma(1/2)\Gamma(1/2+n)\over\Gamma(n+1)}\lbrack K(n+1)\rbrack^{1/2},
\label{eqvp2}
\end{eqnarray}
and for $n<-1$ by
\begin{eqnarray}
Z=\sqrt{2}{\Gamma(-n)\Gamma(1/2)\over\Gamma(1/2-n)}\lbrack -K(n+1)\rbrack^{1/2}.
\label{eqvp3}
\end{eqnarray}
For $n\rightarrow +\infty$, we recover the isothermal distribution
\begin{eqnarray}
f(\theta,v)=\left (\frac{\beta}{2\pi}\right )^{1/2}\rho(\theta)\, e^{-\beta \frac{v^2}{2}}.
\label{eqvp1max}
\end{eqnarray}
Other useful expressions of the polytropic distribution function are given in \cite{cc}.

\subsection{Homogeneous phase: Jeans-type instability}
\label{sec_hpjt}

It is convenient to define the polytropic temperature by
\begin{eqnarray}
\Theta=K\left (\frac{1}{2\pi}\right )^{\gamma-1}.
\label{hpjt2}
\end{eqnarray}
In the homogeneous phase ($M=0$), using Eqs. (\ref{eqvp1}) and (\ref{hpjt2}), the polytropic (Tsallis) distributions
can be written
\begin{eqnarray}
f(v)=C\left \lbrack 1-\frac{v^2}{2(n+1)\Theta}\right \rbrack_+^{n-1/2},
\label{hpjt1}
\end{eqnarray}
where $C$ is given for $n\ge 1/2$ by
\begin{eqnarray}
C=\frac{1}{2\pi}\frac{\Gamma(n+1)}{\Gamma(1/2)\Gamma(1/2+n)}\frac{1}{\sqrt{2(n+1)\Theta}},
\label{eqvp2b}
\end{eqnarray}
and for $n<-1$ by
\begin{eqnarray}
C=\frac{1}{2\pi}\frac{\Gamma(1/2-n)}{\Gamma(1/2)\Gamma(-n)}\frac{1}{\sqrt{-2(n+1)\Theta}}.
\label{eqvp3c}
\end{eqnarray}
For $n\ge 1/2$, the maximum velocity is $v_{max}=\sqrt{2(n+1)\Theta}$. For $n\rightarrow +\infty$, Eq. (\ref{hpjt1})
reduces to the isothermal (Maxwell) distribution
\begin{eqnarray}
f(v)=\left (\frac{\beta}{2\pi}\right )^{1/2}\rho\, e^{-\beta \frac{v^2}{2}}.
\label{eqvp1maxhom}
\end{eqnarray}
On the other hand, using Eqs. (\ref{mfa9}), (\ref{pes5}) and (\ref{hpjt2}), we find that the energy can be written as
\begin{equation}
E=\frac{1}{2}\Theta+\frac{1}{2}.
\label{hpjt4}
\end{equation}
Therefore, in the homogeneous phase, the kinetic temperature coincides with the polytropic temperature:
\begin{equation}
T_{kin}=\Theta.
\label{hpjt4b}
\end{equation}
In addition, the caloric curve $\Theta(E)$ is the same for all indices $n$. It is defined for $T_{kin}=\Theta\ge 0$ and
$E\ge 1/2$. The specific heat is $C=dE/d\Theta=dE/dT_{kin}=1/2$. The equality (\ref{hpjt4b}) is not true anymore in the inhomogeneous phase.

For the polytropic equation of state (\ref{pes5}), the square of the velocity of sound in the homogeneous phase is given by
\begin{eqnarray}
c_{s}^{2}=K\gamma\rho^{\gamma-1}=\gamma \Theta.
\label{hpjt6}
\end{eqnarray}
It can be shown that a spatially homogeneous distribution function $f(v)$ with a single maximum at $v=0$ is dynamically
stable if, and only, if
\begin{eqnarray}
1+\pi\int_{-\infty}^{+\infty} {{f'(v)}\over v}dv>0.
\label{sh2}
\end{eqnarray}
For a distribution function of the form $f=f(v^2)$ with $f'(v^2)<0$, the stability criterion can also be written as
\begin{eqnarray}
c_{s}^2>\frac{1}{2}.
\label{corrg4}
\end{eqnarray}
These two stability criteria are equivalent and they determine the spectral, linear and formal stability of the
distribution \cite{cvb,ccstab,yamaguchi,cd}. According to these stability criteria, the homogeneous phase of a
polytropic distribution is stable if \cite{cvb,cc}:
\begin{eqnarray}
\Theta>\Theta_c\equiv \frac{1}{2\gamma},\qquad E>E_c=\frac{1}{4\gamma}+\frac{1}{2},
\label{hpjt7}
\end{eqnarray}
and unstable otherwise. For isothermal distributions ($n=\infty$, $\gamma=1$, $q=1$), we recover the critical
values $(T_c,E_c)=(1/2,3/4)$. For the waterbag distribution ($n=1/2$, $\gamma=3$, $q=+\infty$), we get
$(\Theta_c,E_c)=(1/6,7/12)$. For the semi-ellipse ($n=1$, $\gamma=2$, $q=3$), we obtain $(\Theta_c,E_c)=(1/4,5/8)$.
As shown in previous works \cite{ik,inagaki,cvb,cd}, the instability below $E_c$ is similar to the Jeans instability
in astrophysics \cite{bt}.

\subsection{Inhomogeneous phase: complete and incomplete polytropes}
\label{sec_cip}

We now consider spatially inhomogeneous polytropic distributions. We can assume without loss of generality that the
distribution is symmetrical with respect to the $x$-axis ({\it i.e.} with respect to the angle $\theta=0$). In that case,
the potential can be written $\Phi(\theta)=1-M\cos\theta$
where $M=\int_{0}^{2\pi}\rho(\theta)\cos\theta\, d\theta$ is the magnetization ($M_y=0$ and $M=M_x$).
The density profile (\ref{pp1}) can be rewritten \cite{cc}:
\begin{equation}
\rho(\theta)=A\left (\kappa+\frac{x}{n+1}\cos\theta\right )_{+}^{n}.
\label{cip6}
\end{equation}
where we have noted
\begin{equation}
x=\frac{M}{KA^{1/n}}.
\label{cip5}
\end{equation}
We must consider two cases $\kappa=\pm 1$. As realized in \cite{hmfq1}, the case $\kappa=-1$ was forgotten
in \cite{cc}. However, this forgetfulness does not alter the study of phase transitions made in \cite{cc} since
the branch $\kappa=-1$ only completes the caloric curve up to the ground state $E=0$.

For $x>0$, the density profile is concentrated around $\theta=0$ and for $x<0$, we get a symmetrical density profile
concentrated around $\theta=\pi$. We can therefore restrict ourselves to $x\ge 0$. For $x=0$, we recover the homogeneous
distribution $\rho=1/(2\pi)$. For $x>0$, the density profile is monotonically decreasing. Let us first assume
$n\ge 1/2$ (i.e. $1\le \gamma\le 3$) and $\kappa=+1$. If $x<x_c\equiv n+1=\gamma/(\gamma-1)$, the density is strictly
positive on $-\pi\le\theta\le\pi$ (incomplete polytrope). If $x>x_c$, the density has a compact support
(complete polytrope). In that case, it vanishes for $|\theta|\ge \theta_c=\arccos(-x_c/x)$. We note that
$\theta_c\ge \pi/2$. For   $x\rightarrow +\infty$, the density profile is given by
$\rho(\theta)=\Gamma((2+n)/2)/\lbrack \sqrt{\pi}\Gamma((1+n)/2)\rbrack \cos^n\theta$ \cite{cc}. We now assume
$n\ge 1/2$ and $\kappa=-1$. The central density is defined only for $x>x_c$. In that case, the polytrope is complete:
the density vanishes for $|\theta|\ge \theta_c=\arccos(x_c/x)$. We note that $\theta_c\le \pi/2$. For $x\rightarrow x_c$,
the distribution tends to a Dirac peak $\rho(\theta)=\delta(\theta)$. Finally, we consider the case $n<-1$
(i.e. $0\le \gamma\le 1$) and $\kappa=+1$.  The central density is defined only for $x<x_c=|n+1|=|\gamma/(\gamma-1)|$
and the polytrope is incomplete. Some typical density profiles are represented below in Figures \ref{profilesN1} and \ref{profiles}.

The amplitude $A$ of the density profile  (\ref{cip6}) is determined by the normalization condition $\int\rho\, d\theta=1$,
leading to
\begin{equation}
A=\frac{1}{2\pi I_{\gamma,0}(x)}.
\label{mag6}
\end{equation}
where we have introduced the $\gamma$-deformed modified Bessel functions \cite{cc}:
\begin{equation}
I_{\gamma,m}(x)=\frac{1}{2\pi}\int_{-\pi}^{\pi}\left (\kappa+\frac{\gamma-1}
{\gamma}x\cos\theta\right )_{+}^{\frac{1}{\gamma-1}}\cos(m\theta)\, d\theta.
\label{mag3}
\end{equation}
On the other hand, substituting Eq. (\ref{cip6}) in Eq. (\ref{cev5}) for the magnetization, we obtain the self-consistency relation
\begin{equation}
M=\frac{I_{\gamma,1}(x)}{I_{\gamma,0}(x)}.
\label{mag7}
\end{equation}
Combining Eqs. (\ref{cip5}), (\ref{mag6}) and (\ref{mag7}), we find that the polytropic temperature (\ref{hpjt2}) is given by
\begin{equation}
{\Theta}=\frac{1}{x}\frac{I_{\gamma,1}(x)}{I_{\gamma,0}(x)^{2-\gamma}}.
\label{mag9}
\end{equation}
Finally, after some calculations \cite{cc}, we can show that the kinetic temperature and the
energy are given by
\begin{eqnarray}
T_{kin}=\frac{\gamma-1}{\gamma}M^2+\kappa\frac{M}{x}.
\label{tpp1}
\end{eqnarray}
\begin{eqnarray}
E=\frac{1}{2}T_{kin}+\frac{1-M^2}{2}.
\label{tpp2}
\end{eqnarray}
Equation (\ref{tpp1}) is equivalent to equation (109) of \cite{cc} after a straightforward simplification. From equations
(\ref{mag7})-(\ref{tpp2}), we can obtain the curves $\Theta(E)$, $M(\Theta)$, $M(E)$, and $T_{kin}(E)$ in parametric form
with parameter $x$.  Some of these curves have been analyzed in detail in \cite{cc}. This analysis will be completed
in Section \ref{sec_phy}.

Close to the bifurcation point ($E_c,\Theta_c)$, i.e. for $x\rightarrow 0$, using the asymptotic expansions given in \cite{cc},
we find that the thermodynamical specific heat $C={dE}/{d\Theta}$ is given by
\begin{eqnarray}
C=-\frac{2\gamma^2-5\gamma-2}{2(2-\gamma)}.
\label{ze10}
\end{eqnarray}
Let us introduce the critical indices
\begin{eqnarray}
\gamma_*=\frac{5+\sqrt{41}}{4}\simeq 2.8507811...
\label{ze7}
\end{eqnarray}
\begin{eqnarray}
n_*=\frac{4}{1+\sqrt{41}}\simeq 0.54031242...
\label{ze8}
\end{eqnarray}
The specific heat of the inhomogeneous phase at the bifurcation point is positive for $n>1$, negative for $n_*<n<1$ and
positive again for $1/2\le n<n_*$. It is infinite for $n=1$ and vanishes for $n=n_*$. For $n<-1$, the specific heat is
always positive. These results explain the behavior of the caloric curve $\Theta(E)$ close to the bifurcation
point \cite{cc}. For the waterbag distribution ($n=1/2$, $\gamma=3$, $q=+\infty$), we get $C=1/2$ like in the
homogeneous phase \cite{hmfq1}. For the semi-ellipse ($n=1$, $\gamma=2$, $q=3$), we obtain $C=\infty$. On the other
hand, close to the critical point, the magnetization is given as a function of the energy by
\begin{eqnarray}
M^2=\frac{8\gamma}{2\gamma^2-5\gamma-2}(E-E_c).
\label{ze8bis}
\end{eqnarray}

For isothermal distribution ($n\rightarrow +\infty$), the density profile (\ref{cip6}) takes the form
\begin{equation}
\rho(\theta)=\frac{1}{2\pi I_{0}(x)}e^{\beta M\cos\theta},
\label{mag6b}
\end{equation}
with $x={M}/{T}$. The self-consistency relation (\ref{mag7}) reduces to
\begin{equation}
M=\frac{I_{1}(x)}{I_{0}(x)}.
\label{mag7b}
\end{equation}
Finally, the temperature (\ref{mag9}) or (\ref{tpp1}) and the energy (\ref{tpp2}) become
\begin{eqnarray}
T=T_{kin}=\frac{M}{x},
\label{tpp1b}
\end{eqnarray}
\begin{eqnarray}
E=\frac{1}{2}T+\frac{1-M^2}{2}.
\label{tpp2b}
\end{eqnarray}
Close to the bifurcation point ($E_c,T_c)$=($1/2$, $3/4$), the specific heat is $C={dE}/{dT}=5/2$ and the magnetization
is $M=\sqrt{8/5}(E_c-E)^{1/2}$. This returns the well-known results of the Boltzmann thermodynamical
analysis (see, e.g., \cite{cvb}).

\subsection{The physical caloric curve}
\label{sec_phy}

In our previous paper \cite{cc}, we have plotted the {\it thermodynamical caloric curves} giving the thermodynamical
temperature $T$, or the polytropic temperature $\Theta$, as a function of the energy $E$. This is the correct way to
determine the stability of the system in canonical and microcanonical ensembles and study phase
transitions\footnote{We recall again that we are actually considering the Vlasov dynamical stability of
polytropic distributions using a thermodynamical analogy.}. However, the temperature that is directly accessible to
the experiments or to the numerical simulations is the kinetic temperature $T_{kin}$. In general,
$T_{kin}\neq T=1/\beta$ in a QSS. For comparison with the numerical
results of Sections \ref{sec_simul}-\ref{sec_waterbagM1}, it is useful to study the {\it physical caloric curve}
$T_{kin}(E)$ as a function of the
polytropic index $n$. This study has been only partly done in our previous paper (see Figures 6 and 23 of \cite{cc})
and it is here systematically continued.

\begin{figure}
\begin{center}
\includegraphics[clip,scale=0.3]{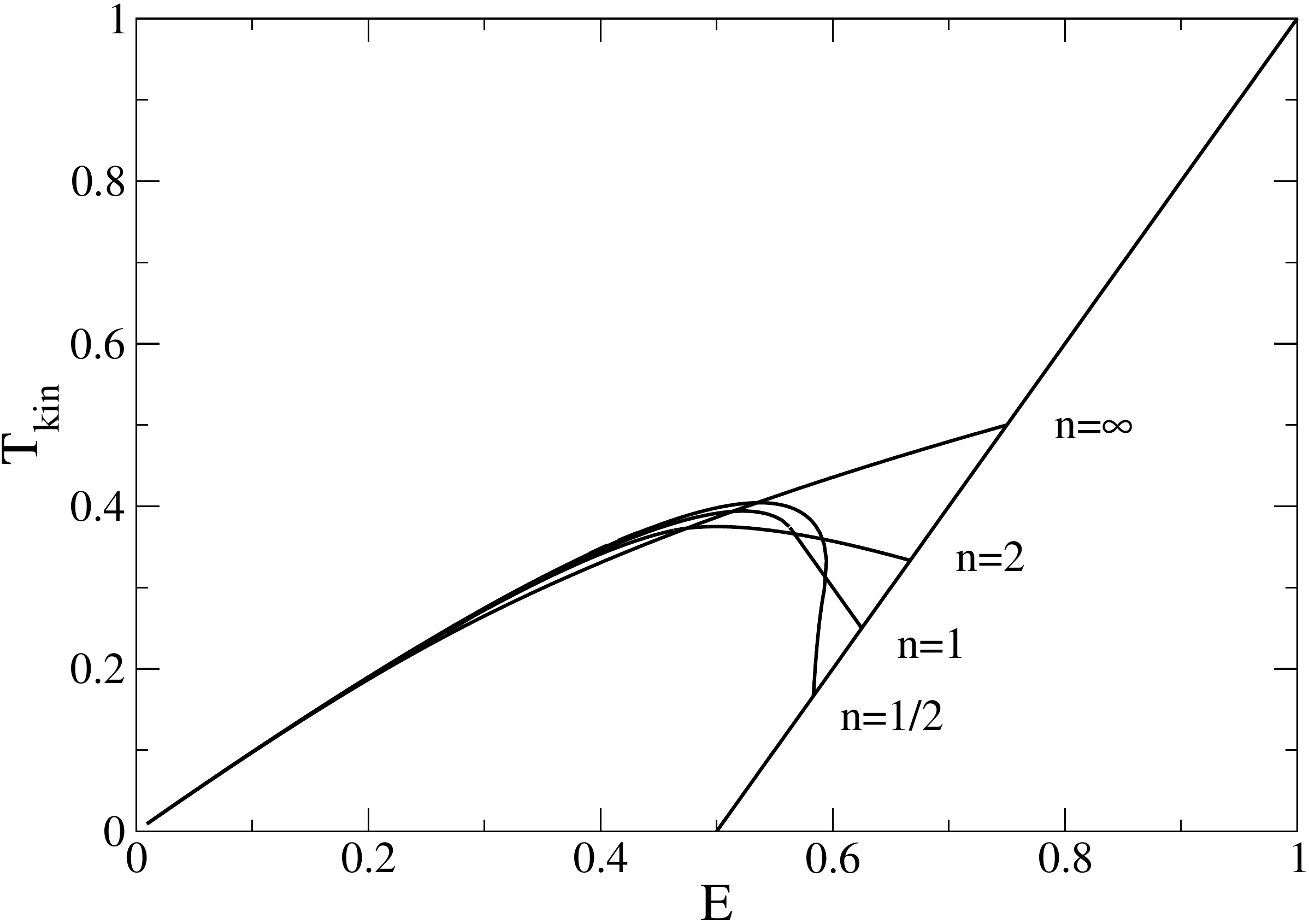}
\caption{The physical caloric curve for different values of $n$.}
\label{caloTkinTOTAL}
\end{center}
\end{figure}

Close to the bifurcation point ($E_c,T_{kin}^c)$, i.e. for $x\rightarrow 0$, using the asymptotic expansions given
in \cite{cc}, we find that the physical (or kinetic) specific heat $C_{kin}={dE}/{dT_{kin}}$ is given by
\begin{eqnarray}
C_{kin}=\frac{2\gamma^2-5\gamma-2}{2(2\gamma^2-\gamma-2)}.
\label{phy5}
\end{eqnarray}
Let us introduce the critical indices
\begin{eqnarray}
\gamma_0=\frac{1+\sqrt{17}}{4}\simeq 1.2807764...
\label{phy1}
\end{eqnarray}
\begin{eqnarray}
n_0=\frac{4}{\sqrt{17}-3}\simeq 3.5615528...
\label{phy2}
\end{eqnarray}
The kinetic specific heat of the inhomogeneous phase at the bifurcation point is positive for $n>n_0$, negative for
$n_*<n<n_0$ and positive again for $1/2\le n<n_*$. It is infinite for $n=n_0$ and vanishes for $n=n_*$. For $n<-1$, the
kinetic specific heat is always positive. These results explain the
behavior of the physical caloric curve $T_{kin}(E)$ close to the
bifurcation point (see  Figure \ref{caloTkinTOTAL}). We note that for $1<n<n_0$, the solutions close to the bifurcation
point have negative kinetic specific heat  $C_{kin}=dE/dT_{kin}<0$ although they have positive thermodynamical specific
heat $C=dE/dT>0$ \cite{cc}. Therefore, one may observe negative kinetic specific heats although the ensembles are equivalent. A
similar observation has been made in \cite{stfcn} for the Lynden-Bell distribution.

\begin{figure}
\begin{center}
\includegraphics[clip,scale=0.3]{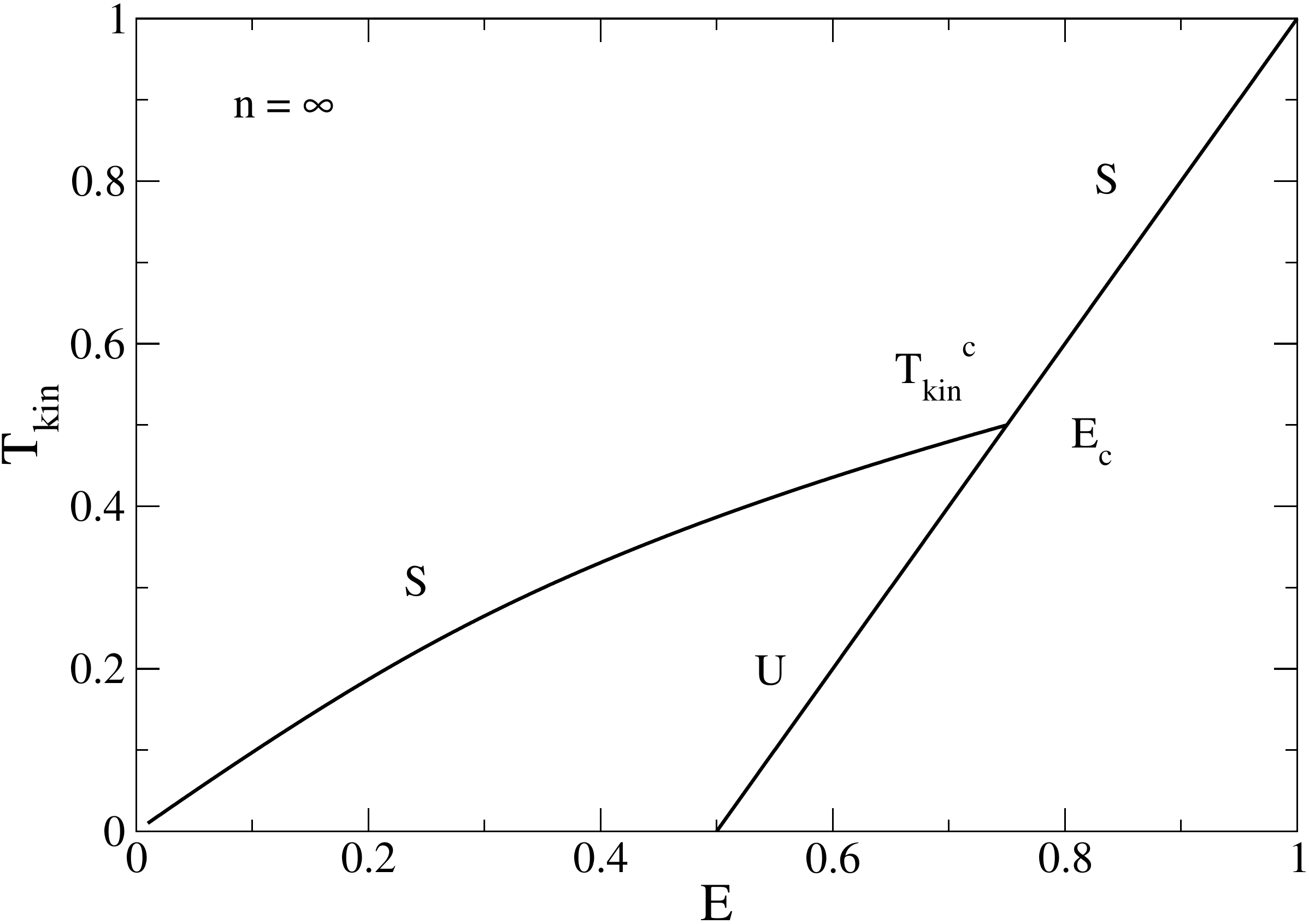}
\caption{The physical caloric curve for $n>n_0$ (specifically $n=\infty$). The kinetic specific heat is positive. The physical
caloric curve exhibits a second order phase transition at $E_c$.}
\label{caloTkinNinfini}
\end{center}
\end{figure}

\begin{figure}
\begin{center}
\includegraphics[clip,scale=0.3]{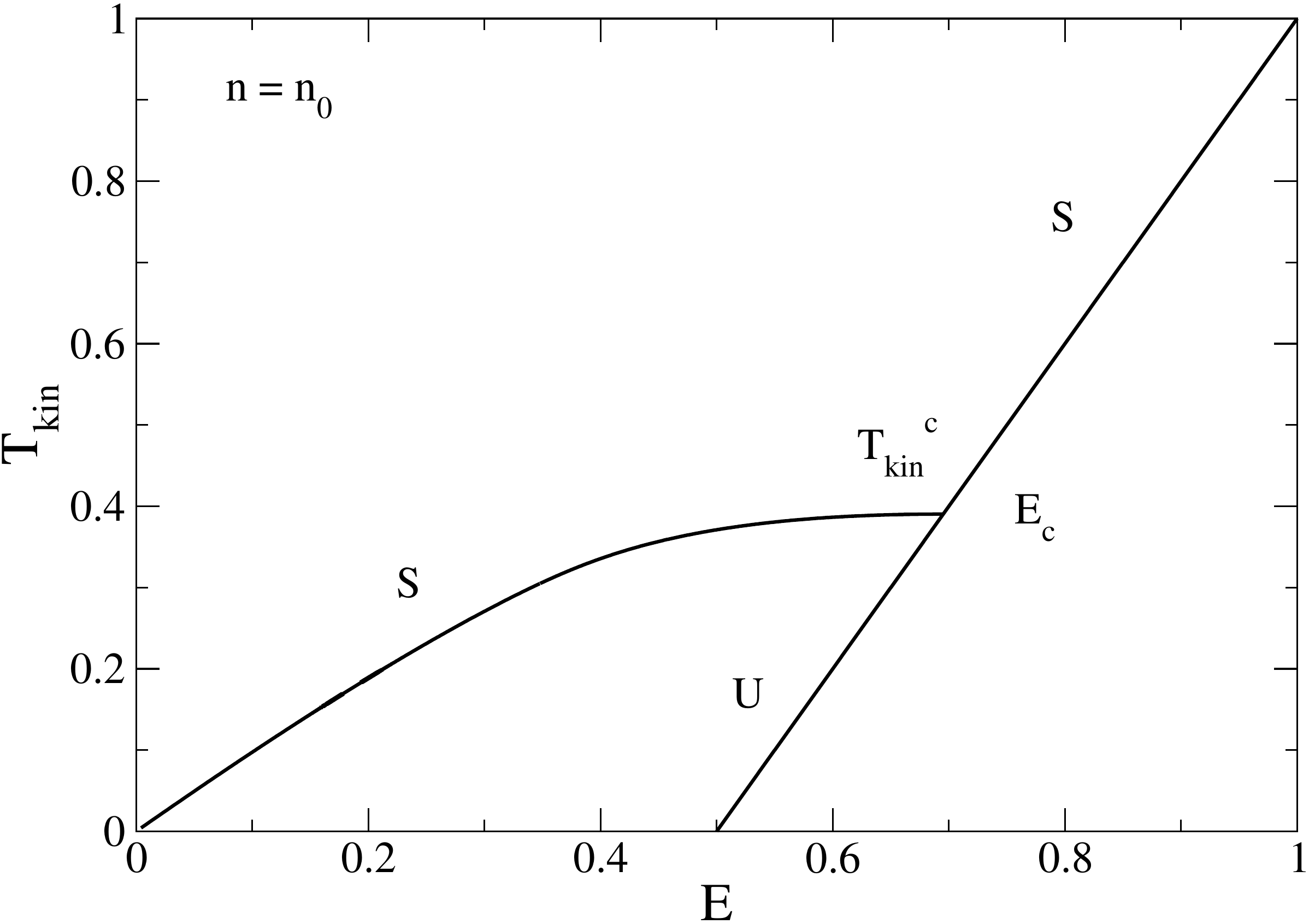}
\caption{The physical caloric curve for $n=n_0\simeq 3.56$. The kinetic specific heat close to the bifurcation point is infinite.}
\label{caloTkinN3p5615528}
\end{center}
\end{figure}

\begin{figure}
\begin{center}
\includegraphics[clip,scale=0.3]{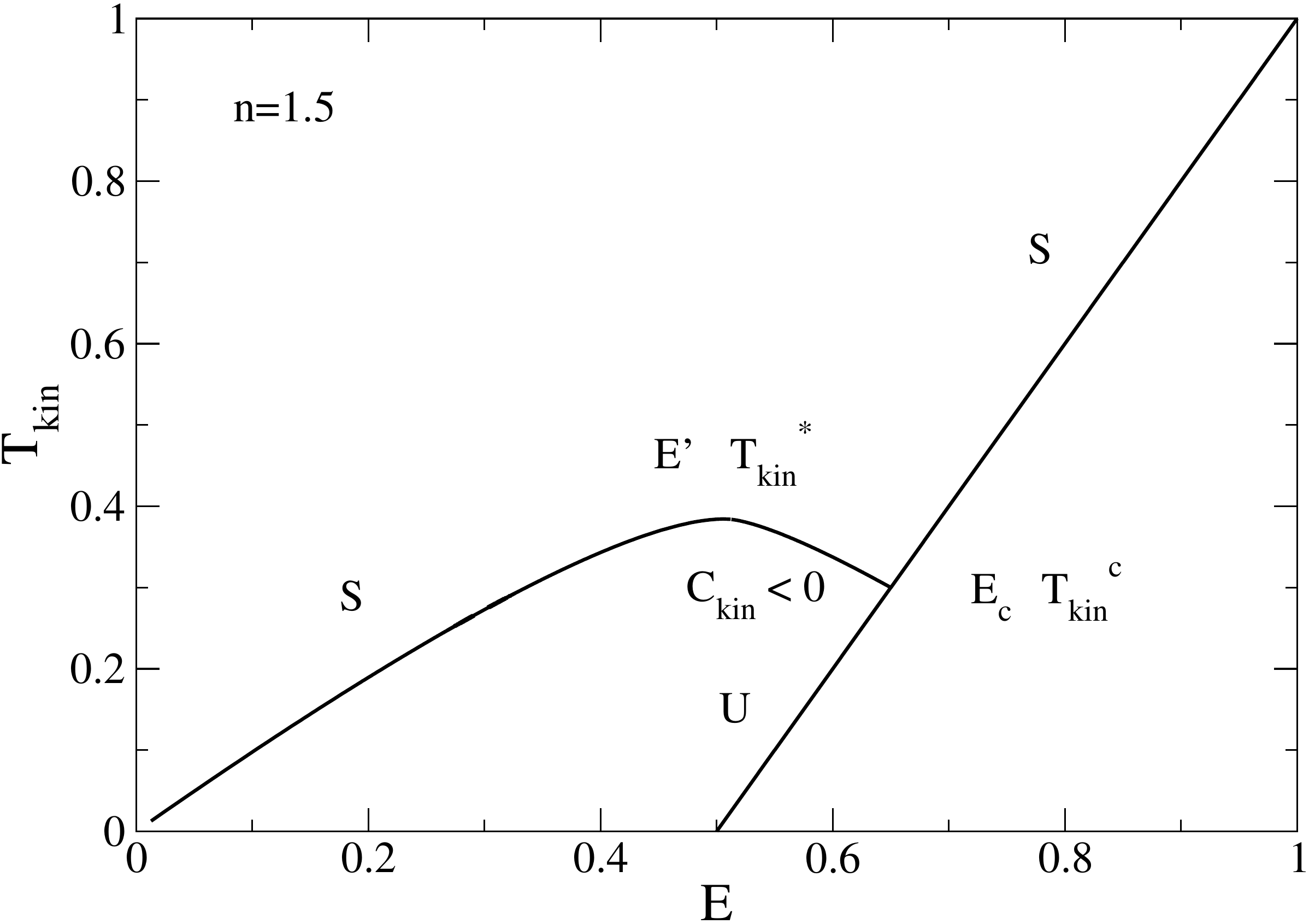}
\caption{The physical caloric curve for $n_{MCP}\simeq 0.68<n<n_0$ (specifically $n=1.5$). The kinetic specific heat close to
the bifurcation point is negative although the thermodynamical specific heat is positive \cite{cc}.}
\label{caloTkinN1p5}
\end{center}
\end{figure}

\begin{figure}
\begin{center}
\includegraphics[clip,scale=0.3]{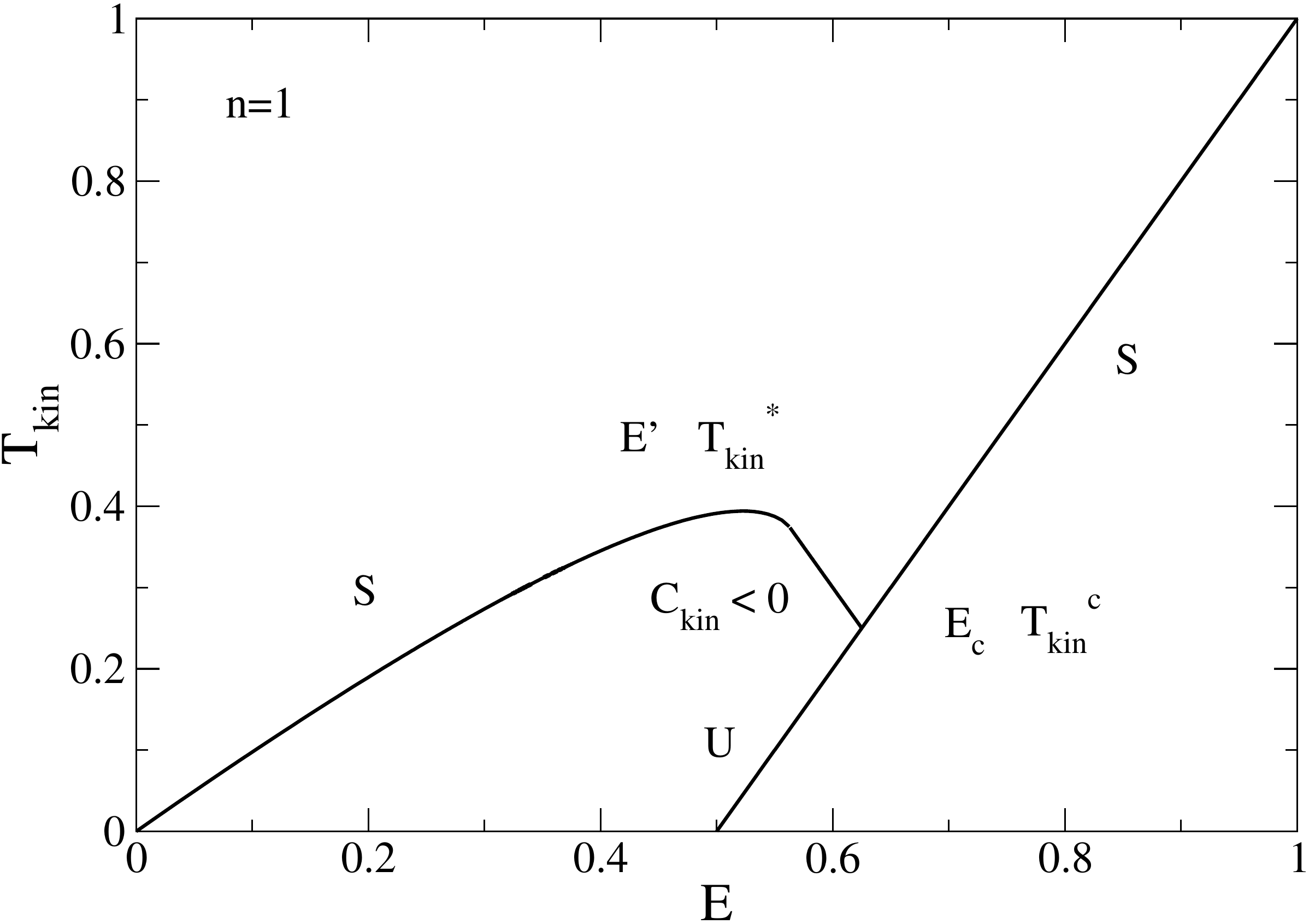}
\caption{The physical caloric curve for $n=1$. The kinetic specific heat close to the bifurcation point 
is negative ($C_{kin}=-1/2$).
For $9/16\le E\le E_c=5/8$, the physical caloric curve is a straight line given by $T_{kin}=3/2-2E$. 
The thermodynamical specific heat is infinite \cite{cc}.}
\label{caloTkinN1}
\end{center}
\end{figure}

\begin{figure}
\begin{center}
\includegraphics[clip,scale=0.3]{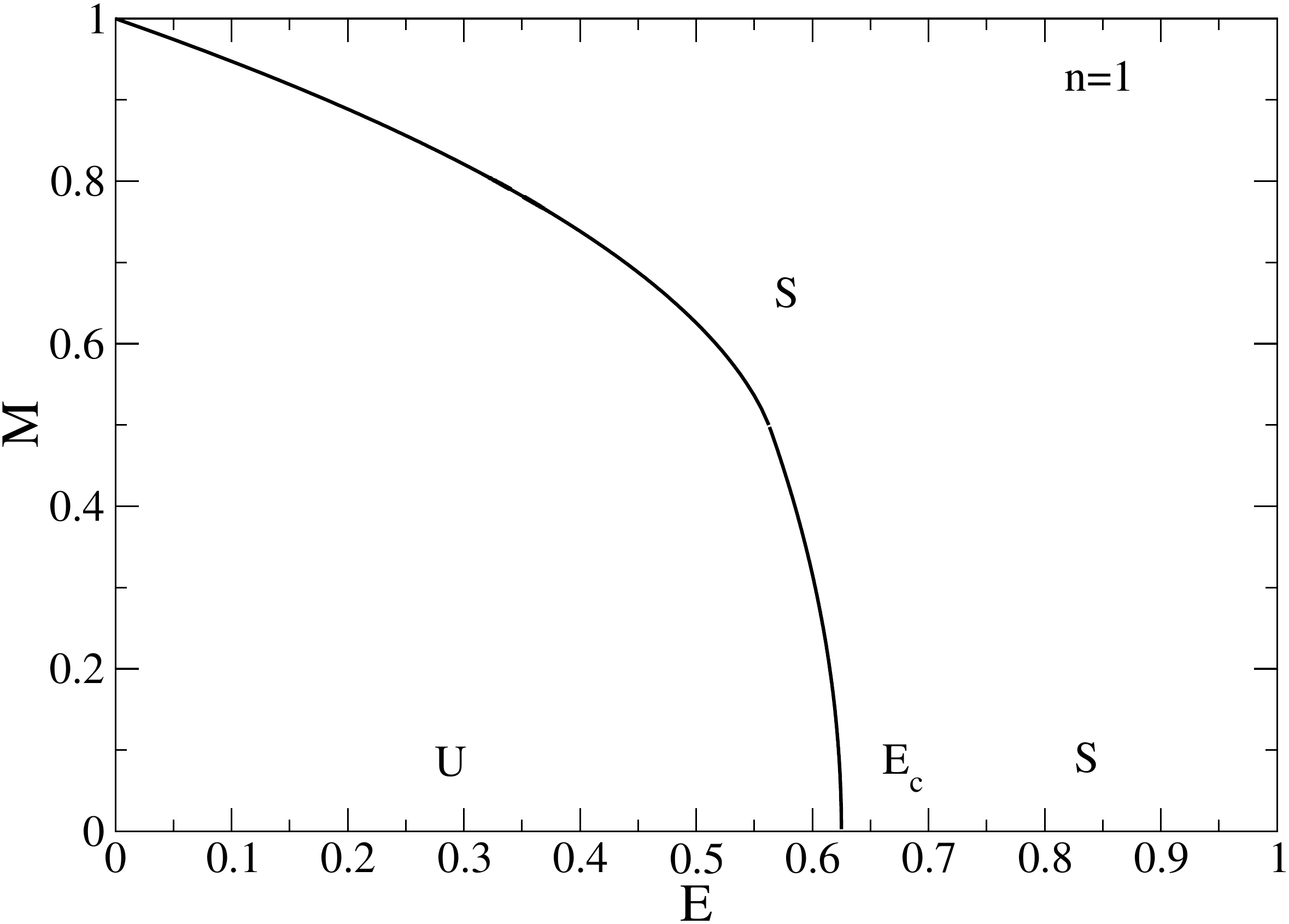}
\caption{The magnetization curve for $n>n_{MCP}$ (specifically $n=1$). For $9/16\le E\le E_c=5/8$, we have $M=2(E_c-E)^{1/2}$.}
\label{magnN1}
\end{center}
\end{figure}

\begin{figure}
\begin{center}
\includegraphics[clip,scale=0.3]{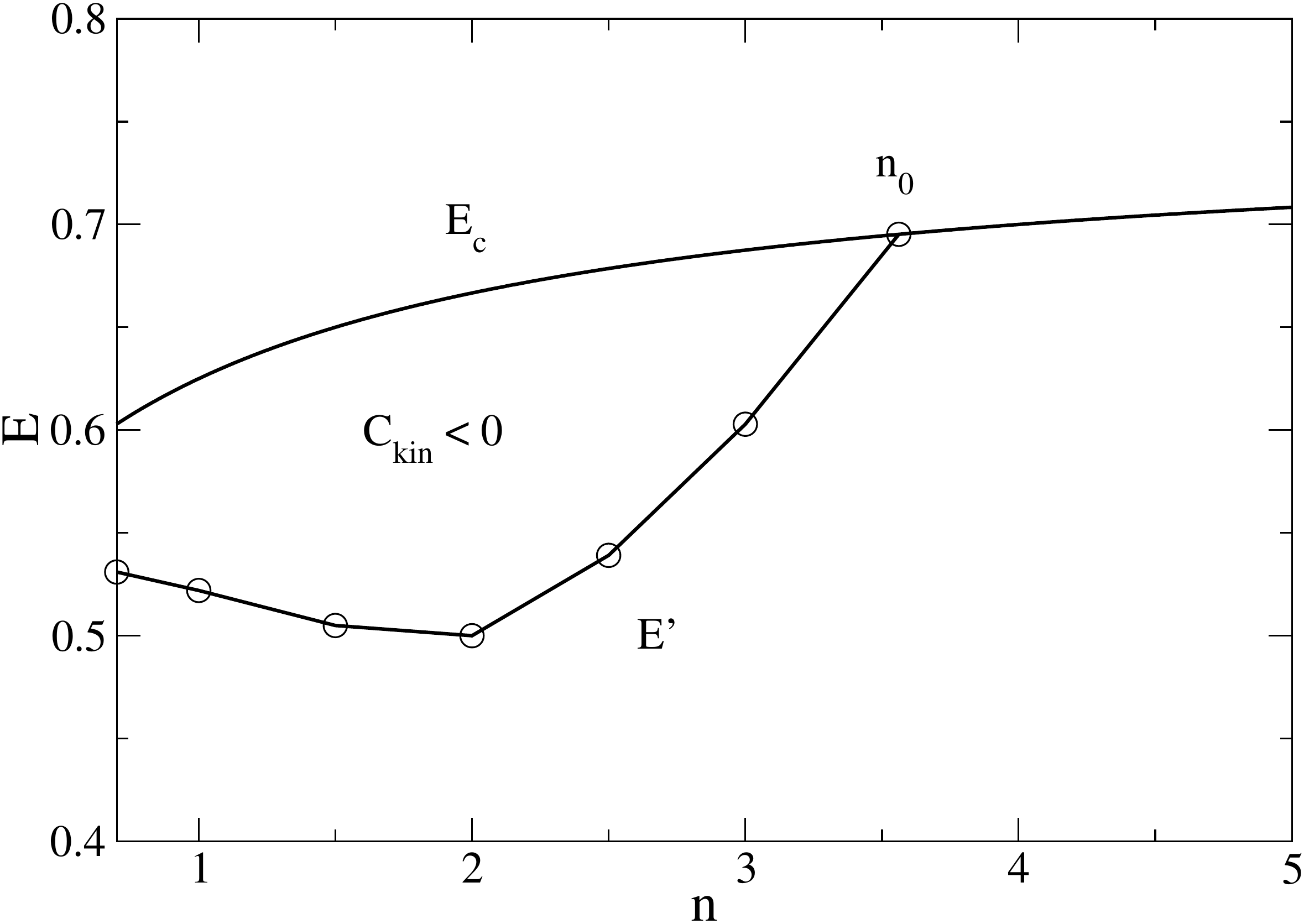}
\caption{Microcanonical phase diagram for $n>n_{MCP}\simeq
0.68$. It shows a second order phase transition at $E_c$ and a region
of negative kinetic specific heat between $E'$ and $E_c$ for
$n_{MCP}<n<n_0$.  }
\label{phasemicro}
\end{center}
\end{figure}

\begin{figure}
\begin{center}
\includegraphics[clip,scale=0.3]{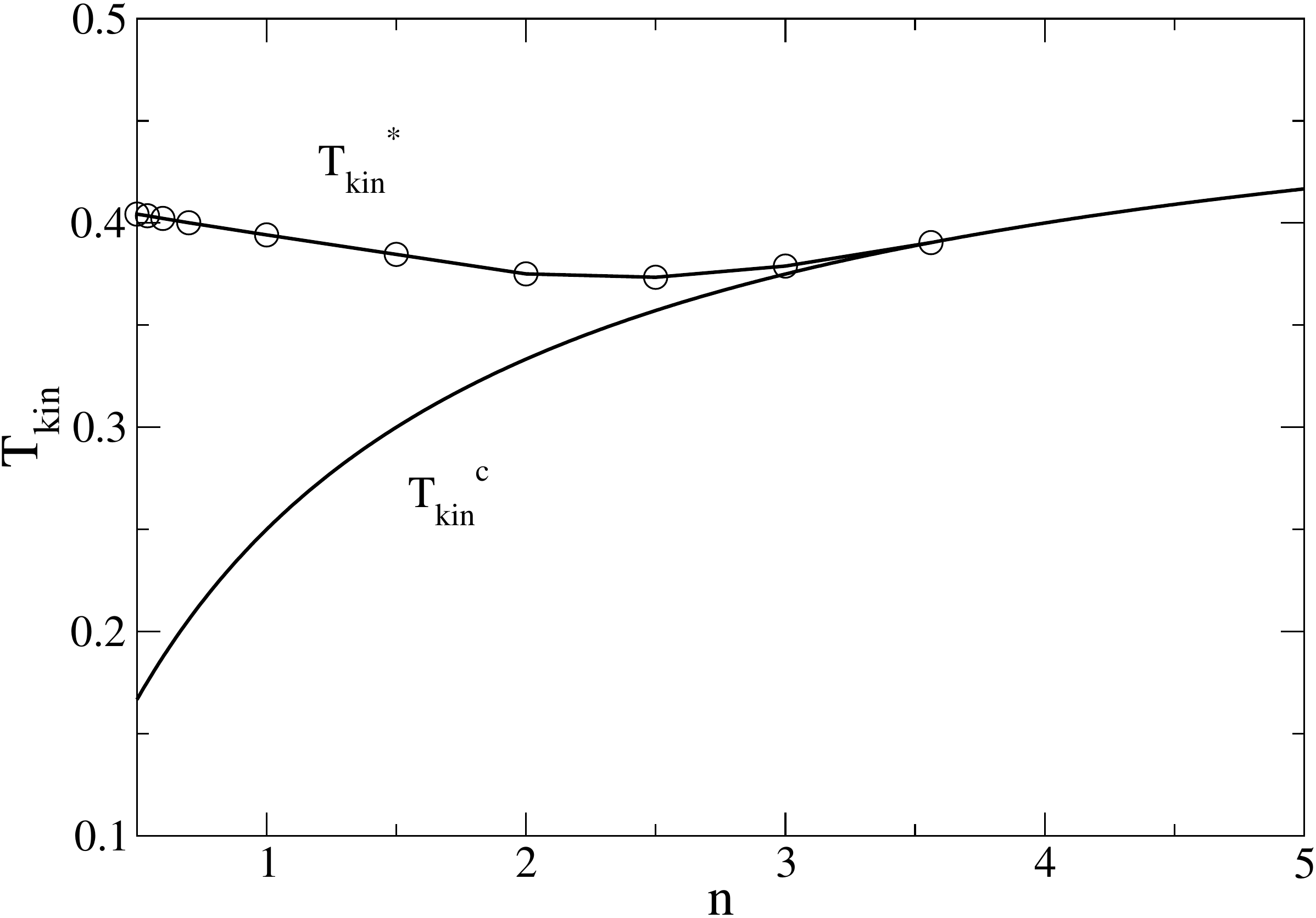}
\caption{Evolution of $T_{kin}^c$ and $T_{kin}^*$ as a function of $n$.  }
\label{phasecano}
\end{center}
\end{figure}

We must be careful that we cannot deduce any stability result in the canonical ensemble (the optimization problem
(\ref{tte4})) from the physical caloric curve $T_{kin}(E)$ since $T_{kin}$ is not the correct variable (the correct
variable in the canonical ensemble is $T$ or $\Theta$ \cite{cc}). Therefore, we restrict ourselves to the microcanonical
ensemble (the optimization problem (\ref{tte3})). Actually, this is the most appropriate ensemble since the natural control
parameter is the energy. Since phase transitions for polytropic distributions have already been discussed in our
previous paper (from the thermodynamical caloric curve $\Theta(E)$), our discussion here will be more concise. We
shall denote global entropy maxima by (S), local entropy maxima by (M) and minima or saddle points of entropy by (U).
In the thermodynamical analogy, these symbols correspond to stable, metastable\footnote{For systems with long-range
interaction, the metastable states are extremely robust. In practice, they should be considered on the same footing
as stable states. However, in our theoretical analysis, we shall distinguish between fully stable and metastable states.}
and unstable states. Coming back to the dynamical interpretation, stable states (S) and metastable states (M) 
are dynamically Vlasov stable. We cannot definitely conclude that unstable
states (U) are Vlasov unstable, except for homogeneous distributions, since the optimization problem (\ref{tte3})
provides just a {\it sufficient} condition of dynamical stability \cite{ccstab}.

\begin{figure}
\begin{center}
\includegraphics[clip,scale=0.3]{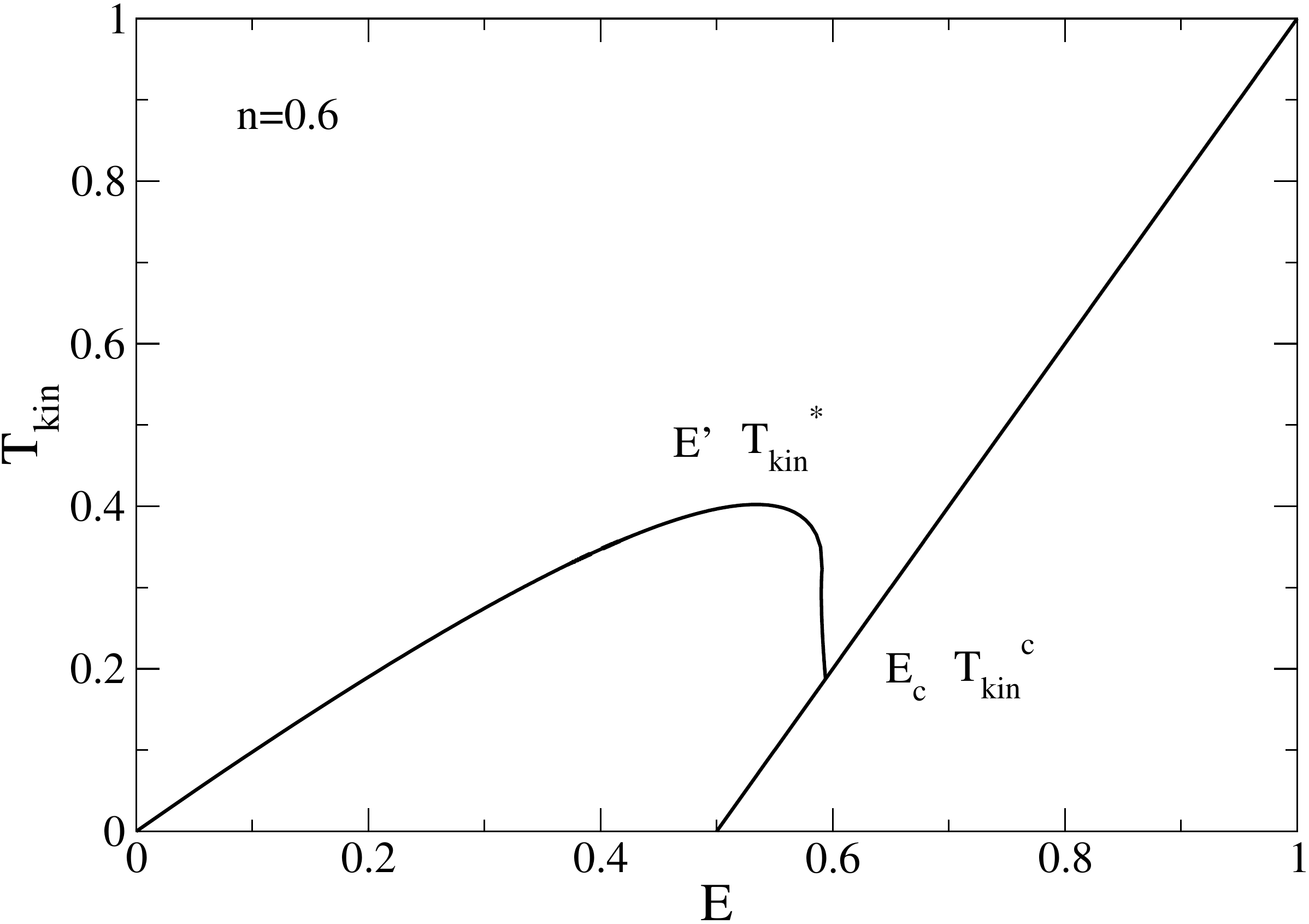}
\caption{The physical caloric curve for $n_{MTP}\simeq 0.563<n<n_{MCP}$ (specifically $n=0.6$). }
\label{caloTkinN0p6}
\end{center}
\end{figure}

\begin{figure}
\begin{center}
\includegraphics[clip,scale=0.3]{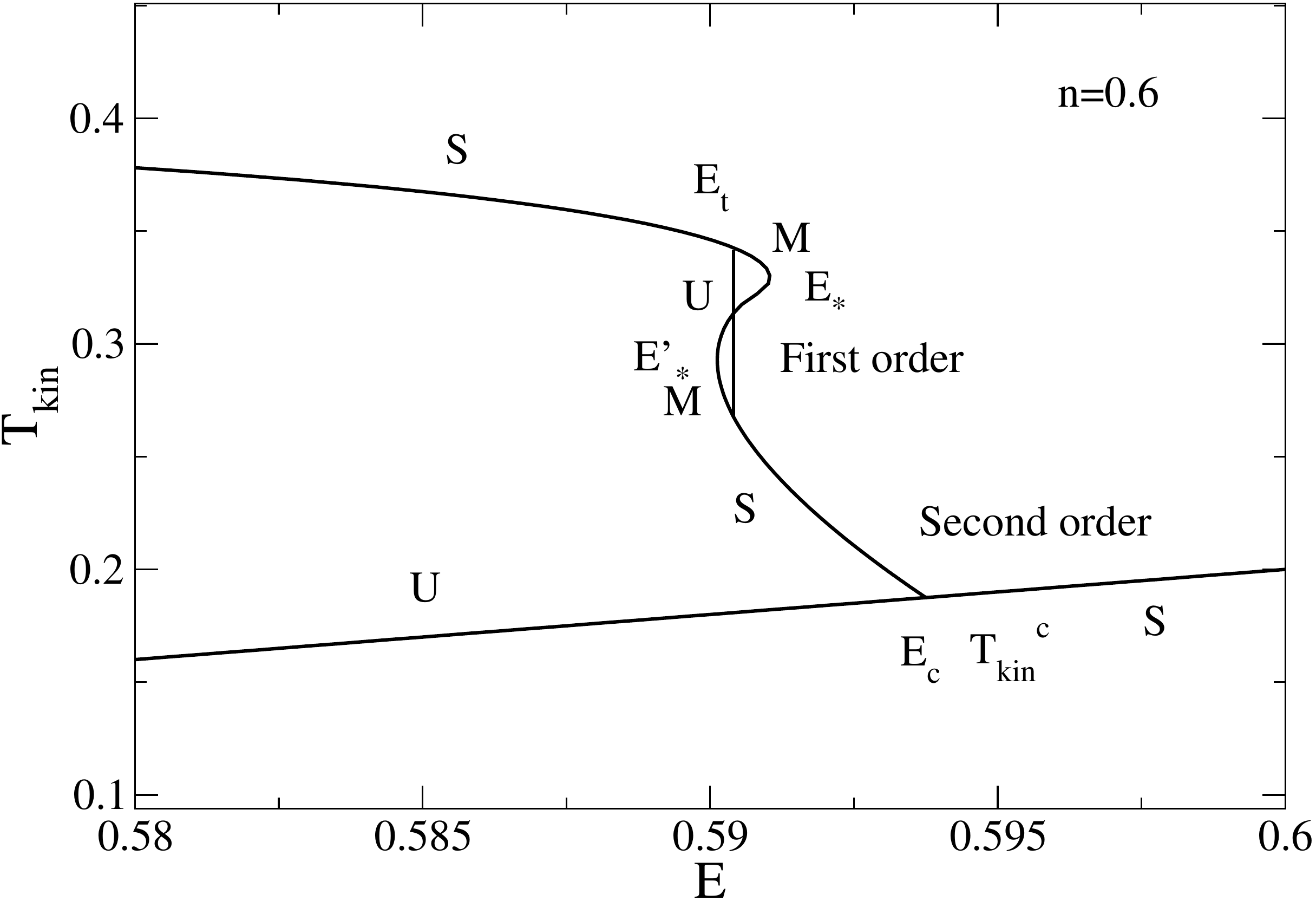}
\caption{Enlargement of the physical caloric curve for $n_{MTP}<n<n_{MCP}$ (specifically $n=0.6$). The kinetic specific
heat close to the bifurcation point is negative. The kinetic caloric curve displays a second order phase transition
at $E_c$ and a first order phase transition at $E_t$. The transition energy $E_t$ has been determined from the
entropy curve $S(E)$ as explained in \cite{cc}. Note that the  Maxwell construction cannot be performed on the physical
caloric curve $T_{kin}(E)$ since $T_{kin}\neq T=1/\beta$.}
\label{caloTkinN0p6ZOOM}
\end{center}
\end{figure}

\begin{figure}
\begin{center}
\includegraphics[clip,scale=0.3]{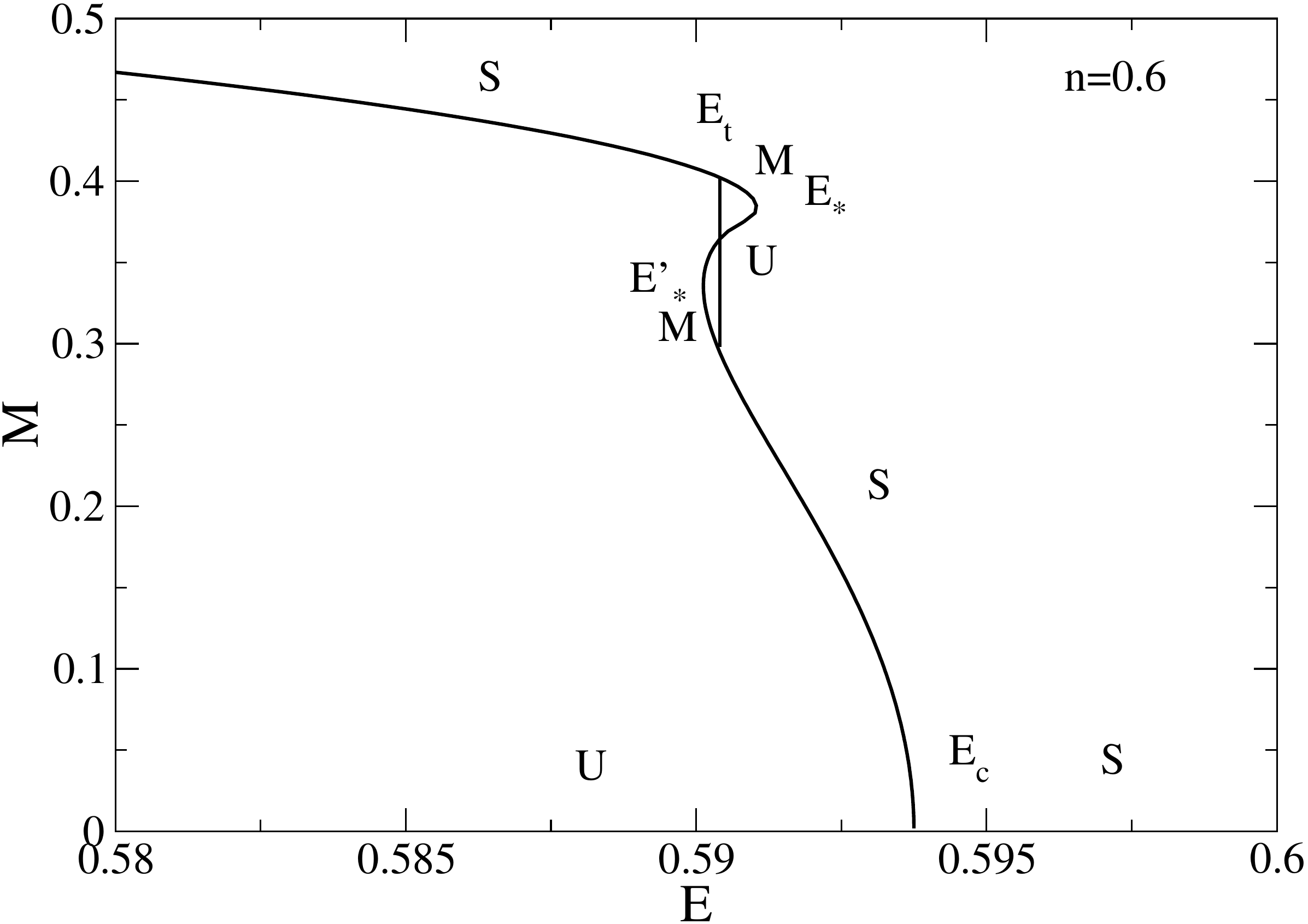}
\caption{The magnetization curve for $n_{MTP}<n<n_{MCP}$ (specifically $n=0.6$).}
\label{magnN0p6}
\end{center}
\end{figure}

\begin{figure}
\begin{center}
\includegraphics[clip,scale=0.3]{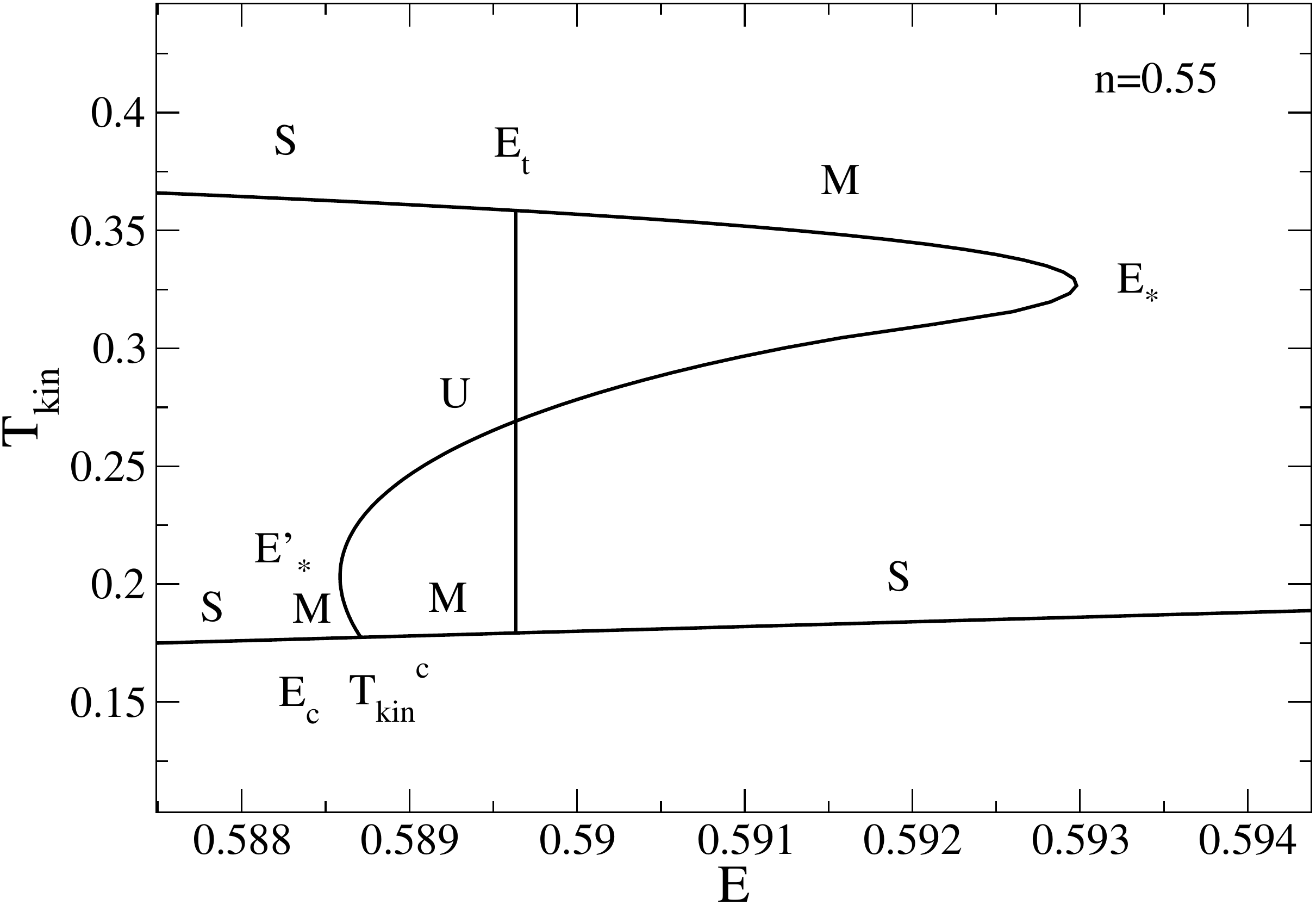}
\caption{The physical caloric curve for $n_{*}<n<n_{MTP}$ (specifically $n=0.55$). The kinetic specific heat close to
the bifurcation point is negative. There is a first order phase transition at $E=E_t$. The metastable branch exhibits a
second order phase transition at $E_c$.}
\label{caloTkinN0p55ZOOM}
\end{center}
\end{figure}

\begin{figure}
\begin{center}
\includegraphics[clip,scale=0.3]{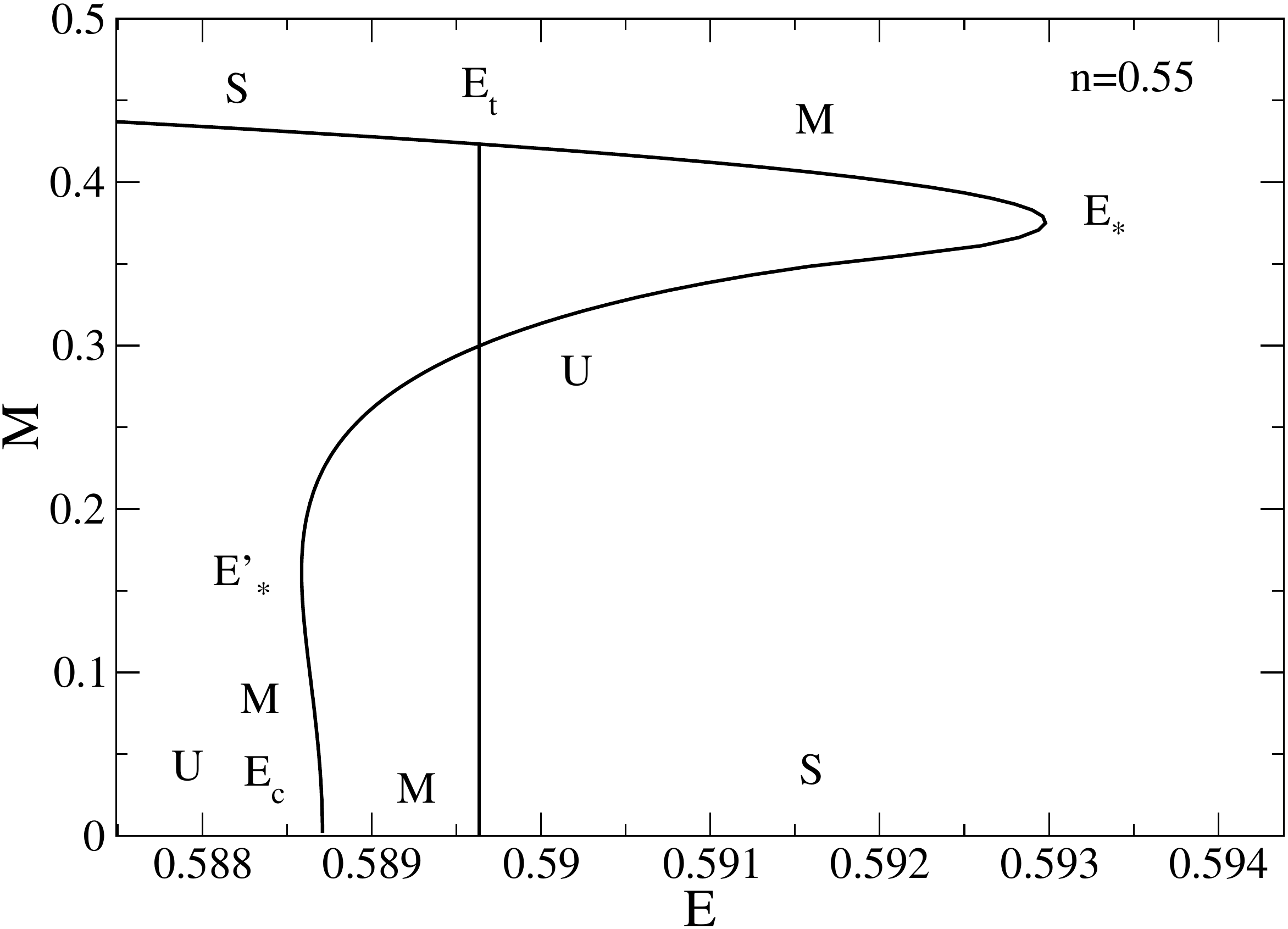}
\caption{The magnetization curve for $n_{*}<n<n_{MTP}$ (specifically $n=0.55$).}
\label{magnN0p55}
\end{center}
\end{figure}

\begin{figure}
\begin{center}
\includegraphics[clip,scale=0.3]{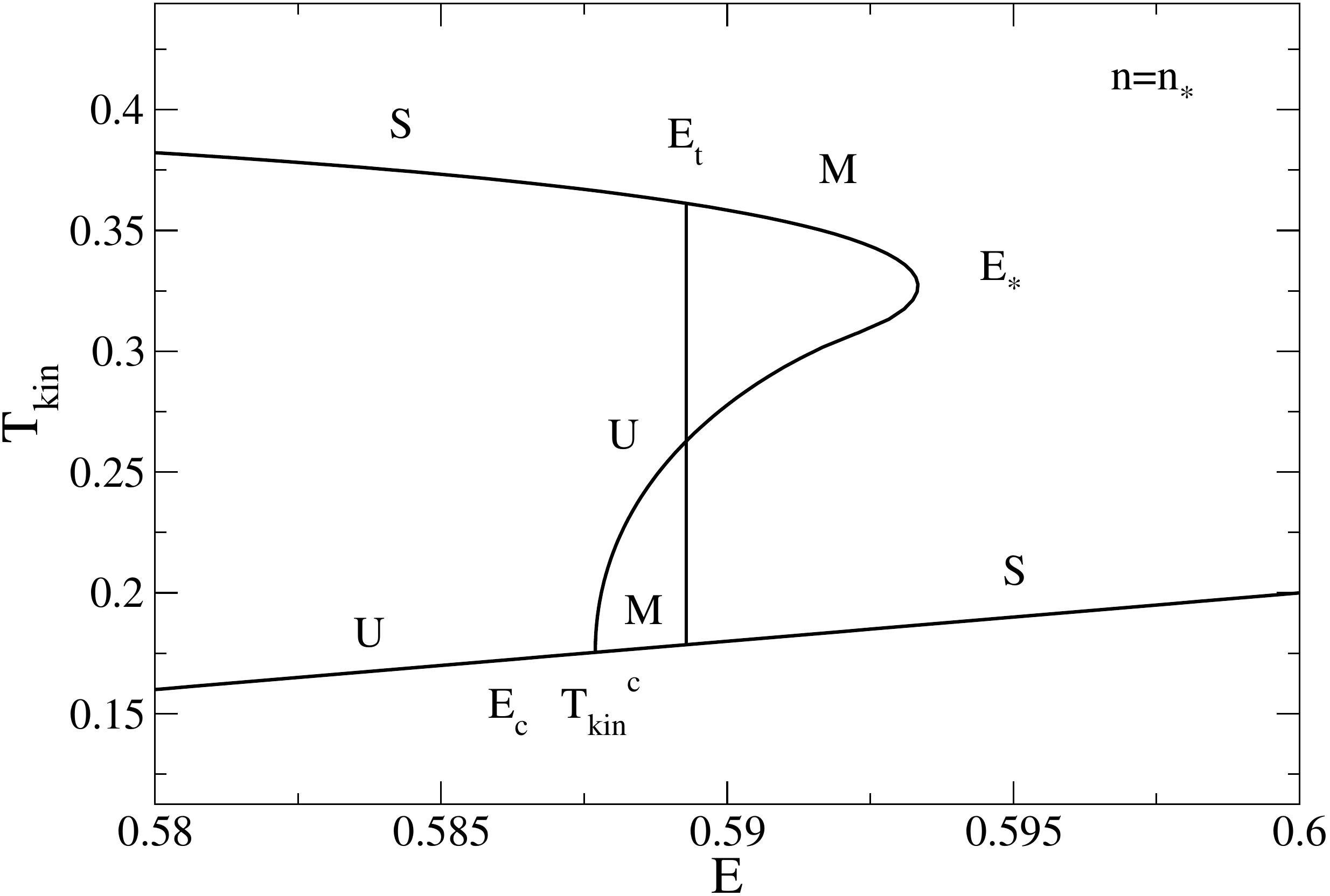}
\caption{The physical caloric curve for $n=n_*\simeq 0.54$. The kinetic specific heat close to the bifurcation point vanishes.}
\label{caloTkinN0p54031242ZOOM}
\end{center}
\end{figure}

\begin{figure}
\begin{center}
\includegraphics[clip,scale=0.3]{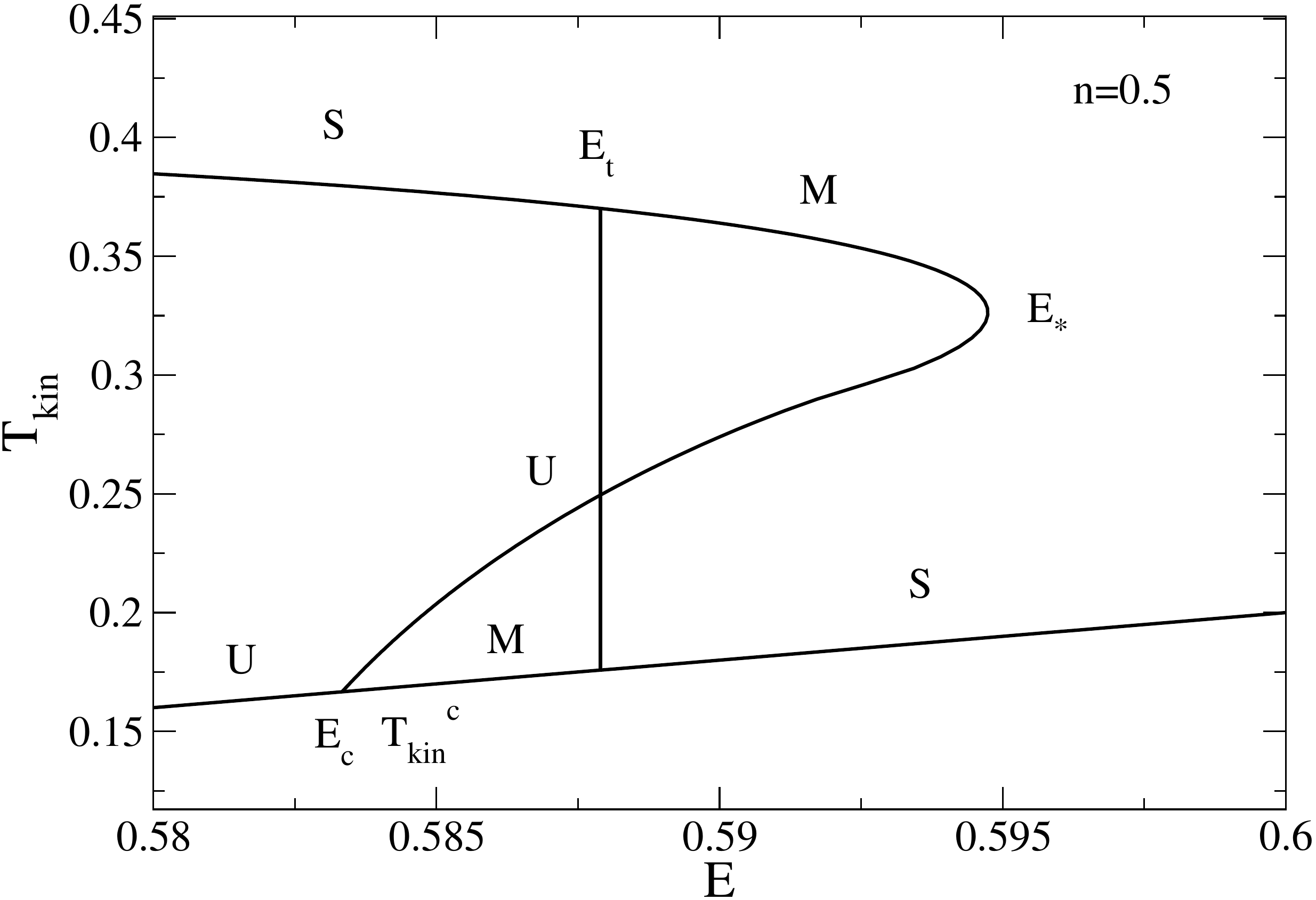}
\caption{The physical caloric curve for $1/2\le n<n_*$ (specifically $n=0.5$). The kinetic specific heat close to the
bifurcation point is positive. There is just a first order phase transition.}
\label{caloTkinN0p5ZOOM}
\end{center}
\end{figure}

\begin{figure}
\begin{center}
\includegraphics[clip,scale=0.3]{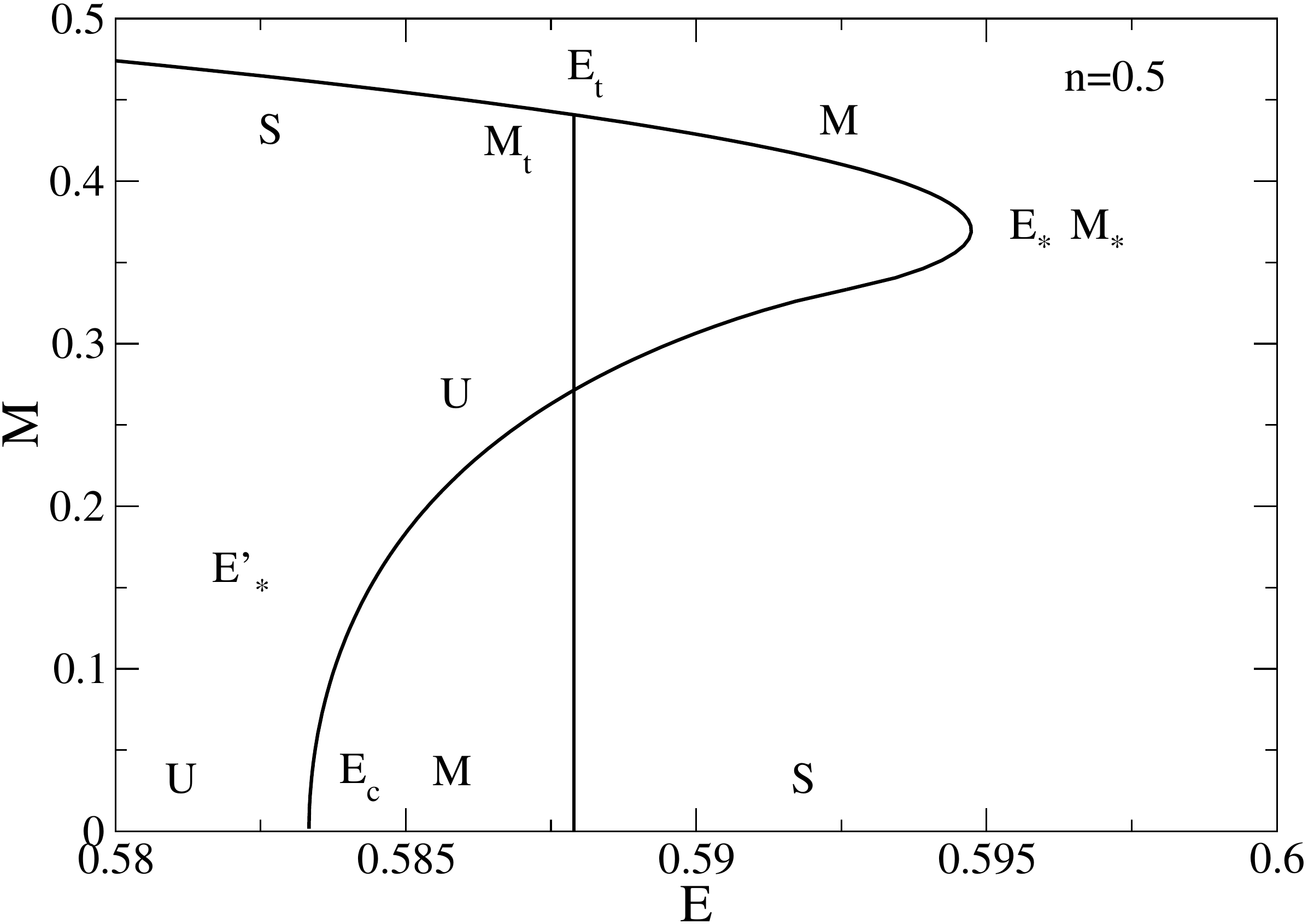}
\caption{The magnetization curve for $1/2\le n<n_{*}$ (specifically $n=0.5$).}
\label{magnN0p5}
\end{center}
\end{figure}

\begin{figure}
\begin{center}
\includegraphics[clip,scale=0.3]{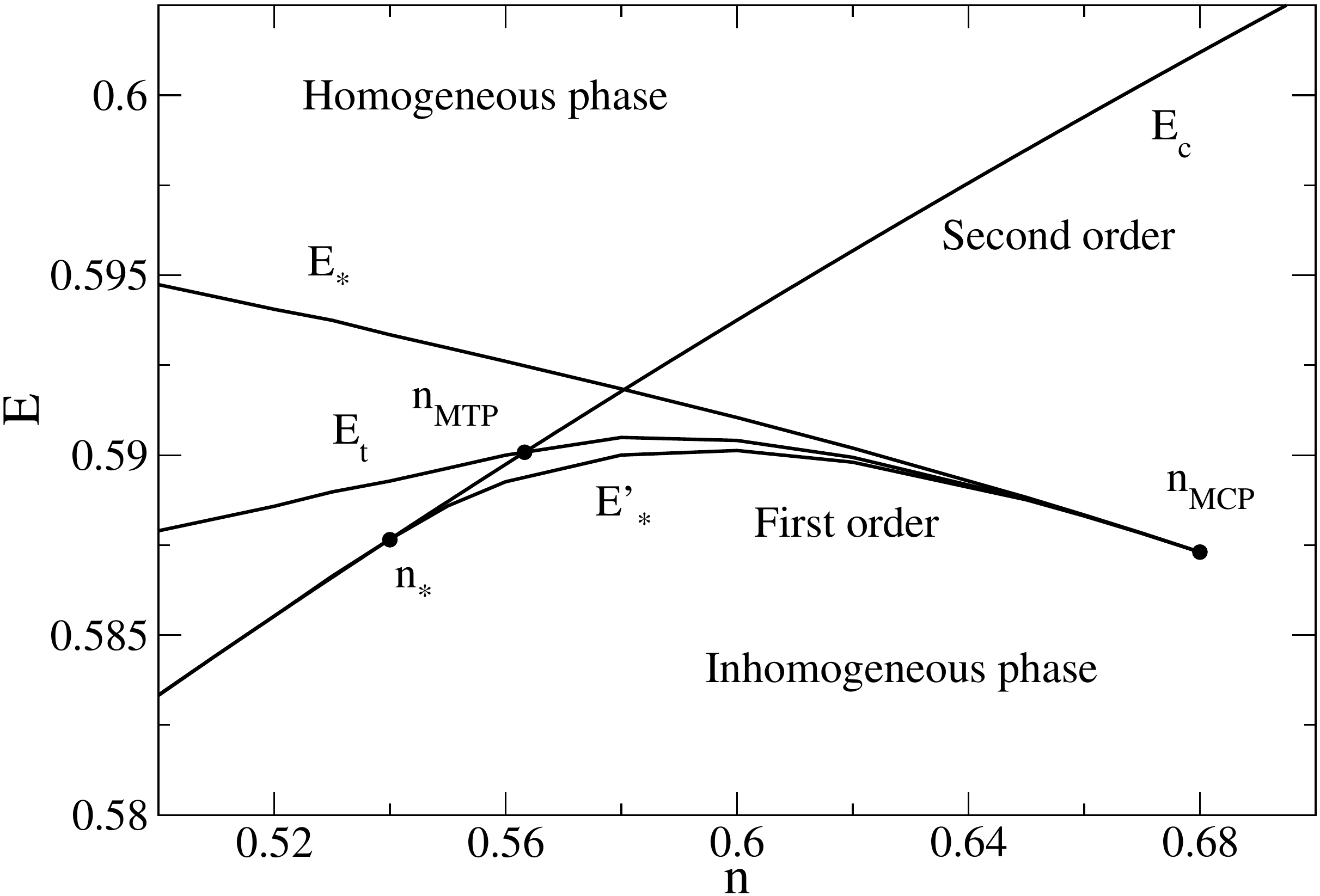}
\caption{Microcanonical phase diagram.}
\label{phasemicroADD}
\end{center}
\end{figure}

For $n>n_{MCP}\simeq 0.68$, the physical caloric curve $T_{kin}(E)$ exhibits a second order
phase transition marked by the discontinuity of $dT_{kin}/dE$ at
$E=E_c$ (see Figures \ref{caloTkinNinfini}, \ref{caloTkinN3p5615528}, \ref{caloTkinN1p5}, and \ref{caloTkinN1}). An
example of magnetization curve is given in Figure \ref{magnN1}.
For $n>n_{0}\simeq 3.56$, the kinetic specific heat is positive (see
Figure \ref{caloTkinNinfini}). For $n_{MCP}<n<n_{0}$, there is a region of
negative kinetic specific heat between
$E'$ and $E_c$ (see Figures \ref{caloTkinN1p5} and \ref{caloTkinN1}). The
microcanonical phase diagram in the $(n,E)$ plane for $n>n_{MCP}$ is represented in Figure \ref{phasemicro}. For
completeness, we have also plotted $T_{kin}^c$ and  $T_{kin}^*$ (the latter being the 
kinetic temperature corresponding to $E'$) as a function of $n$ in Figure \ref{phasecano},
although $T_{kin}$ should not be regarded as a control parameter.  For $n_{MTP}\simeq 0.563<
n<n_{MCP}$, there is a very interesting
situation, already noted in \cite{cc},  in which the caloric curve exhibits a second order phase
transition at $E_c$ between the homogeneous phase and the
inhomogeneous phase (as before) and a first order phase transition at
$E_t$ between two inhomogeneous phases (see
Figures \ref{caloTkinN0p6}-\ref{magnN0p6}).  The first order
phase transition is marked by the discontinuity of $T_{kin}$ at
$E=E_t$ and the existence of metastable
branches. Therefore, $n_{MCP}\simeq 0.68$ and $E_{MCP}\simeq
0.5873$ is a microcanonical critical
point marking the appearance of the first order phase transition. At
$n=n_{MTP}$, the energies of the first and second order
phase transitions coincide ($E_t=E_c$) and, for $1/2\le
n<n_{MTP}$, there is only a first order phase transition
at $E=E_t$ between homogeneous and inhomogeneous
states (see Figures \ref{caloTkinN0p55ZOOM}, \ref{magnN0p55}, \ref{caloTkinN0p54031242ZOOM}, and \ref{caloTkinN0p5ZOOM}).
Therefore, $n_{MTP}\simeq 0.563$ and $E_{MTP}\simeq 0.5901$
is a microcanonical tricritical point
separating first and second order phase transitions. For $n_*\simeq
0.54<n<n_{MTP}$, there remains a sort of second order
phase transition at $E=E_c$ for the {\it metastable}
states (see Figures \ref{caloTkinN0p55ZOOM} and \ref{magnN0p55}). Between $E'_*$ and
$E_c$, the kinetic specific heat is negative. At $n=n_*$, the kinetic specific heat close to the critical point vanishes
and the energies $E'_*$ and $E_c$ coincide (see Figure \ref{caloTkinN0p54031242ZOOM}). For $1/2\le n<n_*$, the kinetic
specific heat close to the critical point is positive and the ``metastable'' second order phase transition disappears
(the inhomogeneous states close to the bifurcation point are unstable). In that
case, there is only a first order phase transition (see
Figures \ref{caloTkinN0p5ZOOM} and \ref{magnN0p5}).  The microcanonical
phase diagram in the $(n,E)$ plane summarizing all these results is represented in Figure \ref{phasemicroADD}.

In conclusion, for $n>n_{MCP}$ we have second order phase transitions, for  $n_{MTP}<n<n_{MCP}$ we have first and second
order phase transitions, and for $1/2\le n<n_{MTP}$ we have first order phase transitions.

{\it Remark 1:} As a corollary, we emphasize that for $n>n_{MCP}\simeq 0.68$, the inhomogeneous polytropes are always
entropy maxima at fixed energy and normalization so they are dynamically Vlasov stable. This is an important
theoretical result of \cite{cc}.

\subsection{The polytrope $n=1$}
\label{sec_crit}

The polytropic index $n=1$ (i.e. $\gamma=2$, $q=3$) is
particular because it corresponds to a canonical tricritical
point \cite{cc}. Furthermore, for this index, the
algebra greatly simplifies and analytical results can be
obtained. This is because the relationship (\ref{pp1}) between the
density and the potential is linear\footnote{Actually, except for these nice mathematical properties, it is not clear whether
the polytrope $n=1$ plays a special role in the physics of the problem. As we shall see in the numerical part of the paper,
other polytropic distributions are relevant as well. This soften the claim made in our previous paper \cite{cc}.}.

The spatially homogeneous distribution is
\begin{eqnarray}
f(v)=\frac{1}{2\pi^2\sqrt{\Theta}}\left (1-\frac{v^2}{4\Theta}\right )^{1/2},
\label{sevp}
\end{eqnarray}
if $|v|\le v_{max}=2\sqrt{\Theta}$ and $f=0$ otherwise. It corresponds to what has been called ``semi-ellipse''
in \cite{campa1}. The homogeneous phase is dynamically stable if, and only, if $\Theta\ge \Theta_c=1/4$ or $E\ge E_c=5/8$.

\begin{figure}
\begin{center}
\includegraphics[clip,scale=0.3]{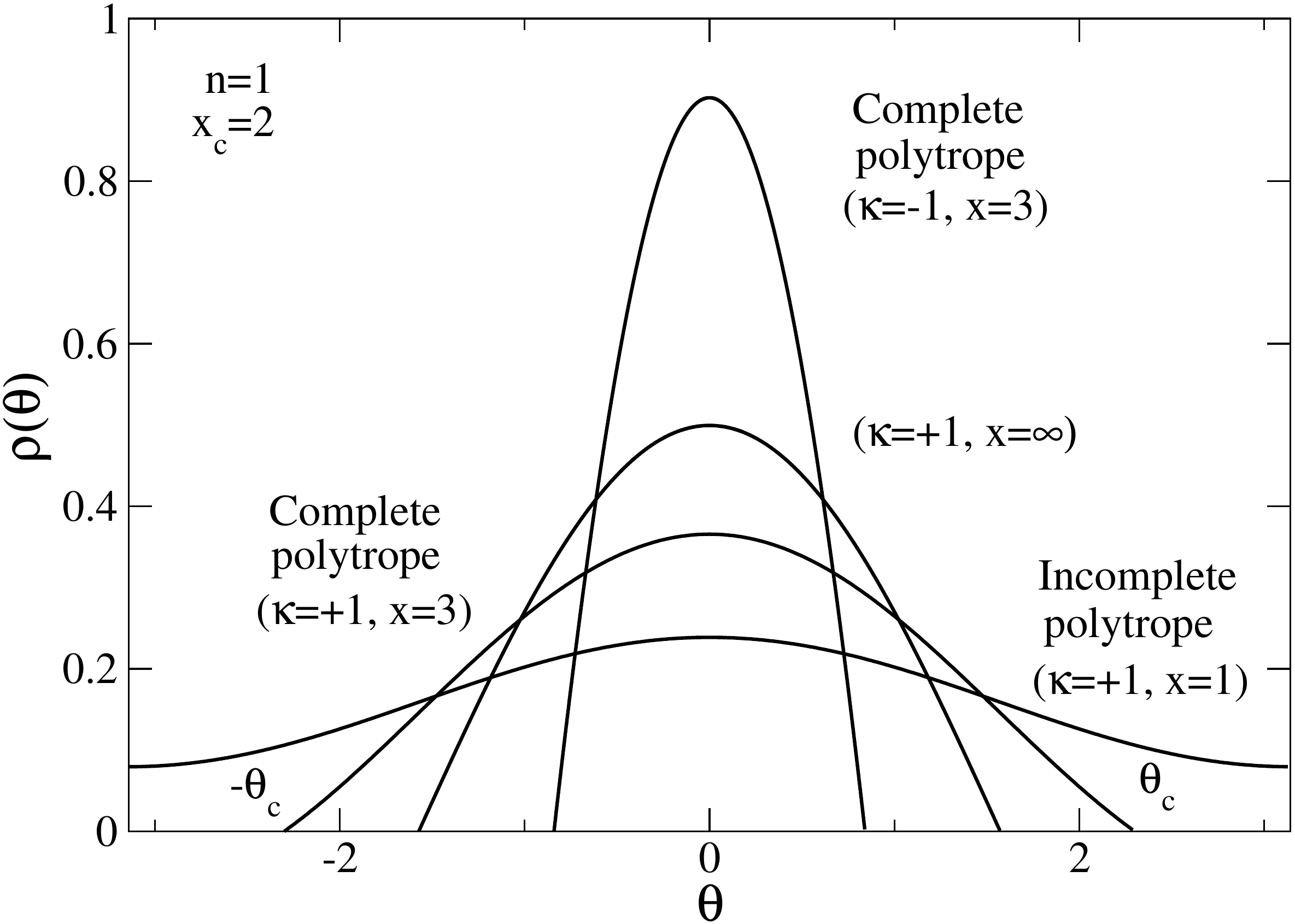}
\caption{Density profile of $n=1$ polytropes for different values of $x$. We have added the case $\kappa=-1$ that was
forgotten in \cite{cc}.}
\label{profilesN1}
\end{center}
\end{figure}

The density profile of $n=1$ polytropes is
\begin{equation}
\rho(\theta)=A\left (\kappa+\frac{x}{2}\cos\theta\right )_{+}.
\label{ae1}
\end{equation}
Some density profiles are plotted in Figure \ref{profilesN1}. For $\kappa=+1$ and $x<x_c=2$ (incomplete polytropes),
the deformed Bessel functions take the simple form $I_{2,0}(x)=1$ and $I_{2,1}(x)={x}/{4}$. Then, we get
\begin{equation}
A=\frac{1}{2\pi},\quad \Theta=\frac{1}{4},\quad M=\frac{x}{4},
\label{ae3}
\end{equation}
\begin{eqnarray}
E=\frac{5}{8}-\frac{x^2}{64},\quad T_{kin}=\frac{1}{4}+\frac{x^2}{32}.
\label{ae4}
\end{eqnarray}
The spatially inhomogeneous distributions with $x<x_c$ have the same polytropic temperature  $\Theta_c=1/4$ but different
energies ranging from $9/16$ to $E_c=5/8$. Their magnetizations are in the range $0 \le M\le 1/2$, and they are related
to the energy by $M=2\sqrt{E_c-E}$ (see Figure \ref{magnN1}).  The thermodynamical caloric curve  forms a
plateau\footnote{These solutions are stable in the microcanonical ensemble and metastable in the canonical ensemble \cite{cc}.
This corresponds to a situation of {\it partial ensemble equivalence} \cite{ellis}.}  $\Theta(E)=\Theta_c$ in the
range $9/16\le E\le 5/8$, so it has an infinite specific heat $C=dE/d\Theta=\infty$ (see Figure 12 of \cite{cc}). By
contrast, in the same range of energies, the physical caloric curve  (see Figure \ref{caloTkinN1}) is given by
\begin{eqnarray}
T_{kin}=\frac{3}{2}-2E,
\label{ae5}
\end{eqnarray}
and it has a  constant specific heat
\begin{eqnarray}
C_{kin}=\frac{dE}{d T_{kin}}=-\frac{1}{2},
\label{ae6}
\end{eqnarray}
which turns out to be negative. Therefore, for $n=1$, the thermodynamical and physical caloric curves are very different.

For complete polytropes, we can get analytical expressions of the thermodynamical parameters by using the following
expressions of the $\gamma$-deformed  modified Bessel functions:
\begin{equation}
I_{2,0}(x)=\frac{1}{2\pi}\left\lbrack \sqrt{x^2-4}+2\kappa\arccos\left (-\frac{2\kappa}{x}\right )\right\rbrack,
\label{ae7}
\end{equation}
\begin{equation}
I_{2,1}(x)=\frac{1}{2\pi}\left\lbrack \frac{\kappa}{x}\sqrt{x^2-4}+\frac{x}{2}\arccos\left (-\frac{2\kappa}{x}\right )\right\rbrack.
\label{ae8}
\end{equation}
Other analytical results are given in \cite{cc}.

\subsection{The waterbag distribution (polytrope $n=1/2$)}
\label{sec_waterbag}

A detailed study of the possibly inhomogeneous waterbag distribution that is a steady state of the Vlasov
equation has been performed in \cite{hmfq1}. Here, we only
recall the main results of this study that will be needed to interpret the numerical simulations of
Section \ref{sec_waterbagnum}. The waterbag distribution
is defined by
\begin{eqnarray}
f=\Biggl\lbrace \begin{array}{cc}
f_0   & \qquad (\epsilon< \epsilon_{F}), \\
0  & \qquad (\epsilon\ge \epsilon_{F}).
\end{array}
\label{polyw1}
\end{eqnarray}
It is similar to the Fermi distribution in quantum mechanics where $f_0$ plays the role of the maximum value of the
distribution function fixed by the Pauli exclusion principle and $\epsilon_F$ plays the role of the Fermi energy.
Comparing Eq. (\ref{polyw1}) with Eq. (\ref{ps4}), we see that the waterbag distribution corresponds to a polytrope of
index $n=1/2$ (i.e. $\gamma=3$, $q=+\infty$). This correspondence can also be obtained by determining the equation of
state associated with the waterbag distribution. To that purpose, we rewrite Eq. (\ref{polyw1}) in the form
\begin{eqnarray}
{f}=\Biggl\lbrace \begin{array}{cc}
f_0  & \qquad (v< v_{F}(\theta)), \\
0  & \qquad (v\ge v_{F}(\theta)),
\end{array}
\label{polyw2}
\end{eqnarray}
where $v_{F}(\theta)=\sqrt{2(\epsilon_F-\Phi(\theta))}$ is the local maximum velocity (the analogous of the Fermi velocity
in quantum mechanics). The density is given by $\rho(\theta)=2 f_0 v_{F}(\theta)$ and the pressure
by $p(\theta)=(2/3) f_0 v_{F}^{3}(\theta)$. Eliminating $v_{F}(\theta)$ between these two
expressions, we find that the equation of state of the waterbag distribution is
\begin{eqnarray}
\label{polyw3}
p=\frac{1}{12 f_0^{2}}\rho^{3}.
\end{eqnarray}
This is the equation of state of a polytrope of index $n=1/2$ and polytropic constant $K=1/(12 f_0^2)$. Therefore, $f_0$ is
related to the polytropic temperature $\Theta=K/(2\pi)^2$ by
\begin{eqnarray}
\label{polyw4}
f_0=\frac{1}{4\pi\sqrt{3\Theta}}.
\end{eqnarray}
The velocity of sound is given by
\begin{eqnarray}
\label{polyw5}
c_s(\theta)=\frac{1}{2 f_0}\rho(\theta)=v_F(\theta),
\end{eqnarray}
and it coincides with the maximum local velocity (the Fermi velocity in quantum mechanics).

The spatially homogeneous waterbag distribution is the step function
\begin{eqnarray}
\label{polyw6}
f=f_0, \quad {\rm if}\quad |v|\le v_F,\nonumber\\
f=0,\quad {\rm if}\quad |v|> v_F,
\end{eqnarray}
where $v_F$ is determined by the normalization condition (\ref{mfa12}) leading to
\begin{eqnarray}
\label{polyw7}
v_F=\frac{1}{4\pi f_0}=\sqrt{3\Theta}.
\end{eqnarray}
The kinetic temperature is given by  $T_{kin}=\Theta=v_F^2/3$, so the total energy can be written as
\begin{eqnarray}
\label{polyw8}
E=\frac{1}{6}v_F^2+\frac{1}{2}.
\end{eqnarray}
The velocity of sound in the homogeneous phase is $c_s=v_F$. According to the general criteria (\ref{sh2}) and
(\ref{corrg4}), the homogeneous phase is dynamically stable if
\begin{eqnarray}
\label{polyw9}
E>E_c=\frac{7}{12},\qquad \Theta>\Theta_c=\frac{1}{6},
\end{eqnarray}
\begin{eqnarray}
\label{polyw10}
f_0<(f_0)_c=\frac{1}{2\pi\sqrt{2}},\qquad v_F>(v_F)_c=\frac{1}{\sqrt{2}},
\end{eqnarray}
and unstable otherwise. These stability criteria can also be directly obtained from the dispersion relation
$\epsilon(\omega)=0$ which can be solved analytically for the waterbag distribution (see, e.g., \cite{cvb}).

\begin{figure}
\begin{center}
\includegraphics[clip,scale=0.3]{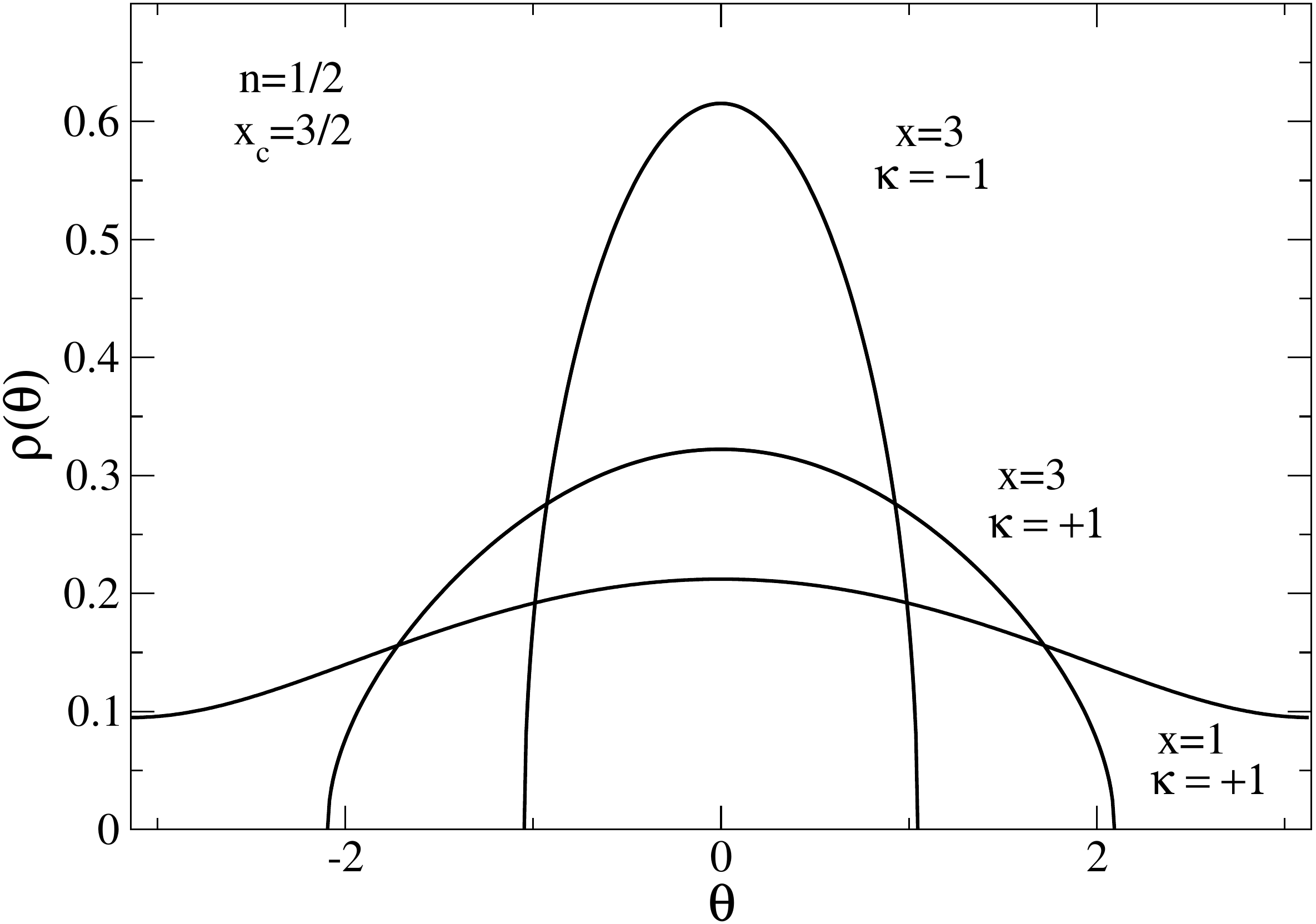}
\caption{Density profile of $n=1/2$ polytropes for different values of $x$. For $\kappa=+1$ we have taken $x=1$
and $x=3$. For $\kappa=-1$ we have taken $x=3$.}
\label{profiles}
\end{center}
\end{figure}

\begin{figure}
\begin{center}
\includegraphics[clip,scale=0.3]{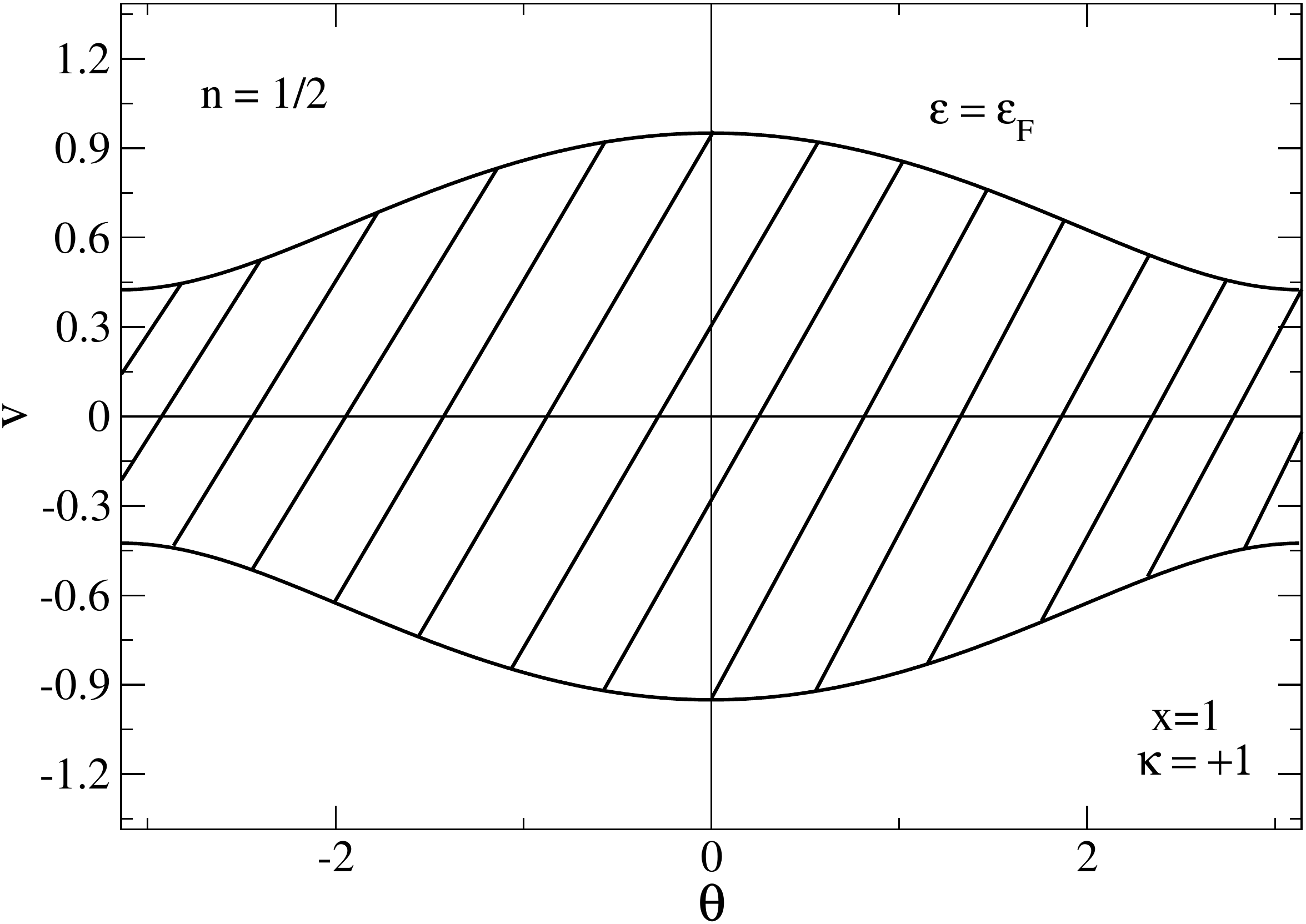}
\caption{Phase space portrait of $n=1/2$ incomplete ($x=1$) polytropes with $\kappa=+1$. The distribution function has
the uniform value $f=\eta_0$ in the shaded area ($\epsilon <\epsilon_F$) and $f=0$ outside ($\epsilon >\epsilon_F$).}
\label{phasespaceINCOMPLET}
\end{center}
\end{figure}

\begin{figure}
\begin{center}
\includegraphics[clip,scale=0.3]{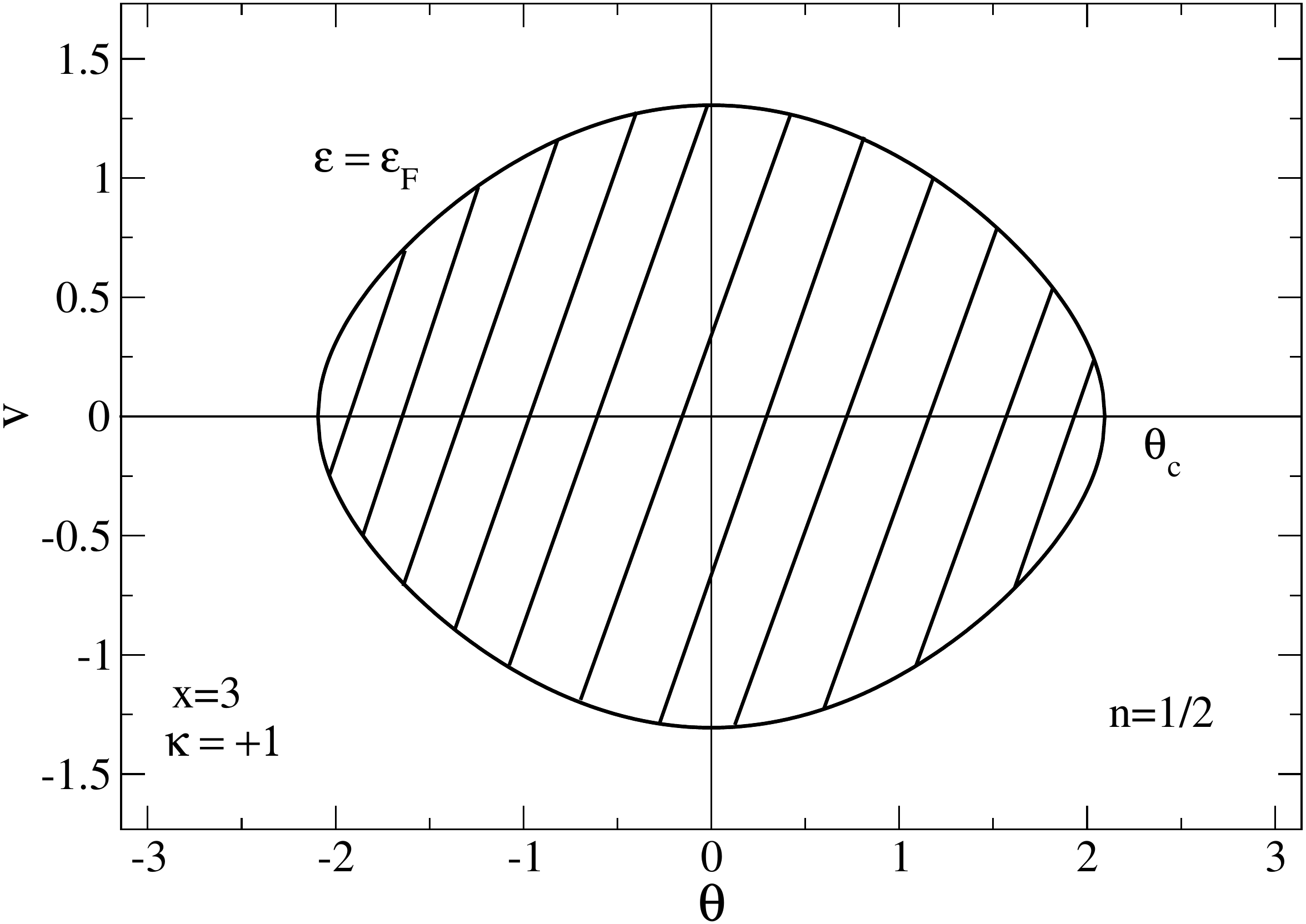}
\caption{Phase space portrait of $n=1/2$ complete ($x=3$) polytropes with $\kappa=+1$. The distribution function has
the uniform value $f=\eta_0$ in the shaded area ($\epsilon <\epsilon_F$) and $f=0$ outside ($\epsilon >\epsilon_F$).}
\label{phasespaceCOMPLET}
\end{center}
\end{figure}

Using the results of Section \ref{sec_cip}, the density profile of the spatially inhomogeneous waterbag distribution is given by
\begin{eqnarray}
\label{polyw11}
\rho(\theta)=\frac{1}{2\pi I_{3,0}(x)}\left (\kappa+\frac{2}{3}x\cos\theta\right )_{+}^{1/2},
\end{eqnarray}
with $\kappa={\rm sgn}(\epsilon_F-1)$. The polytrope is incomplete for $\kappa=+1$ and $x<x_c=3/2$, and complete
otherwise. Some representative density profiles are plotted in
Figure \ref{profiles}.  The local maximum velocity is related
to the local density by $v_F(\theta)=\rho(\theta)/(2f_0)$, so it is proportional to the density profile. Combining
Eqs. (\ref{mag9}), (\ref{polyw4}), and (\ref{polyw11}),  we obtain
\begin{eqnarray}
\label{polyw12}
v_F(\theta)=\sqrt{\frac{3 I_{3,1}(x)}{x I_{3,0}(x)}}\left (\kappa+\frac{2}{3}x\cos\theta\right )_{+}^{1/2}.
\end{eqnarray}
The waterbag distribution is represented in phase space in Figures \ref{phasespaceINCOMPLET} and \ref{phasespaceCOMPLET}.
The velocity distribution $\phi(v)$ can be obtained analytically \cite{hmfq1}.

The dynamical stability of the waterbag distribution has been studied in \cite{hmfq1}. The magnetization curve $M(E)$ is
reproduced in Figure \ref{magnN0p5} and the physical caloric curve $T_{kin}(E)$ is plotted in
Figure \ref{caloTkinN0p5ZOOM}. There is a first order phase transition at $E_t=0.588$ marked by the discontinuity of the
magnetization passing from $M_t=0.44$ to zero\footnote{As discussed in detail in \cite{hmfq1}, the thermodynamical
caloric curve $\Theta(E)$  of the waterbag distribution is very particular since the temperature jump at the transition
point shrinks to zero as $n\rightarrow 1/2$. It is therefore more convenient to study the phase transition on the
magnetization curve $M(E)$ or on the physical caloric curve  $T_{kin}(E)$ than on the thermodynamical caloric
curve $\Theta(E)$.}. The homogeneous phase is stable for $E>E_t$,
metastable for $E_c=7/12<E<E_t$ and unstable for $0\le E<E_c$. The  inhomogeneous phase is stable for
($0\le E<E_t$, $M_t<M\le 1$), metastable for ($E_t<E\le E_*=0.595$, $M_*=0.37\le M<M_t$), and unstable
for ($E_c<E<E_*$, $0<M<M_*$).

{\it Remark 2:} The magnetization curve  $M(E)$ of the pure waterbag distribution (polytrope $n=1/2$) reported
in Figure \ref{magnN0p5} (see also \cite{hmfq1}) may be compared with Figure 2 of
Pakter \& Levin \cite{levin} for a core-halo state in which the core is a polytrope $n=1/2$. They both exhibit a first
order phase transition.

{\it Remark 3}: As discussed in \cite{hmfq1}, the waterbag distribution corresponds to the minimum energy state in
the Lynden-Bell theory, when the initial condition has only two levels $0$ and $f_0$. This has been used to construct
the curve $E_{ground}(f_0)$ in Appendix A of \cite{staniscia2}.

\section{Numerical simulations}
\label{sec_simul}

We have performed a number of simulations of the HMF model that we analyze in this and in the following two
Sections. We have considered different types of initial conditions. In this Section, we study the characteristics
of the QSS that are reached by the system after a violent relaxation from a spatially homogeneous initial state
which is Vlasov unstable. In Section \ref{sec_rising}, we study the evolution of the system initially prepared
in a Vlasov stable spatially homogeneous state. In that case, the system is in a QSS from the beginning, and the slow
evolution is due to finite size effects (``collisions''). We have already analyzed this evolution in the non-magnetized
regime in Ref. \cite{campa2}, and here
we extend our analysis to the magnetized regime. In Section \ref{sec_waterbagM1}, we treat the case
in which the system is initially in an unsteady inhomogeneous state with magnetization $M_0=1$ and isotropic waterbag distribution.

In this Section, the initial distributions $f_{0}(v)$ of the velocities are of several types: Gaussian, i.e.,
$f_{0}(v) \propto {\rm e}^{-av^2}$, semi-elliptical, i.e., $f_{0}(v) \propto \sqrt{v_0^2 - v^2}$, and
waterbag, i.e., $f_{0}(v) \propto \theta(v_m - |v|)$, with the value of the parameters $a$, $v_0$ and
$v_m$ determining the energy $E$. These three cases correspond to homogeneous polytropic distributions
with $n = \infty$, $n=1$ and $n={1}/{2}$, respectively. In all cases, the energy $E$ is chosen smaller than the
critical energy $E_c$ for Vlasov stability of the corresponding case. The critical energy, computed from
Eq. (\ref{hpjt7}), can be written in terms of $n$ as
\begin{equation}
\label{hpjt7_b}
E_c = \frac{3n + 2}{4n + 4} \,\, .
\end{equation}
Substituting the values of $n$ we get that $E_c = 0.75$ for the Gaussian distribution, $E_c = {5}/{8}$
for the semi-elliptical distribution, and $E_c = {7}/{12}$ for the waterbag distribution. 
The state is unstable if $E<E_c$. Then,
after a rapid
relaxation, the system settles down in a QSS. In this Section, our purpose is to analyze the physical caloric curves
of the QSSs and to compare them with those of the polytropic distribution functions.

The simulations have been performed with $N=2^{17}$ particles. However, they are representative of a system with
twice the number of particles, i.e., $2^{18}$, gaining in this way a factor of $2$ in the computer time needed.
In fact, we have exploited the following symmetry property of the HMF model. If, in the initial conditions, for
each particle with $(\theta,v)$ there is a particle with $(-\theta,-v)$, this property is conserved throughout
the dynamics, keeping at the same time $M_y \equiv 0$. In fact, choosing the numbering of the particles in such
a way that in the initial conditions we have $\theta_{i+\frac{N}{2}} = -\theta_i$ and $v_{i+\frac{N}{2}} = -v_i$,
the equations of motion (\ref{heeqm}) give:
\begin{equation}
\label{eqn2}
\frac{d^2 \theta_{i+\frac{N}{2}}}{dt^2} = - M_x \sin \theta_{i+\frac{N}{2}} =
M_x \sin \theta_i = -\frac{d^2 \theta_i}{dt^2} \,\, ,
\end{equation}
that proves the statement. Furthermore, it is sufficient to follow (and thus to represent in the computer) the
dynamics of only the first $\frac{N}{2}$ particles. We remark that extracting the initial condition from
a distribution that is invariant under the inversion $(\theta,v) \to (-\theta,-v)$ should not introduce
any peculiarity, since we expect that in the thermodynamic limit this invariance is satisfied.

We should note also the following point. The length of our simulations is sufficient to follow the rapid
relaxation to the QSS and its initial stages. With $2^{18}$ particles the lifetime of the QSS is
expected to be very large, since the collisional processes of the dynamics, due to finite size effects,
are very slow. It is true that the lifetime of magnetized QSSs, as the ones we are going to analyze, is
expected to be proportional to $N$ \cite{jstat}, and not to a higher power of $N$ as for homogeneous QSSs \cite{campa1};
however, such high values of $N$ lead to lifetimes that are much larger than the times we have considered.
In conclusion, we have observed only the settling of the QSS in its inital form, before the slow
collisional evolution has caused any significant variation in the direction of the BG equilibrium.

\subsection{Gaussian initial conditions}
\label{sec_gauss}

\begin{figure}
\begin{center}
\includegraphics[clip,scale=0.3]{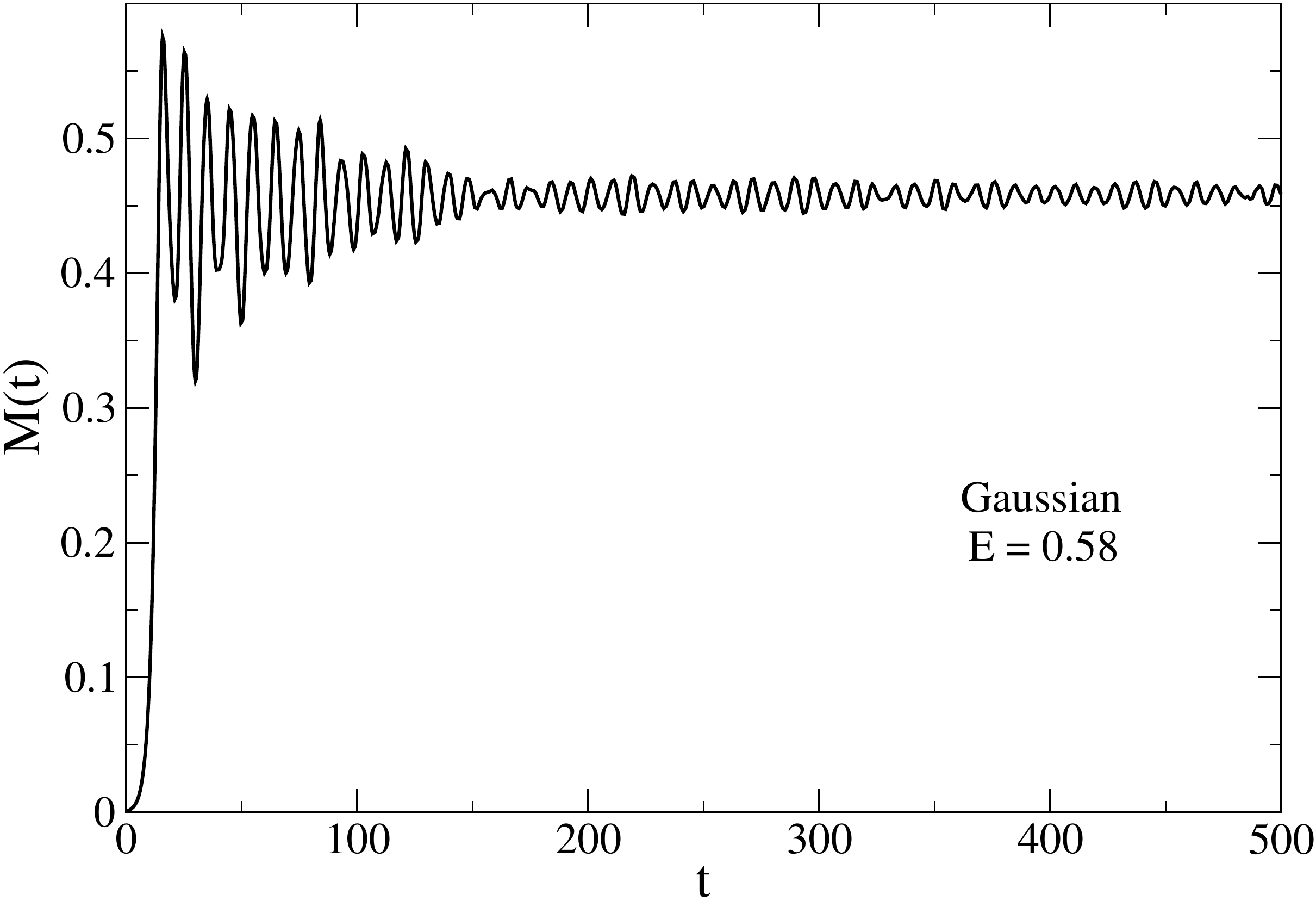}
\caption{Magnetization as a function of time $M(t)$ for $E=0.58$. It shows damped oscillations. They may be related to
Landau damping for inhomogeneous QSS.}
\label{magnetisation058gauss}
\end{center}
\end{figure}

In this case, for a given energy $E$, the initial velocities are extracted from the distribution
\begin{equation}
\label{gaussinit}
f_{0}(v) = \frac{1}{\sqrt{\pi \left( 4E -2 \right) }} \exp \left[ -\frac{v^2}{4E-2} \right] \,\, .
\end{equation}
The critical energy for Vlasov stability is $E_c = 0.75$. We have performed simulations at
several energies smaller than $E_c$, and precisely at the energy values $0.51$, $0.55$, $0.58$, $0.60$,
$0.62$, $0.65$ and $0.69$. We note that the smallest possible energy for an initial condition with $M=0$
is $0.5$ (see Eq. (\ref{mfa7})). In all cases, after the rapid relaxation, the system reaches a QSS with a
magnetization smaller than the equilibrium value, and thus also with an average kinetic temperature
smaller than at equilibrium (again from Eq. (\ref{mfa7}) one clearly sees that, for a given energy $E$, a
decrease of $M$ implies a decrease of $T_{kin}$). In Figure \ref{magnetisation058gauss}
we show a representative plot of the magnetization vs time, $M(t)$, for the run at $E=0.58$.


It appears that, after the rapid increase of $M$ and after a
short transient with large oscillations, the system settles to a state with small persistent
magnetization oscillations. The almost regularity of the oscillations indicates that they are not due
to finite size noise, but that they are an intrinsic property of the QSS. Therefore, strictly speaking,
we should consider the following two possibilities for the state reached by the system after the
rapid relaxation. Either the state is not a stationary state of the Vlasov equation, but rather a
quasi-periodic state; or the oscillations are due to a sort of Landau damping of an inhomogeneous
QSS, and they will eventually die out on a time scale large with respect to the one simulated here.
The first case is reminiscent of the results found by Morita and
Kaneko \cite{mk}, that studied initial conditions different from ours, and in some cases they obtained
almost periodic oscillations of the magnetization.

A similar picture holds for the other energies studied, in some case with oscillations of a somewhat larger
amplitude. Whatever the explanation of the oscillations, we can use the average value of the
magnetization (and consequently the average value of the kinetic temperature) to characterize the state
of the system; since the oscillations are small we can approximately consider the state to be
stationary.

As already noted in the Introduction, a steady state (stable or unstable) of the Vlasov equation is
characterized by
a one particle distribution function that depends on $\theta$ and $v$ only through the individual
energy $\epsilon = v^2/2+\Phi(\theta)$. Then, in our QSS the coordinates $(\theta,v)$ of the particles
should be distributed according to this property. In our simulations, where $M_y \equiv 0$, the
potential energy $\Phi(\theta)$ is given by $1-M_x \cos \theta$ (see Eq. (\ref{cev4})), where $M_x$ is the
average value of the magnetization mentioned above.
We have therefore divided the one particle phase
space in cells small enough to characterize each cell with a good approximation with a single value of the
individual energy $\epsilon$ (e.g., the value of the central point of the cell), but big enough to contain
a large number of particles. Taking a snapshot of the system configuration in the QSS (at the end of the
run), we have then counted the number of particles in each cell, and plotted this number vs the energy
value attributed to the cells. If the points of this plot are arranged on a continuous line, this means
that the one particle distribution function depends only on $\epsilon$, confirming that the state is
a QSS.

\begin{figure}
\begin{center}
\includegraphics[clip,scale=0.3]{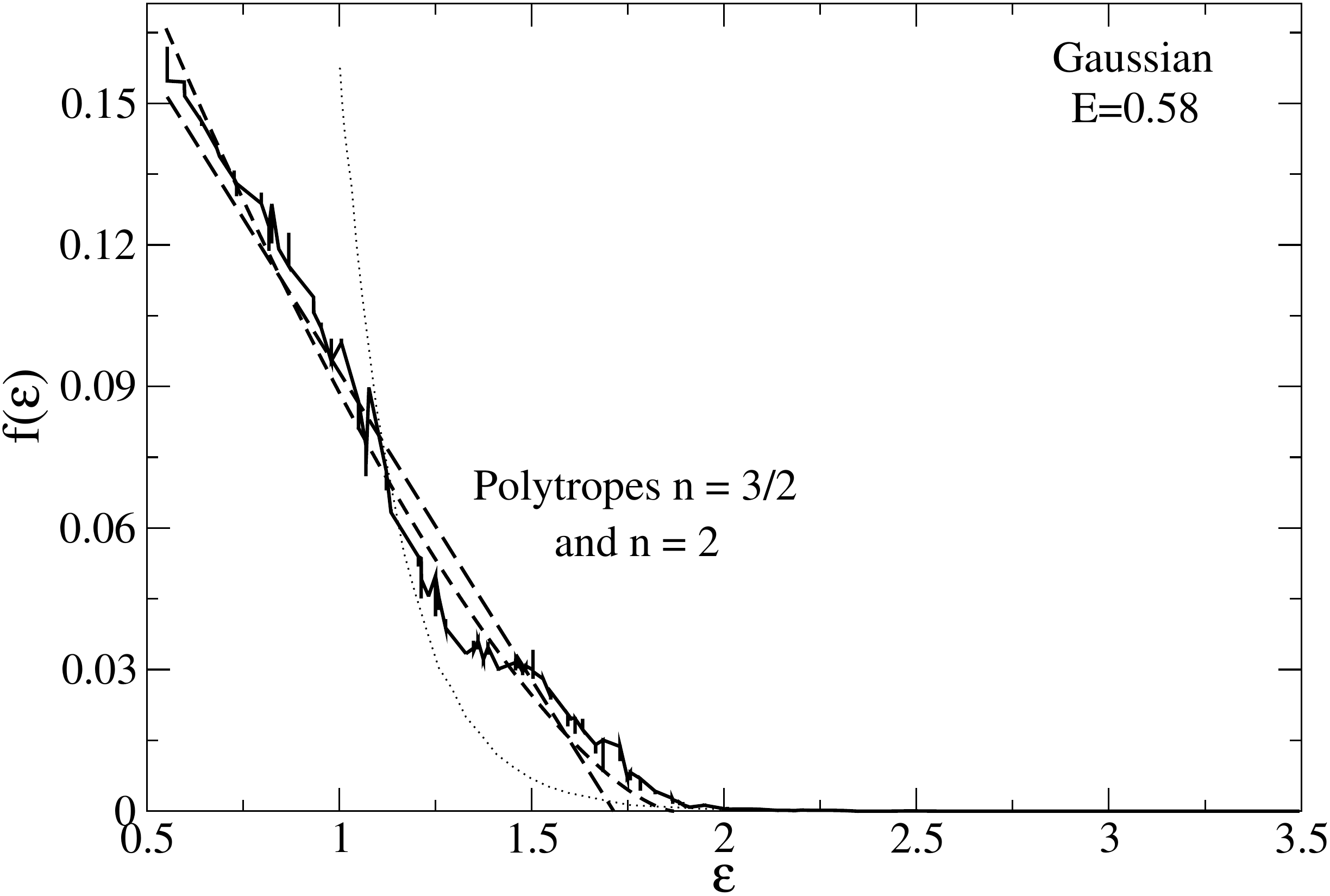}
\caption{The distribution $f(\epsilon)$ has been fitted by a pure polytrope $n=3/2$ (long-dash)
or $n=2$ (short-dash). We have also plotted the initial condition (dot).}
\label{feps058gauss}
\end{center}
\end{figure}

\begin{figure}
\begin{center}
\includegraphics[clip,scale=0.3]{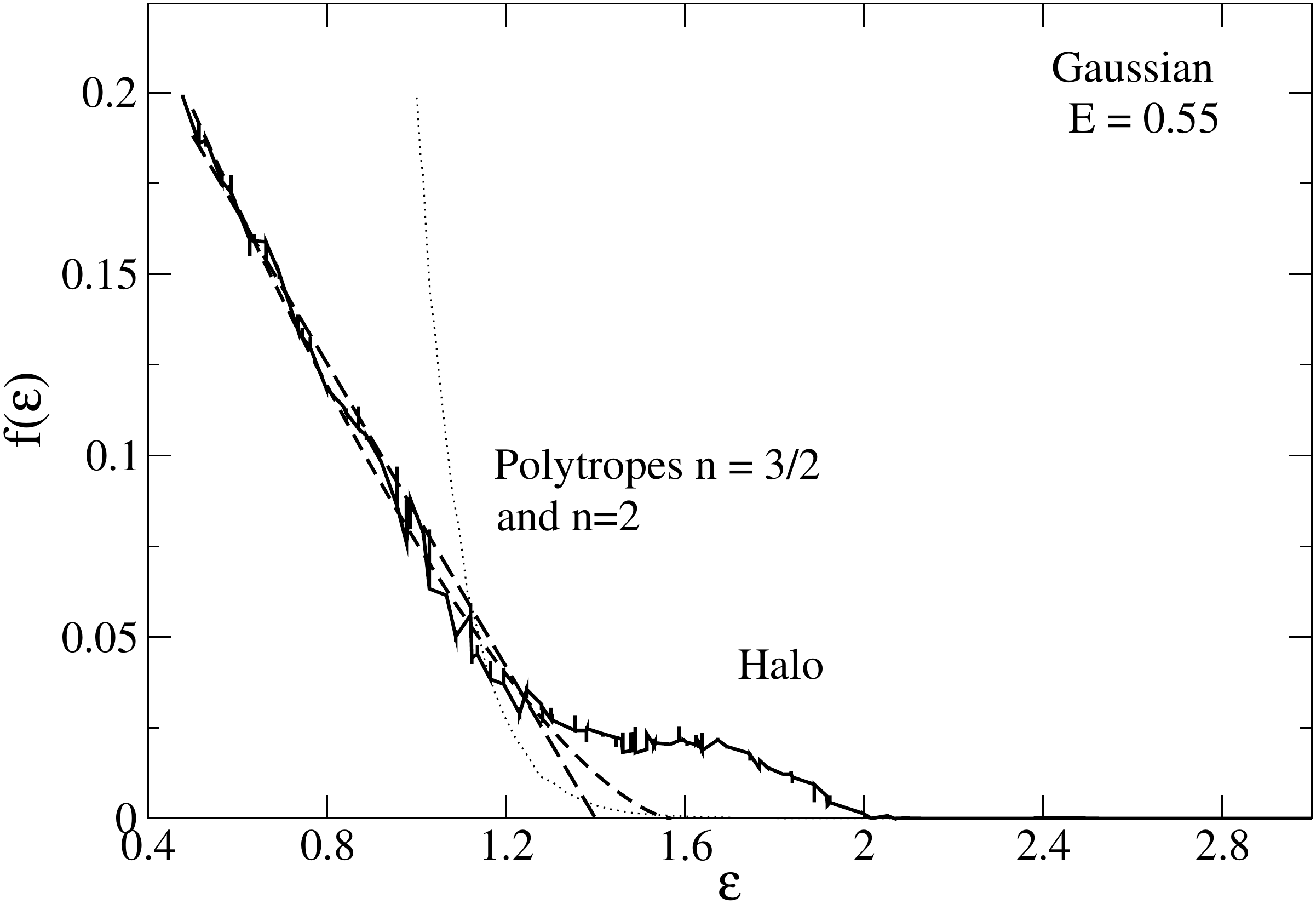}
\caption{The distribution $f(\epsilon)$ has been fitted by a  polytropic core $n=3/2$ (long-dash)
or $n=2$ (short-dash) and a halo. We have also plotted the initial condition (dot).}
\label{feps055gauss}
\end{center}
\end{figure}

In Figures \ref{feps058gauss} and \ref{feps055gauss}   we plot the distribution $f(\epsilon)$ for
the two cases $E=0.58$ and $E=0.55$, respectively. It is clear that, with a very good approximation, the
points are arranged along a single line in both cases, confirming that the state of the system is
a QSS. In particular, we note that the line is practically a straight line in Figure
\ref{feps058gauss}, while it is a straight line in Figure \ref{feps055gauss} up to a certain
energy $\epsilon$, above which it deviates.

Looking at the polytropic distribution, Eq. (\ref{ps2}), and at the
relation between the indices $q$ and $n$ in Eq. (\ref{ps3}), we see
that an $f(\epsilon)$ given by a straight line between the minimum and
maximum values of the individual energy, and zero for higher energies,
corresponds to a polytrope with $n=3/2$.  In Figure
\ref{feps055gauss} and in Figure \ref{feps058gauss} the fit of the
numerical distributions $f(\epsilon)$ with a straight line, i.e., with
a polytrope with $n=3/2$, is reported. However, we report also
the fit with a polytrope with index $n=2$, that, in spite of the
different functional form, is numerically very close to the
$n=3/2$ fit (it is actually slightly better). The use of the $n=2$ polytropes
is in addition justified by the numerical caloric curve that we are
going to discuss shortly.

From the operative point of view the fit is performed by writing Eq. (\ref{ps2}) for $q>1$ in the following form:
\begin{equation}
\label{fepsnum}
f(\epsilon)=A(b-\epsilon)_{+}^{n-1/2}
\end{equation}
with $n \ge {1}/{2}$. Once chosen, from the numerical distribution data, the minimum and maximum
energies within which the fit has to be performed ($\epsilon_{min}$ and $\epsilon_{max}$ respectively),
the parameter $b$ is equal to $\epsilon_{max}$, while $A(b-\epsilon_{min})_{+}^{n-1/2}$ is the value of
the fit at the minimum energy.

The difference between $E=0.55$ and $E=0.58$ is evident:
for $E=0.55$ the part of $f(\epsilon)$ at the higher energies $\epsilon$ is outside the fit. It is due to a halo of particles,
and we have checked that this halo is robust and does not disappear at later times. Later, showing the results
for the waterbag initial conditions, we will present a picture of the location of the particles in the
one particle phase space in another case in which the halo in the distribution function $f(\epsilon)$ is evident.

The numerical distributions $f(\epsilon)$ for the other energies are not plotted here, and we briefly describe
what has been obtained. For the smallest energy, $0.51$, and for the energies $0.60$ and $0.62$, the fit with
$n={3}/{2}$ or $n=2$ is equally good, with the exception of the higher energies $\epsilon$ for $E=0.51$, where a halo
is present as for $E=0.55$. For $E=0.62$ a small bending at the smaller and at the higher energies $\epsilon$ begins to appear
in the numerical distribution, such that its second derivative is negative at small energies and positive
at high enegies. This bending is marked at $E=0.65$ and even more
pronounced at $E=0.69$. For these cases, the fit with a polytrope is not good, especially for $E=0.69$. This can
easily be understood from Eq. (\ref{fepsnum}): the sign of the second derivative of the polytrope is the same in
all the energy range $[\epsilon_{min},\epsilon_{max}]$, i.e., positive for $n > {3}/{2}$ and negative for
$n < {3}/{2}$.

\begin{figure}
\begin{center}
\includegraphics[clip,scale=0.3]{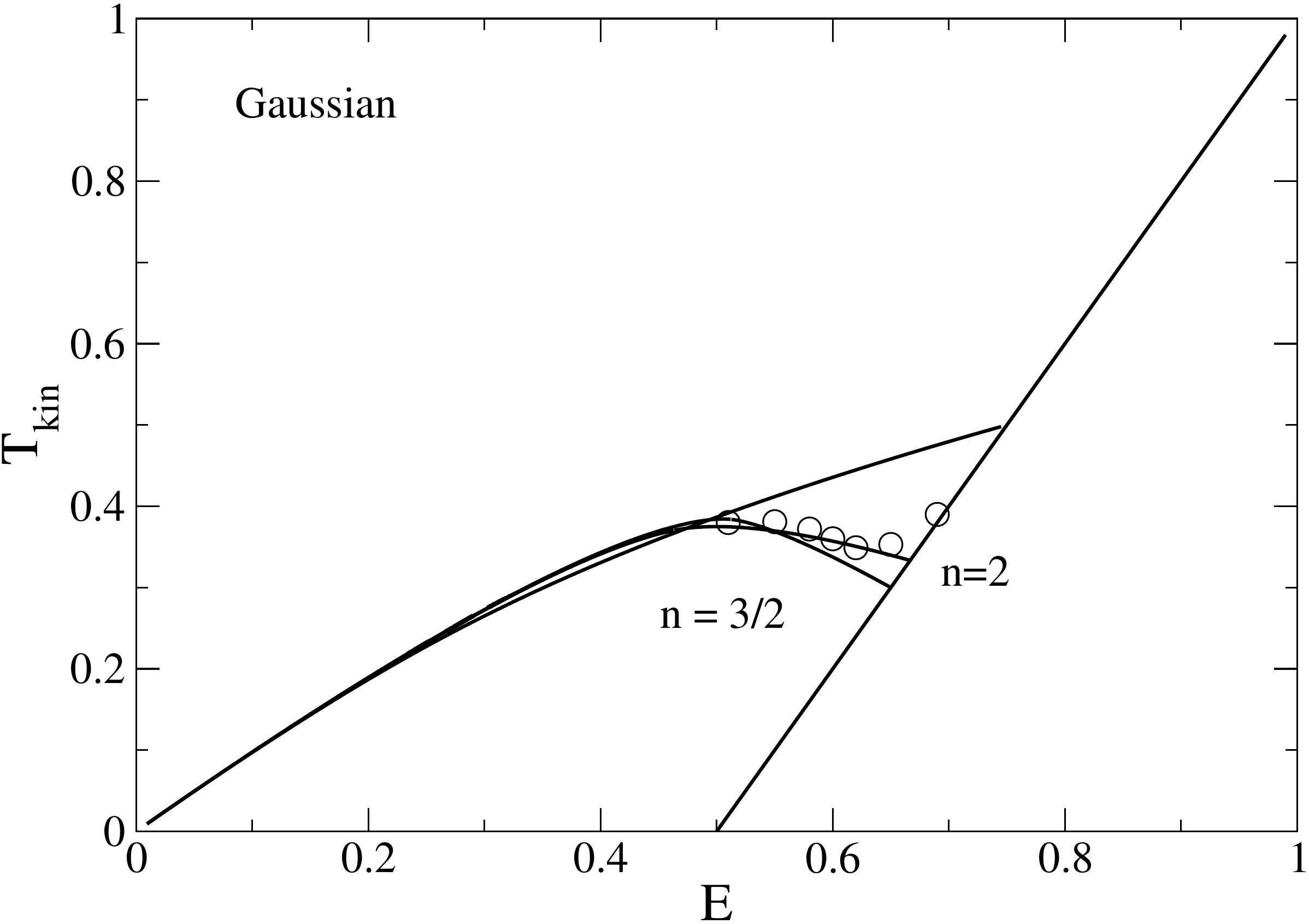}
\caption{The numerical caloric curve displays a region of negative kinetic specific heat $C_{kin}=dE/dT_{kin}<0$. It
has been fitted by a polytrope $n=3/2$ for which $C_{kin}=-43/34$  and by a polytrope $n=2$ for which $C_{kin}=-5/2$.}
\label{gaussian}
\end{center}
\end{figure}

In Figure \ref{gaussian} we plot, with white circles, the kinetic temperature as a function of the energy for the QSS
reached by the system with gaussian initial conditions. Accordingly with the fit performed in 
Figures \ref{feps058gauss} and \ref{feps055gauss}, we fit this numerical caloric curve with the analogous one that holds for
$n={3}/{2}$ and $n=2$ polytropes (the fit by $n=2$ polytropes appears to be very good). Both of them show a negative
kinetic specific heat in that energy range, and so does the numerical caloric curve. The explanation of this negative
specific heat region in terms of polytropic, or close to polytropic distributions, is an important result of our paper.
This is intrinsically due to incomplete relaxation (lack of ergodicity). We ``guess'' that the Lynden-Bell theory, which
assumes complete mixing, would not produce a negative specific heat region\footnote{We have not computed the
prediction of the Lynden-Bell theory which would require to develop a specific algorithm to treat 
multi-levels initial distributions. However, the Lynden-Bell distribution has not a compact support 
so it does not account for observations.}.

We see that at the higher energies, $0.65$ and $0.69$, the fit of the caloric curve is less good, coherently with what
has been found for the fit of the distribution functions. We note that at these energies the kinetic temperature
of the QSS is not much higher than that of the homogeneous branch (equivalently, the magnetization of the QSS is small).
It is then plausible that the relaxation from the homogeneous initial condition to the stable state can be approximately
described by a quasi-linear theory, that treats the inhomogeneity of the distribution function as a perturbation
of the angle-averaged homogeneous distribution. This is left for a future investigation \cite{quasilinear}.

\subsection{Semi-ellipse initial conditions}
\label{sec_ellipse}

Now, for a given energy $E$, the initial velocities are extracted from the distribution
\begin{equation}
\label{ellipseinit}
f_{0}(v) = \frac{1}{\pi \left( 4E -2 \right) } \sqrt{8E -4 -v^2}\,\, ,
\end{equation}
corresponding to a $n=1$ homogeneous polytrope.
The critical energy for Vlasov stability is $E_c = {5}/{8} = 0.625$. We have performed runs at the energies
$0.51$, $0.55$, $0.58$, $0.59$, $0.60$ and $=0.61$. As before, after a rapid relaxation the system reaches
a QSS with a magnetization and a kinetic temperature smaller than the respective equilibrium values.

\begin{figure}
\begin{center}
\includegraphics[clip,scale=0.3]{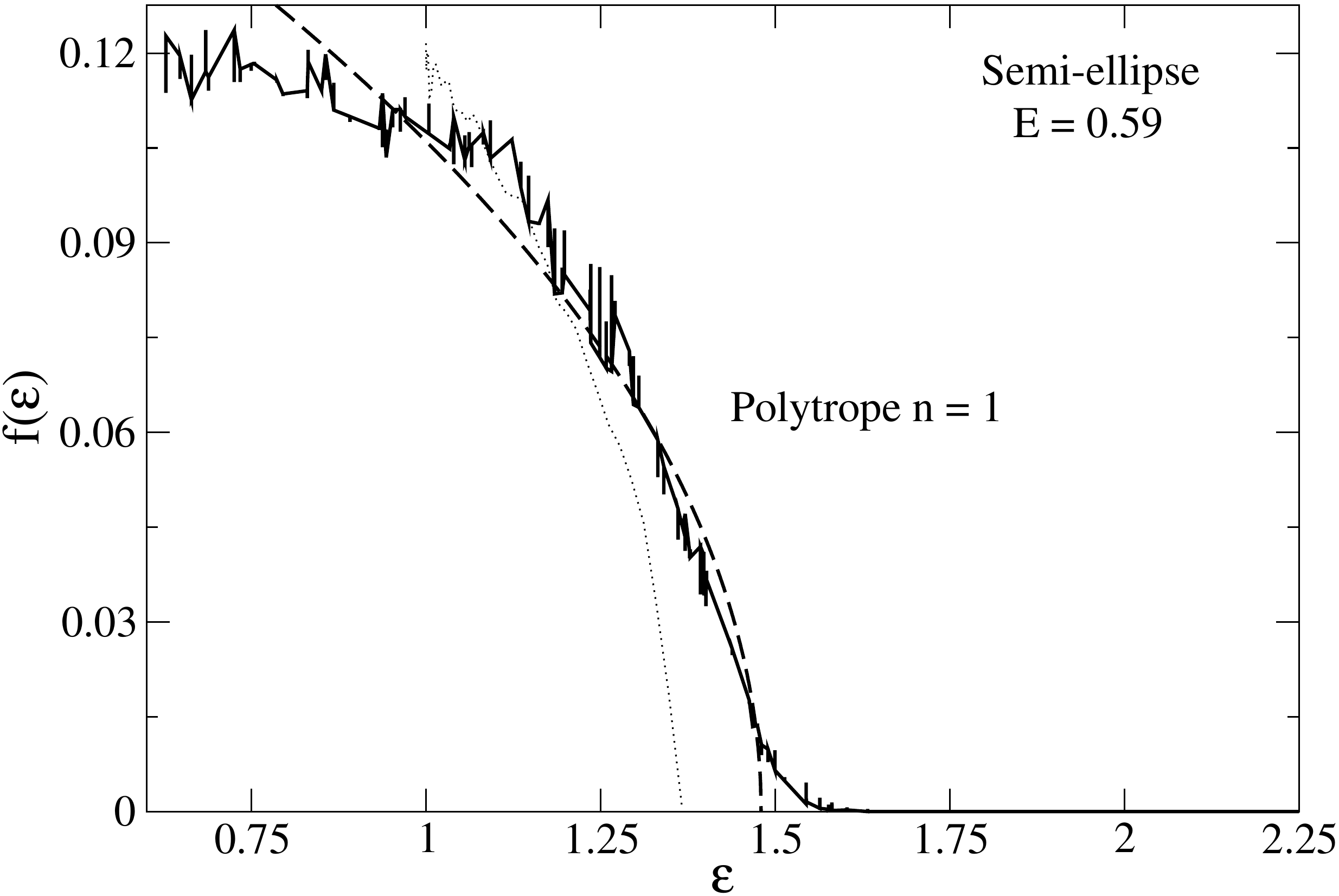}
\caption{The distribution $f(\epsilon)$ has been fitted by a pure  polytrope $n=1$ (dash). 
We have also plotted the initial condition (dot).}
\label{feps059ellipse}
\end{center}
\end{figure}

\begin{figure}
\begin{center}
\includegraphics[clip,scale=0.3]{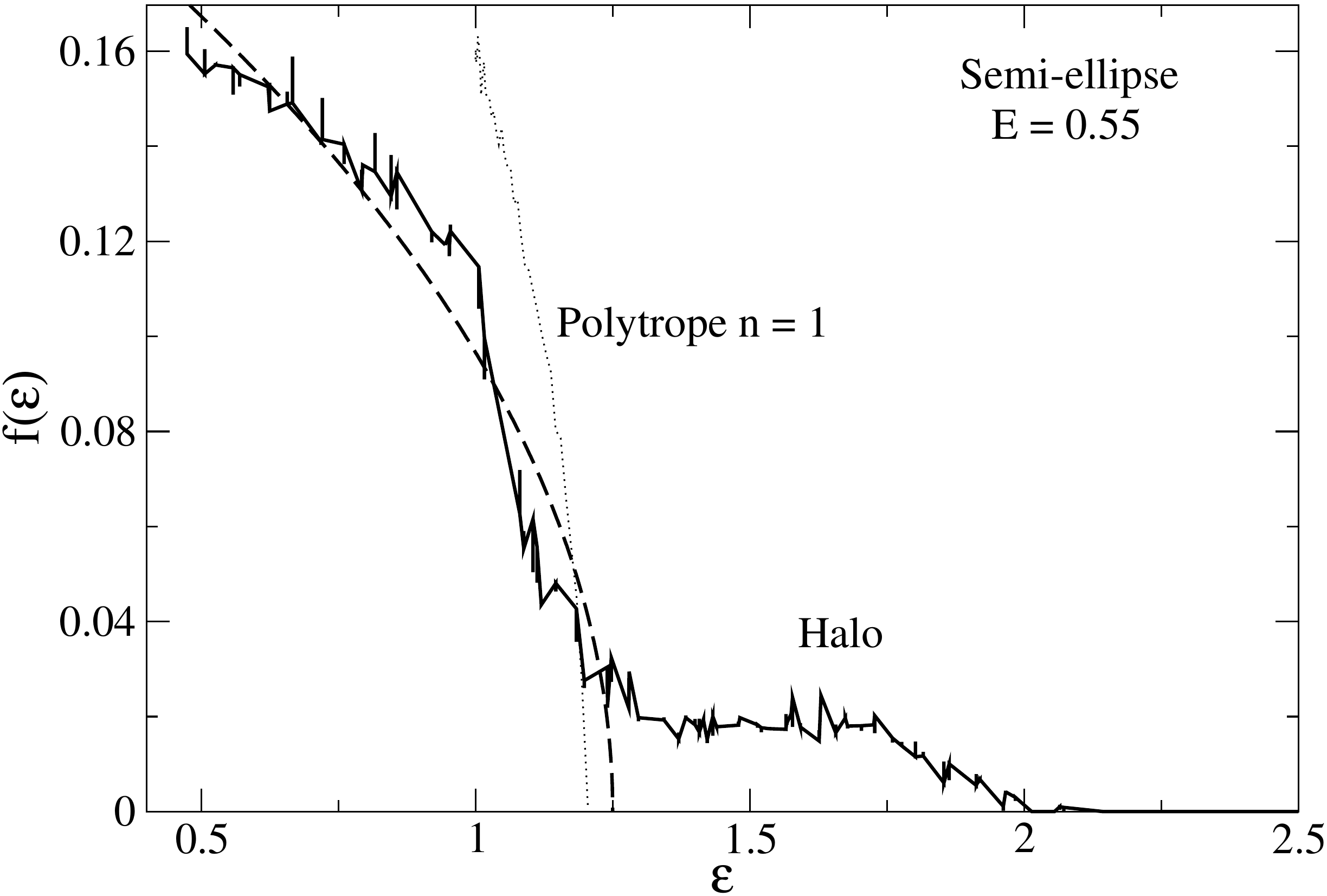}
\caption{The distribution $f(\epsilon)$ has been fitted by a  polytropic core $n=1$ (dash) and a halo.
We have also plotted the initial condition (dot).}
\label{feps055ellipse}
\end{center}
\end{figure}

We do not show analogous plots of the magnetization vs time, since they are very similar to those obtained for
the gaussian initial conditions, with small persistent oscillations. We go directly to the plots of the
numerical distribution functions. We show the two cases $E=0.59$ and $E=0.55$, respectively in Figures
\ref{feps059ellipse} and \ref{feps055ellipse}.

The points of the numerical distribution functions are arranged along a line, as it should be for a QSS.
As for the gaussian case, the distributions at the smaller energies $E$ present a halo. The fit with a polytropic
distribution has been performed with the $n=1$ polytrope. The fit appears rather good in both cases, of course
except for the halo at $E=0.55$. The picture is similar as for the gaussian case, i.e., there is a halo at
small energies (we have one also for $E=0.51$) and the fit with the polytrope worsen at high energies,
here $E=0.60$ and $E=0.61$, when the sign of the second derivative of the numerical distributions is negative
at small energies $\epsilon$  and positive at high energies.
Actually, we found that also at energy $E=0.58$ the fit with a polytrope distribution is not satisfying.

\begin{figure}
\begin{center}
\includegraphics[clip,scale=0.3]{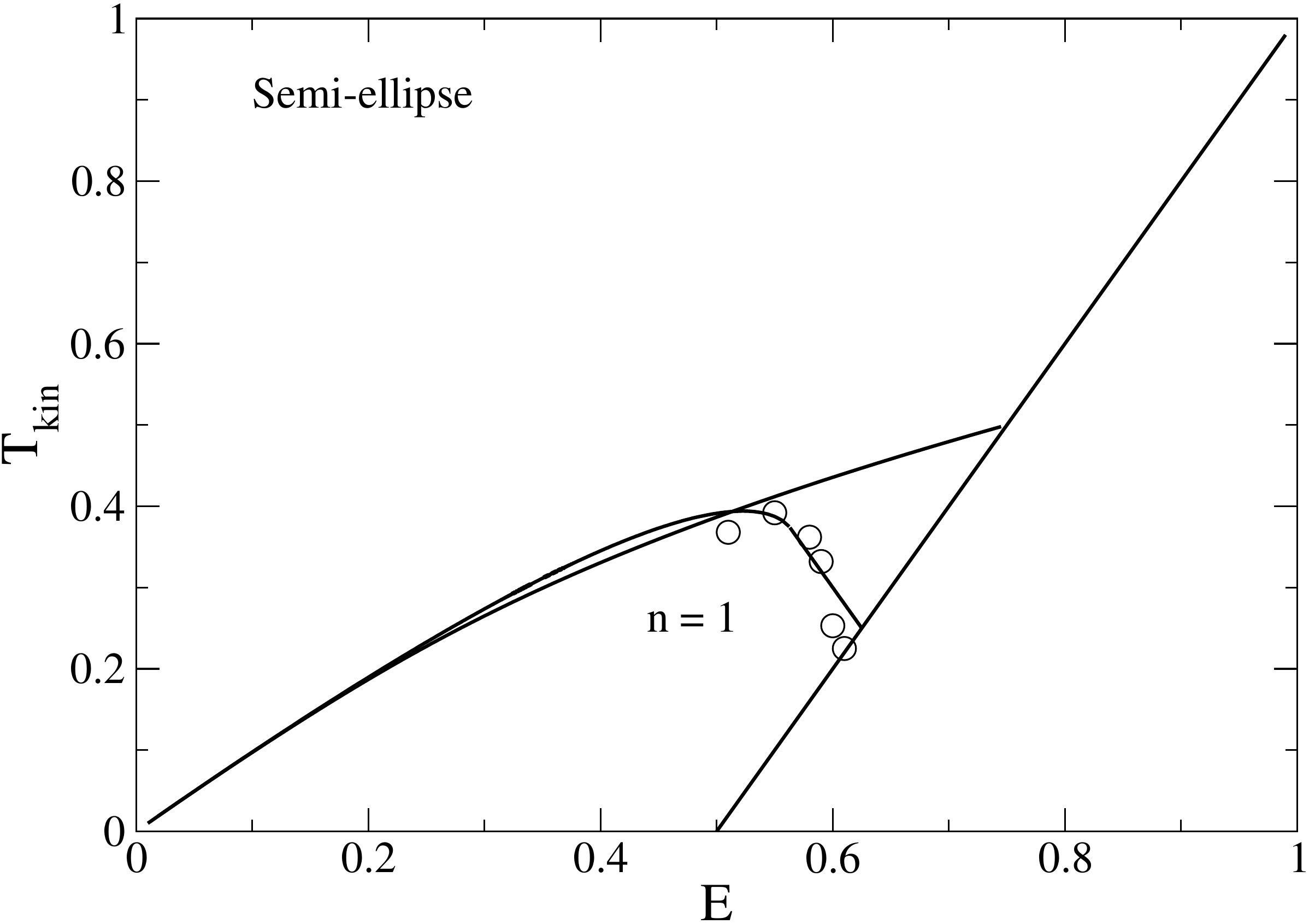}
\caption{The numerical caloric curve displays a region of negative kinetic specific heat $C_{kin}=dE/dT_{kin}<0$. It has
been fitted by a polytrope $n=1$ for which $C_{kin}=-1/2$.}
\label{ellipse}
\end{center}
\end{figure}

In Figure \ref{ellipse} we show the numerical kinetic caloric curve of the QSS reached by the system, fitted by
the $n=1$ polytropic kinetic caloric curve, that has a negative kinetic specific heat in that energy range. 
We see that the polytropic fit is relatively good for most energies.
As for the gaussian case, at the highest energies the fit is less good, and probably the QSS in that case could
be better explained by a perturbative analysis.

We remark that the polytropic index that fits the QSS is the same as that of the initial velocity distribution, i.e. $n=1$.

\subsection{Waterbag initial conditions}
\label{sec_waterbagnum}

\begin{figure}
\begin{center}
\includegraphics[clip,scale=0.3]{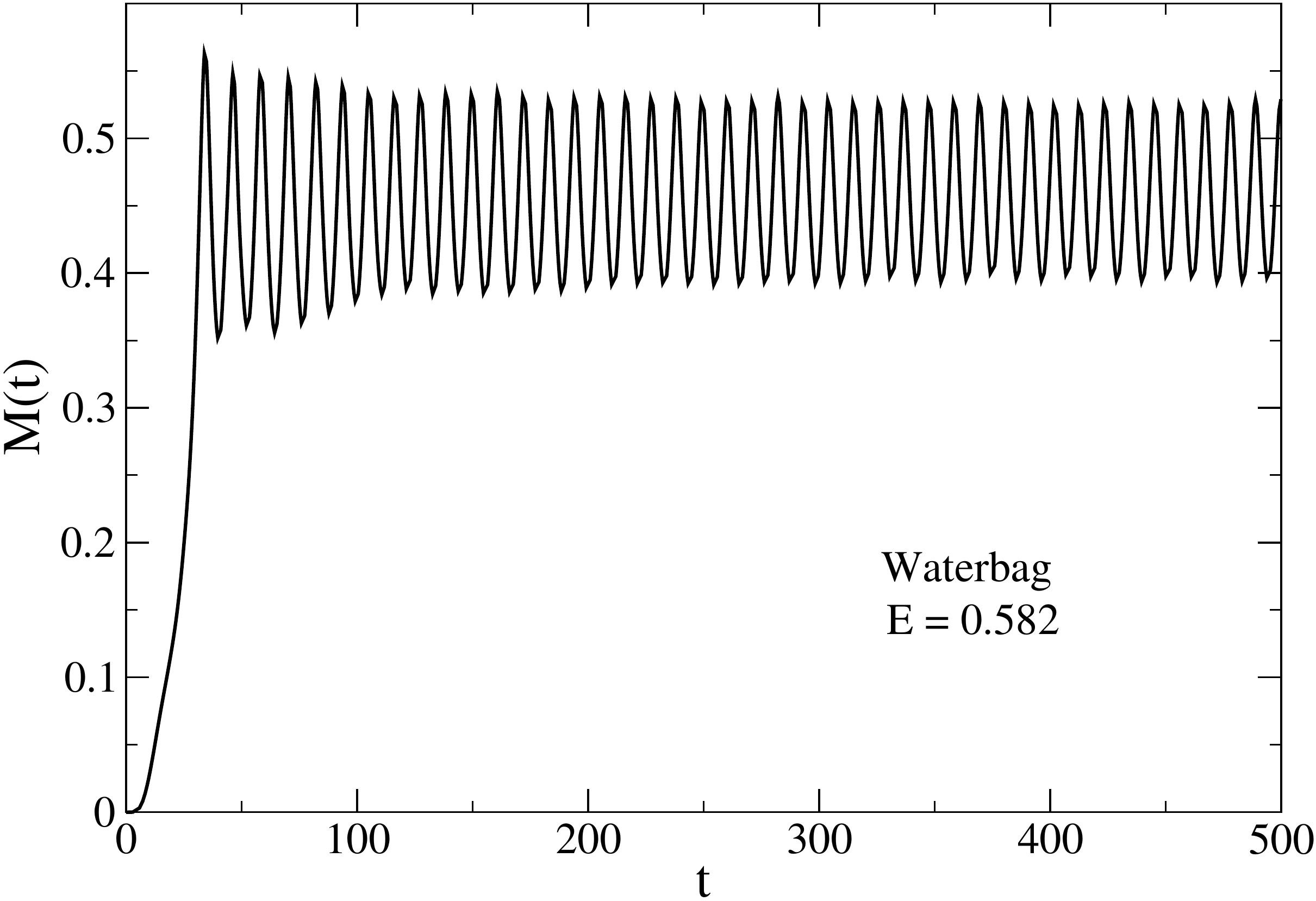}
\caption{Magnetization as a function of time $M(t)$ for $E=0.582$.
It shows undamped (or slowly damped) oscillations. This
may be related to the absence of Landau damping for the QSS ($=$
inhomogeneous waterbag) as for the homogeneous waterbag distribution.}
\label{magnetisation0582gauss}
\end{center}
\end{figure}

\begin{figure}
\begin{center}
\includegraphics[clip,scale=0.3]{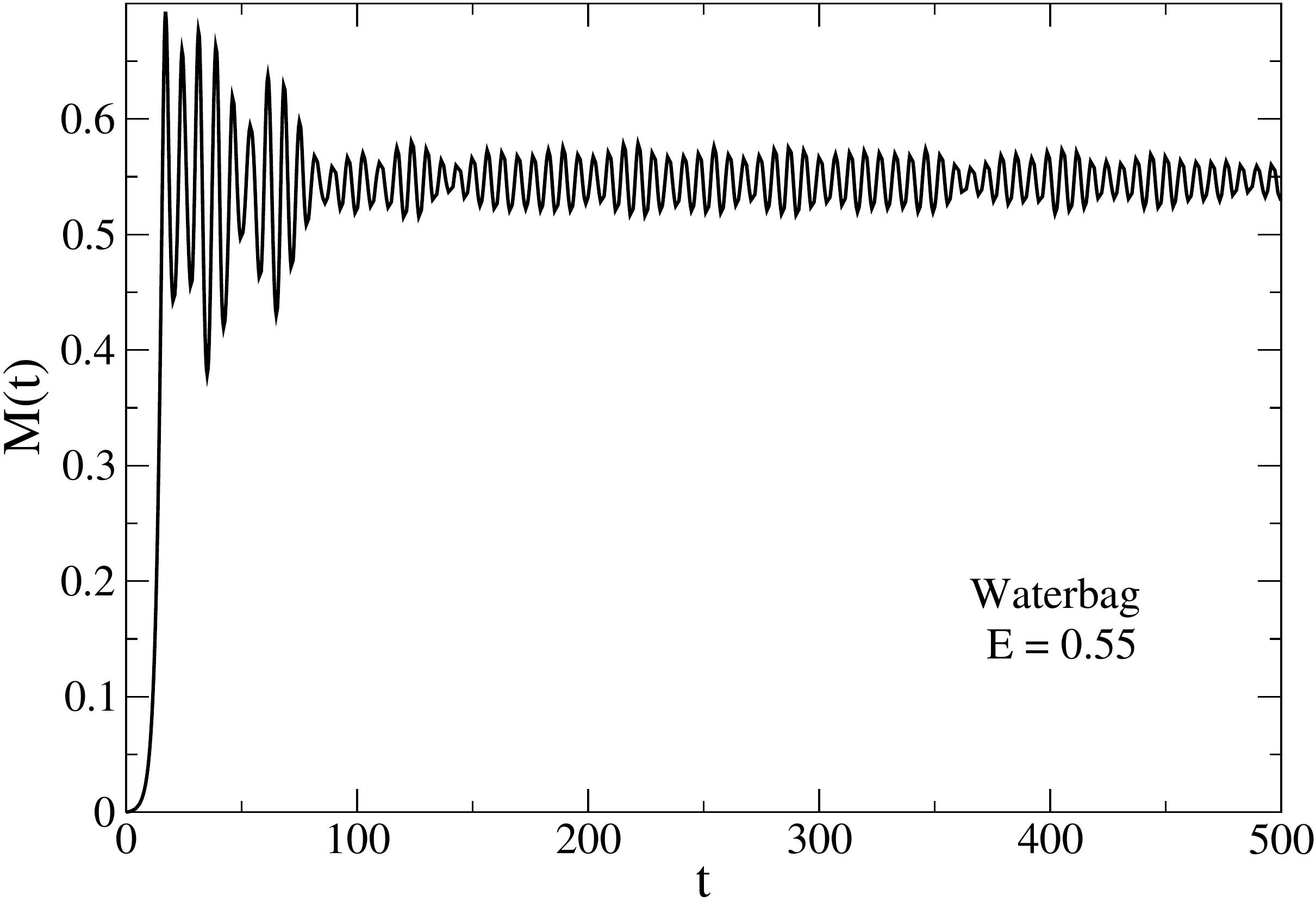}
\caption{Magnetization as a function of time $M(t)$ for $E=0.55$. The amplitude of the oscillations
is smaller than in Figure \ref{magnetisation0582gauss}. This may be due to the presence of the halo. }
\label{magnetisation055waterbag}
\end{center}
\end{figure}

For this class of initial conditions, the initial velocities at a given $E$ are extracted from the distribution
\begin{equation}
\label{waterbaginit}
f_{0}(v) = \frac{1}{2\sqrt{6E -3}} \Theta \left( \sqrt{6E - 3} -|v| \right) \,\, ,
\end{equation}
corresponding to a $n={1}/{2}$ polytrope.
The critical energy for Vlasov stability is $E_c = {7}/{12}$. We have performed runs at the energies
$0.51$, $0.55$, and $0.582$. The last energies is only very slightly smaller than the critical energy.
While sharing the previous general picture, i.e., the rapid approach to a magnetized QSS, there are some new features.
We begin by showng the plots of the magnetization vs time for the two 
energies  $E=0.582$  and $E=0.55$, respectively
in Figures  \ref{magnetisation0582gauss} and \ref{magnetisation055waterbag}.

We note that the magnetization oscillations are more marked than previously, becoming  very pronounced at
$E=0.582$. This is relatively close to the situation reported by Morita \& Kaneko \cite{mk}. We noted before that
one possible explanation of the oscillations is a Landau damping associated
to the inhomogeneous QSS. Now, these large oscillations suggest that this damping could be very weak. It
is suggestive to interpret this as a continuation with what happens for the homogeneous waterbag QSS, for
which we know that Landau damping is absent (this being due to the singularity of the waterbag distribution,
that, contrary to distributions strictly decreasing for increasing $|v|$, admits purely 
real proper frequencies).

\begin{figure}
\begin{center}
\includegraphics[clip,scale=0.3]{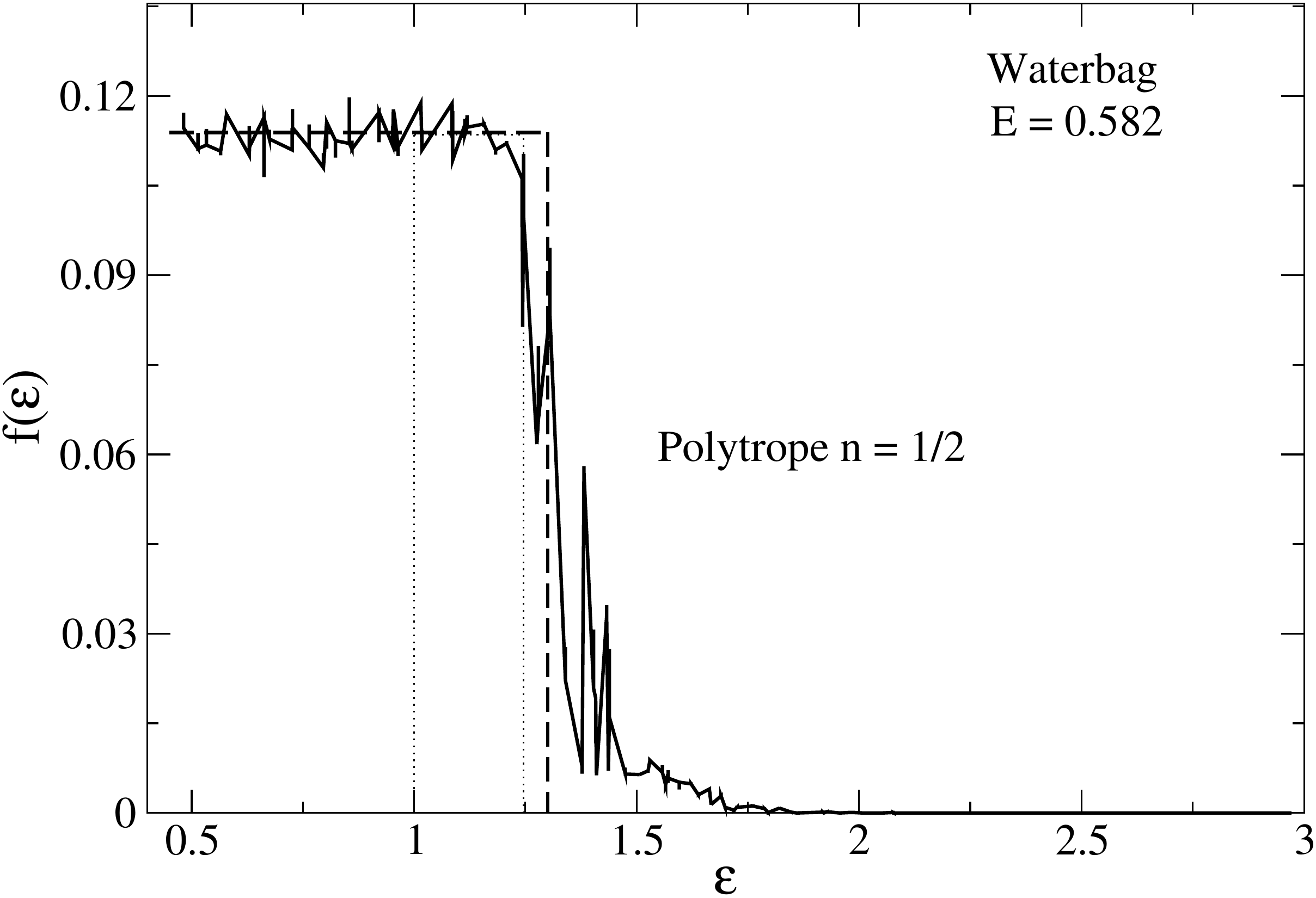}
\caption{The distribution $f(\epsilon)$ has been fitted by a ``pure'' homogeneous distribution (waterbag) corresponding to
a $n=1/2$ polytrope. We have also plotted the initial
homogeneous waterbag distribution (dot).}
\label{feps0582waterbag}
\end{center}
\end{figure}

\begin{figure}
\begin{center}
\includegraphics[clip,scale=0.3]{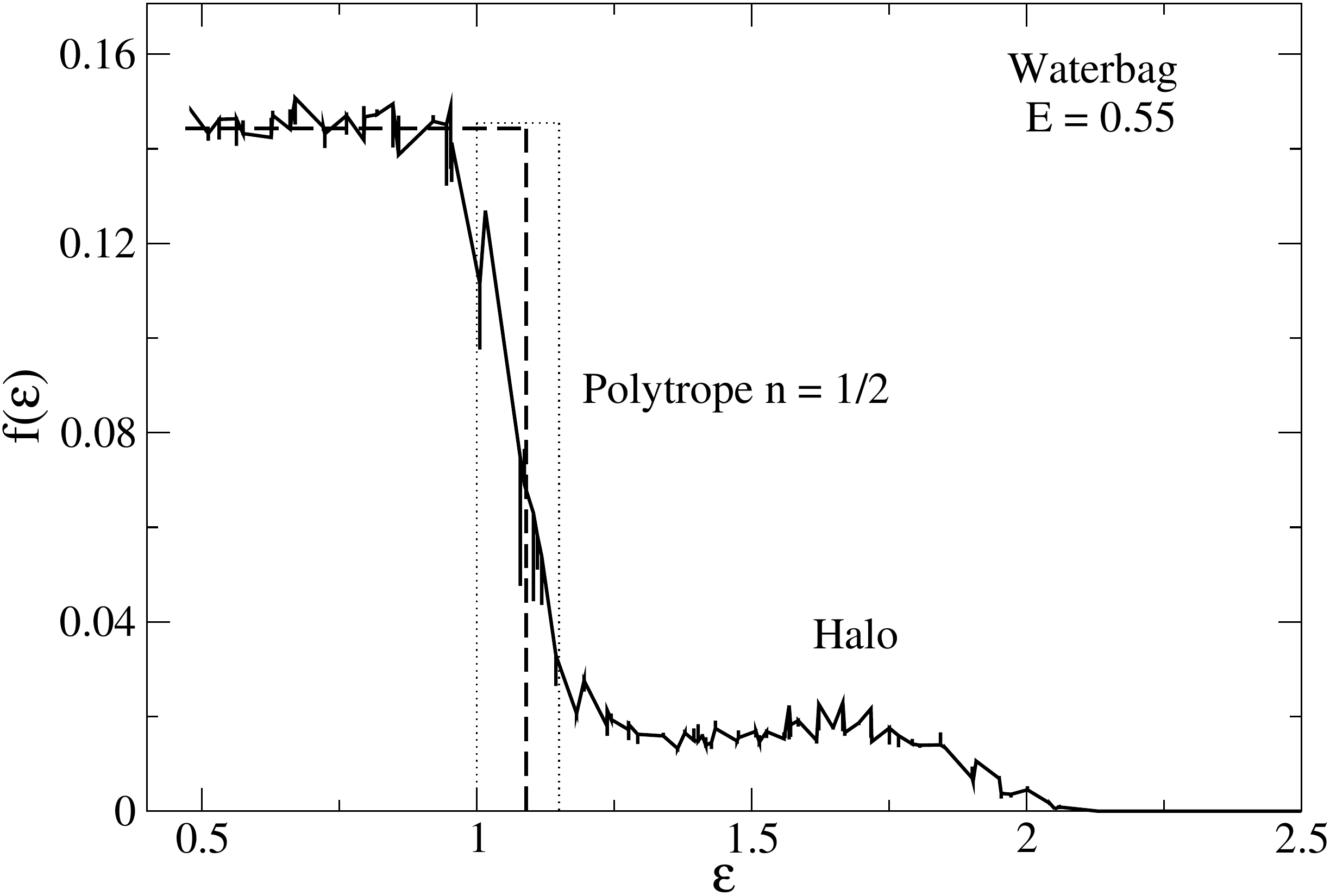}
\caption{The distribution $f(\epsilon)$ has been fitted by  a $n=1/2$ polytrope (core) with a halo. This
``core-halo'' structure is similar to the one found in \cite{levin}. We have also plotted the initial
homogeneous waterbag distribution (dot).}
\label{feps055waterbag}
\end{center}
\end{figure}

\begin{figure}
\begin{center}
\includegraphics[clip,scale=0.3]{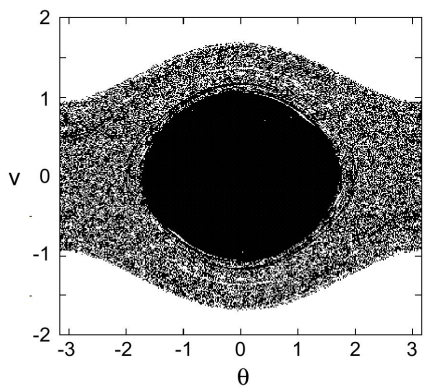}
\caption{Occupation of the one particle phase space in the QSS state at $E=0.55$. The halo, i.e.,
the region less densely populated, is clearly visible.}
\label{halowaterbag}
\end{center}
\end{figure}

Nevertheless, we proceed as before and plot the numerical distributions as a function of the individual
energy, using the average value of $M$ to define the last quantity. In Figures
\ref{feps0582waterbag} and \ref{feps055waterbag}, we plot the numerical distributions for the cases $E=0.582$ and $E=0.55$,
together with the fit with the $n={1}/{2}$ polytrope. 
At $E=0.582$, the QSS is an almost pure inhomogeneous waterbag 
distribution and the phase portrait looks like Fig. \ref{phasespaceCOMPLET}. 
At the smaller energy $E=0.55$ there is a halo that cannot be reproduced by the polytropic
fit. This halo is clearly evident in Figure \ref{halowaterbag} that
shows the location of all the particle in the one particle phase space
in the QSS state at $E=0.55$.  The boundary between the more densely
populated region and the less densely populated region (the halo) is
at the energy where the halo begins in Figure \ref{feps055waterbag}.

\begin{figure}
\begin{center}
\includegraphics[clip,scale=0.3]{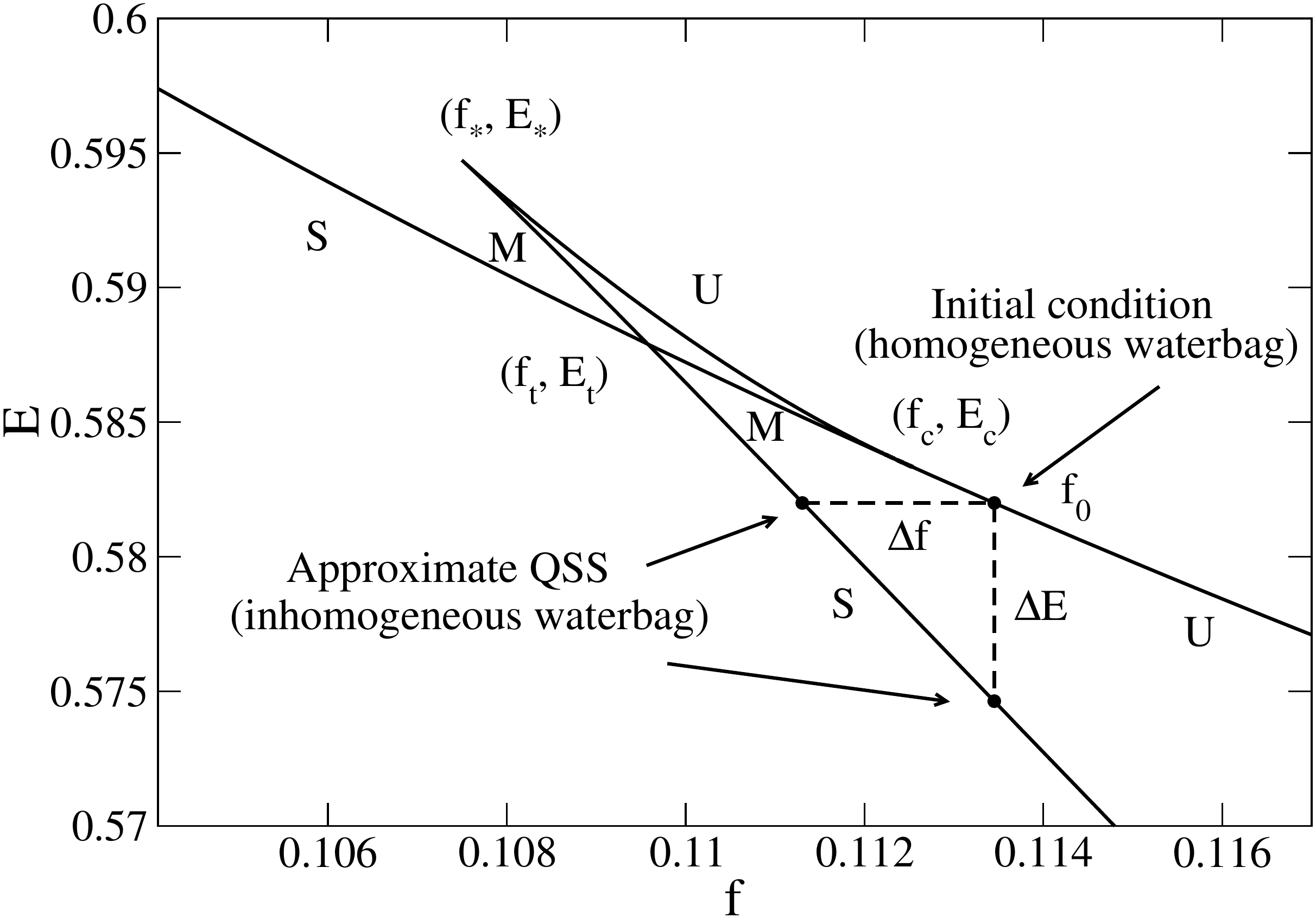}
\caption{Relation between the energy $E$ and the distribution value $f$ of a waterbag distribution
that is a steady state of the Vlasov equation \cite{hmfq1}. Homogeneous and inhomogeneous distributions are represented. 
In the simulations, we start from an unstable homogeneous waterbag distribution with energy $E$ and distribution
value $f_0$. We see that only at $E=E_t$ does the inhomogeneous polytrope have the same $f$ as the 
homogeneous polytrope. In the other cases, a halo is necessary to maintain the initial value $f_0$ in the core of
the QSS (as observed in the simulations) while conserving energy. However, close to $E=E_t$, the effect of the 
halo is weak and a pure waterbag distribution provides a good approximation
of the QSS.}
\label{epsilonVSmu}
\end{center}
\end{figure}

As for the semi-ellipse initial conditions, we remark that the
polytropic index of the QSS is the same as that of the initial velocity distribution, i.e. here $n=1/2$. A difference
with respect to the two previous cases is that the fit appears to be good also very close to the
critical energy $E_c$. Furthermore, the value of the numerical distribution function in the QSS is
practically the same as that of the initial waterbag distribution (see Figures
\ref{feps0582waterbag} and \ref{feps055waterbag}). This means that the core does not 
mix at all. Using this observation, we can explain the presence or the absence 
of the halo. To that purpose,  we plot in Fig. \ref{epsilonVSmu} the relation between 
the uniform distribution value $f$ and the 
energy $E$ of a pure waterbag distribution that is solution of the Vlasov equation 
(the construction of this Figure is explained in 
\cite{hmfq1}). This Figure
shows the first order phase transition between homogeneous and inhomogeneous waterbag distributions
discussed in Sec. \ref{sec_waterbag}. We see that, in
general, the homogeneous and inhomogeneous waterbag distributions {\it with the same $f$} have a different
energy (see the vertical line in Fig. \ref{epsilonVSmu}). Therefore, if $f$ is the same in the initial homogeneous waterbag distribution and in the 
inhomogeneous waterbag QSS (as it turns out to be), there must necessarily exist a halo of particles
in order to satisfy the conservation of energy. The halo should be particularly important at low energies $E$ 
(as in Fig. \ref{feps055waterbag} for $E=0.55$) where $\Delta E$ is large. 
By contrast, at the transition energy $E_t$, the homogeneous and
inhomogeneous waterbag distributions have the same $f$ and $E$ so that the presence of a halo is not 
required. By continuity, close to the transition energy (as in Figure \ref{feps0582waterbag} 
for $E=0.582$), the halo should be modest since $\Delta E$ is small (for $E=0.582$ we find $\Delta E/E=0.01$). These 
arguments are consistent with the observations.

\begin{figure}
\begin{center}
\includegraphics[clip,scale=0.3]{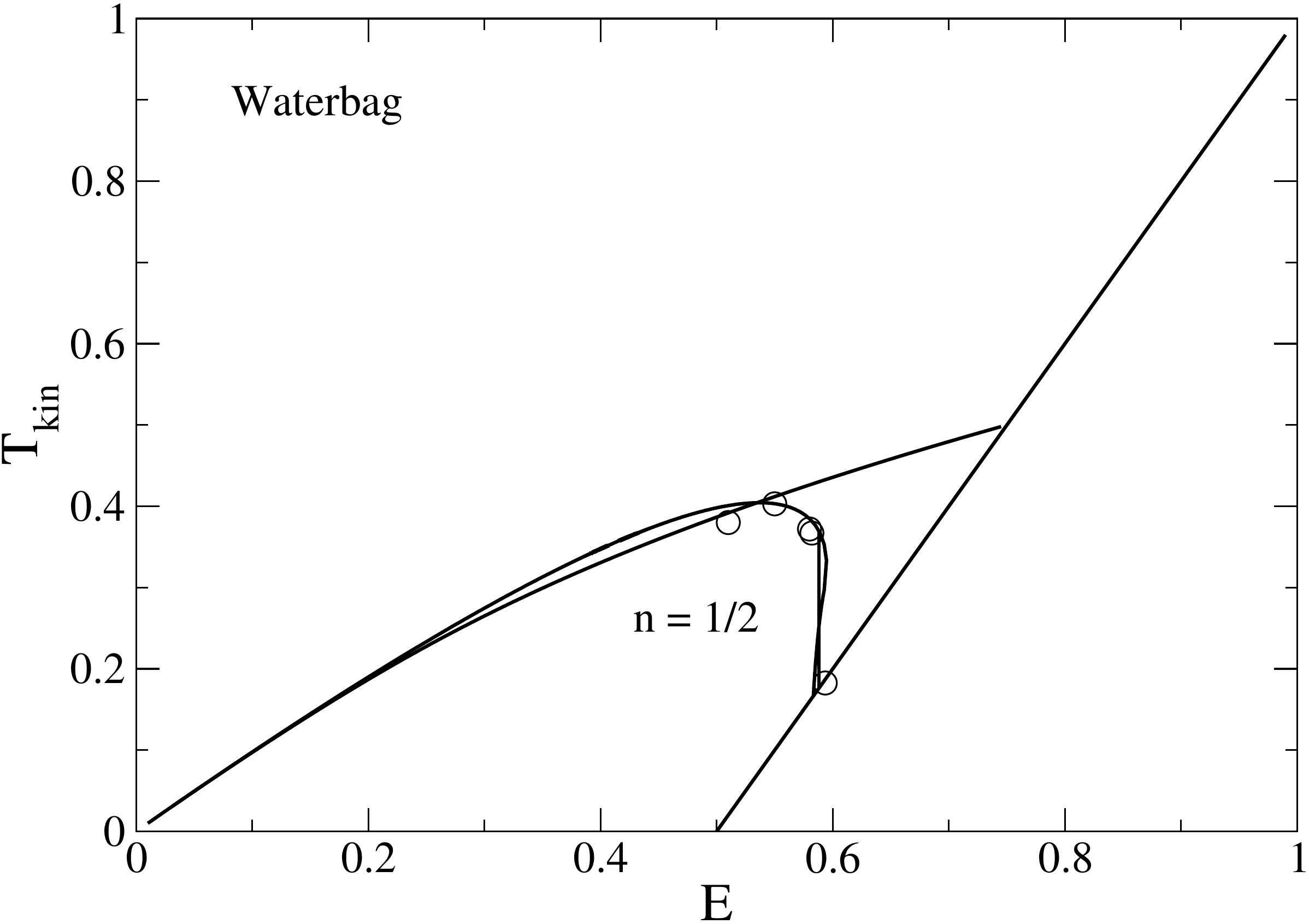}
\caption{The numerical caloric curve has been fitted by a polytrope $n=1/2$. Strictly speaking,
this caloric curve exhibits a first order phase
transition at $E_t=0.588$, leading to
$C_{kin}=0$. However, metastable states are equally relevant as fully
stable states, so the whole series of equilibria has been
plotted.}
\label{waterbag}
\end{center}
\end{figure}

In Figure \ref{waterbag} we plot the numerical kinetic caloric curve, together with the $n={1}/{2}$
kinetic caloric curve. In principle, the $n={1}/{2}$
kinetic caloric curve should display a first order phase
transition as shown in Fig. \ref{caloTkinN0p5ZOOM}. However, for systems with long-range interactions, the metastable
states (local entropy maxima) are extremely robust, and they are as much relevant as fully stable states
(global entropy maxima). For that reason, we have chosen to represent the full series of equilibria, displaying both
global entropy maxima, local entropy maxima, and even saddle points of entropy.

The first thing to note is that, apart from the highest energies, the kinetic temperature is very close to that
of the BG equilibrium. However, as clearly proved from the numerical distribution functions, the state is far
from being the BG one, that has a Boltzmann distribution. The second thing to  note it  that, also in this case, there
is a region of negative kinetic specific heat.

A striking feature of Figure \ref{feps055waterbag} is the core-halo state. This core-halo state,
arising from a waterbag initial condition, was previously observed by Pakter \& Levin \cite{levin},
although they did not explicitly calculate the curve $f(\epsilon)$ (they observed the core-halo state
from the phase space portrait). A new contribution of our work is to show that this core-halo state is
also present at low energies for other types of initial conditions 
(see Secs. \ref{sec_gauss} and \ref{sec_ellipse}). In all the cases considered, the core can be
fitted by a polytrope with a different index ($n=1/2$ for the waterbag initial condition). This generalizes
the results of \cite{levin}.

There are a few differences between our approach and the approach of \cite{levin}. First, we have
considered a waterbag initial condition with $M_0=0$ while they took $M_0=0.40$. For $M_0=0.40$, the
Lynden-Bell theory predicts a second order phase transition (see \cite{prl2} or Figures 2 and 4
of \cite{staniscia1})  which is in clear disagreement with the first order phase transition reported
in \cite{levin}. By contrast, for $M_0=0$, the Lynden-Bell theory predicts a first order phase transition.
It is, however, different from the first order phase transition reported in Figure \ref{waterbag} because
the Lynden-Bell prediction corresponds to a distribution that is {\it partially} degenerate while we find a distribution that is either completely
degenerate, or with a core-halo structure. Therefore, both in \cite{levin} and in the present study,
the Lynden-Bell prediction fails although the situation is a bit different. Secondly,
in order to obtain their theoretical magnetization curve, Pakter \& Levin \cite{levin}
assume that the distribution function in the core of the QSS is equal to the initial distribution
$f_0$ (our numerical distribution functions give further
support to this assumption, as shown in Figures
\ref{feps0582waterbag} and \ref{feps055waterbag}) and determine the properties of
the halo by a semi-analytical approach. While we agree with their
procedure which provides a good prediction of the QSS for all energies, 
we have proceeded differently. We have obtained the theoretical magnetization curve (or kinetic
caloric curve) by assuming that the QSS is a {\it pure} polytrope $n=1/2$. Since we ignore the halo, 
the distribution function $f_{th}$ that we theoretically compute is generally different from the initial distribution 
function $f_0$ in order  to satisfy the conservation of energy (see the horizontal line 
in Fig. \ref{epsilonVSmu}). This procedure provides 
a reasonable agreement with 
the numerical simulations in the region where
the halo is not pronounced, i.e. close to the transition energy, where $\Delta f/f_0\ll 1$
(for $E=0.582$ we find $\Delta f/f=0.02$). By contrast, it clearly fails 
for lower energies where the halo is significant. This simply reflects the fact that a pure polytrope cannot 
hold for all the energies, and that a core-halo state is required.

\begin{figure}
\begin{center}
\includegraphics[clip,scale=0.3]{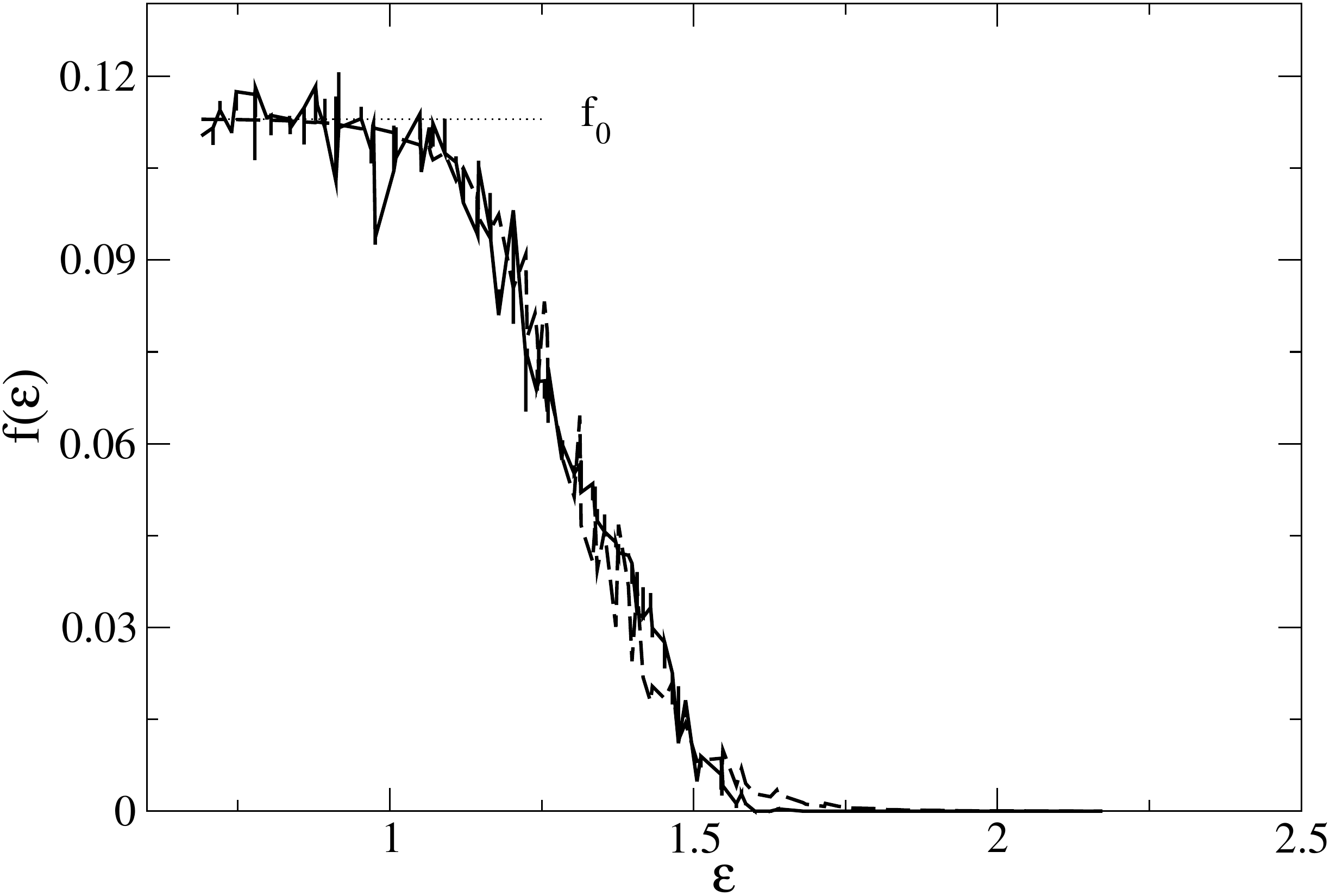}
\caption{Full line: The distribution $f(\epsilon)$ obtained in the QSS reached by the waterbag initial condition
with $E=0.6$ and $M_0=0.1711$ (we have indicated by a dotted line the value of the initial 
distribution function $f_0$). Dashed line: the distribution predicted by the Lynden-Bell theory. We see 
that the agreement is almost perfect. We may also plot $\ln\lbrack \overline{f}/(f_0-\overline{f})\rbrack$ as a function of $\epsilon$, as 
suggested in \cite{staniscia2}, and  check that it is a straight line. In this way, we do not have 
to compute the theoretical Lynden-Bell distribution. The Lagrange multipliers $\alpha$ and $\beta$ 
may be determined from the 
relation $\ln\lbrack \overline{f}/(f_0-\overline{f})\rbrack=-\beta f_0 (\epsilon+\alpha)$.}
\label{distrlynden}
\end{center}
\end{figure}

In this paper, we have chosen to focus on examples where the Lynden-Bell prediction fails. 
However, as recalled in the Introduction, we know that the Lynden-Bell predictions are in 
many cases verified. In order to moderate our message about the inadequacy of the Lynden-Bell distribution
to describe the QSS in certain cases, we give here an example
of a QSS in which the numerical distribution function agrees with that predicted by the Lynden-Bell theory
with extremely good precision. In Figure \ref{distrlynden} we show the numerical distribution function
in the QSS reached by the system initially prepared with a waterbag distribution of the same kind as those
considered in \cite{precommun}; precisely, we considered a rectangular waterbag initial distribution at $E=0.6$
and initial magnetization $M_0=0.1711$ (this initial condition corresponds to the point
$E=0.6$ and $f_0=0.113$ on Figure 10 of \cite{staniscia1}). The Lynden-Bell theory predicts in this case a magnetized
QSS. Not only do we find the magnetization in the simulation  agrees with that of the Lynden-Bell theory 
($M_{QSS}=0.348$), 
as observed previously in \cite{staniscia1},
but we also find that the numerical $f(\epsilon)$ is that predicted by the theory (a feature
that was not explicitly checked in \cite{staniscia1}). The Lynden-Bell function
is plotted, in Figure \ref{distrlynden}, together with the numerical distribution function.

\section{The approach to BG equilibium of homogeneous Vlasov stable states}
\label{sec_rising}

In previous works \cite{campa1,campa2}, we have studied the modalities of the approach to BG equilibrium of the system
prepared in a Vlasov stable homogeneous state at energies below the thermodynamical critical energy $E_c=0.75$. In
that case, we were interested in the lifetime of these homogeneous states,
whose slow evolution is governed by the finite size effects. We know that this evolution changes slowly the state
of the system, that remains homogeneous, until the distribution becomes Vlasov unstable and the system
begins to magnetize and to approach equilibrium. In Ref. \cite{campa2} we showed that, during the 
slow ``collisional'' evolution, and
for homogeneous distributions, the velocity distribution $f(v,t)$ can be fitted with good approximation by a polytropic
function, whose index $n(t)$
changes with time in correspondence with the change of the distribution. We considered in particular the mostly
studied energy $E=0.69$, preparing the system with velocities extracted from a semi-elliptical $n=1$
polytrope\footnote{In \cite{campa1}, the authors started from a rectangular waterbag distribution with energy $E=0.69$
and vanishing magnetization $M_0=0$. This initial condition is Vlasov stable since $E>E_c=7/12$ so it does not
experience phase mixing and violent relaxation. However, it slowly evolves due to finite $N$ effects. They found
that the system rapidly forms a velocity distribution with a semi-elliptical shape. This can be interpreted as a
polytrope $n=1$ \cite{hb3}. Yamaguchi {\it et al.} \cite{yamaguchi} had previously obtained the same result but they
did not to recognize the polytrope (Tsallis distribution); see discussion in \cite{hb3}. In Ref. \cite{campa2}, we
directly started from the polytrope $n=1$ to accelerate the simulation.}.  This initial state is Vlasov stable, since
$E>E_c = {5}/{8}$. We found that, during the
dynamics, the index $n(t)$ of the fitting polytrope increases. Solving Eq. (\ref{hpjt7_b}) for $n$, we see that for
a given energy $E<{3}/{4}$ the homogeneous polytrope is Vlasov stable if, and only, if  $n<n_c=(4E -2)/(3-4E)$. For $E=0.69$ this
gives $n<n_c={19}/{6}$. We found that when $n$ approaches the critical value $n_c$ the homogeneous
distribution begins to destabilize, and the system begins to magnetize.

In Ref. \cite{campa2} we had not analyzed the  distribution functions after the homogeneous phase becomes Vlasov unstable,
and the system becomes magnetized. Here,  we are interested to know whether the magnetized states $f(\epsilon,t)$ can still
be fitted by inhomogeneous polytropes with a time dependent index $n(t)$. If true, the index will start from about
$n_c=19/6$ and increases towards infinity, which corresponds to the BG state, as time goes on. From the study of \cite{cc},
we know that the inhomogeneous polytropes with index $n>n_{MCP}\simeq 0.68$ are always Vlasov stable (see Remark 1). Therefore, we conclude
that the whole sequence of inhomogeneous polytropes with index larger than $n_c=19/6$  is Vlasov stable.

Here, we have performed a simulation of the system initially prepared, at $E=0.69$, in a homogeneous state with the
velocities distributed according to the semi-elliptical distribution. This is the same initial condition considered
in \cite{campa2}, but now we have studied the one-particle distribution function $f(\epsilon,t)$ in the time range mentioned
before (i.e., in the magnetized phase). We could not use the same number of particles as for the analysis of the QSS reached
from a Vlasov unstable
state (see Sec. \ref{sec_simul}) since this would have required an unmanageable computer time. Our runs have been made
with $2^{11}$ particles,
representative of a system of $2^{12}$ particles, for the same symmetry property as before.

\begin{figure}
\begin{center}
\includegraphics[clip,scale=0.3]{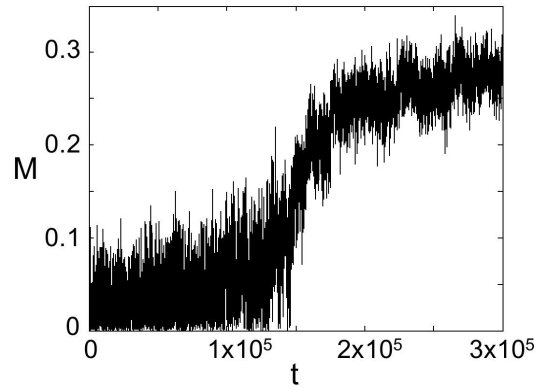}
\caption{Magnetization vs time for homogeneous initial conditions with semi-elliptical velocity distribution,
at $E=0.69$, for a system with $N=2^{12}$ particles.}
\label{ptdynmagn}
\end{center}
\end{figure}

In Figure \ref{ptdynmagn}, we plot the time evolution of the magnetization. We remark that at the end of the run the BG
state has not yet been reached (the equilibrium magnetization is about $0.31$), but the length is sufficient for
our analysis. Here, we consider the one particle distribution functions during the time range in which the
magnetization rises.

\begin{figure}
\begin{center}
\includegraphics[clip,scale=0.3]{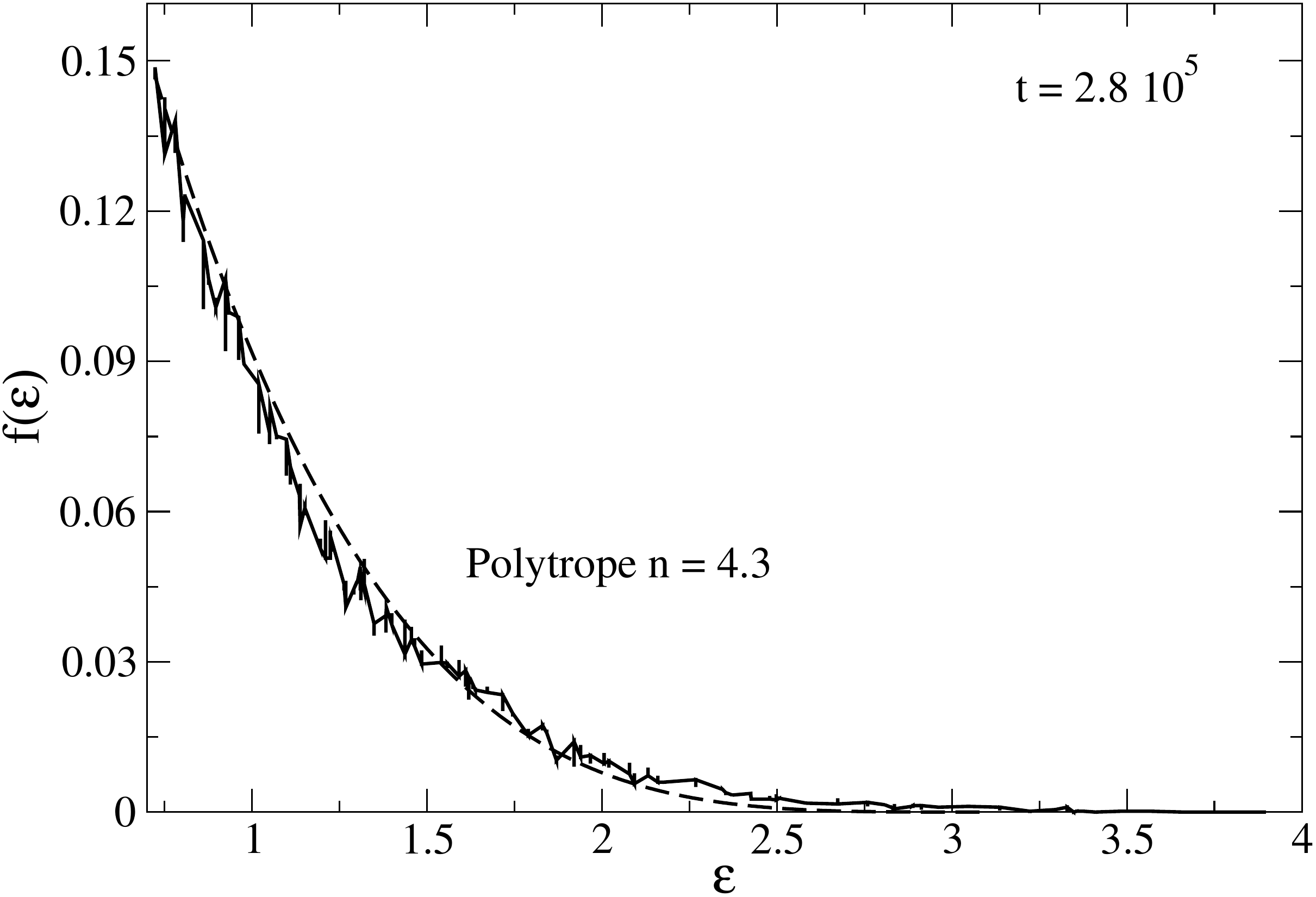}
\caption{One particle distribution function at time $t=2.8\times 10^5$. It has been fitted with a
polytrope with index $n=4.3$.}
\label{ptdynT2p8}
\end{center}
\end{figure}

We have computed the distributions at the times $t=1.5\times 10^5$,
$t=1.7\times 10^5$, $t=2.0\times 10^5$, $t=2.4\times 10^5$,
$t=2.8\times 10^5$. From Figure \ref{ptdynmagn} it can be seen that
these times are all after the destabilization of the homogeneous
phase (the corresponding values of the magnetization, averaged over a short interval of time, 
 are $M=0.17$, $M=0.19$, $M=0.26$, 
$M=0.26$, and $M=0.28$). In Figure  \ref{ptdynT2p8} we show the distribution $f(\epsilon)$ obtained at $t=2.8\times 10^5$. 





In all cases, the points are arranged along a line and therefore at
all times the system is in a stationary state of the Vlasov
equation. Therefore, for all the time range during the approach to BG
equilibrium, the system passes by a sequence of quasi stationary
steady solutions of the Vlasov equation, slowly evolving under the
effect of ``collisions''. This property is due to the scale separation
between the relaxation time (larger than $N$) and the dynamical time
(of order unity).  Furthermore, the system never goes through Vlasov
instability, and  all the dynamics is governed by the collisional
finite size effects.

The distributions have been fitted with polytropic functions, with
index $n=2.5$, $3.0$, $3.5$, $4.0$ and $4.3$, respectively. The fit is
very good in all cases (it is shown in Figure \ref{ptdynT2p8} for
the case $n=4.3$). In addition, the fits suggest an increasing index
with time. This is of course natural since the BG equilibrium
corresponds to a polytrope with index $n=\infty$. We
note that the first index calculated in the magnetized phase is
somewhat smaller than $n_c$, and the second index is very close but
slightly smaller than $n_c$. In principle this can be considered
perfectly legitimate; in fact, even though the numerical distribution function
is all the time very close to a polytrope, small discrepancies can give
rise to small uncertainties in the value of the index $n$, considering
that in the case of inhomogeneous distribution we have to fit
a function $f(\epsilon)$ instead of a function $f(v)$, and there are
more sources of numerical errors (like, e.g., the determination
of the magnetization $M$ in the definition of the individual energy
$\epsilon$). Furthermore, a slight change of the maximum energy of the
fit could correspond to a slight change of the polytropic index $n$.
We remark that for small values of $n$, like those of the previous
Section, this problem of the small uncertainty in the value of $n$ is much
less relevant, because of the steeper decrease of those polytropes. 
It is also possible that a value of $n(t)$ smaller 
than $n_c$ in the inhomogeneous phase is real (i.e. it is not a artifact due
to the fit). We may well imagine that the homogeneous phase destabilizes when 
$n(t)=n_c$ and that the inhomogeneous polytropes just after the transition have 
an index smaller than $n_c$. In other words, the evolution of $n(t)$
could be {\it non-monotonic}.

\section{Waterbag initial condition with $M_0=1$}
\label{sec_waterbagM1}

In this Section, we consider a different class of initial conditions, i.e., a waterbag
distribution for the velocities, and all the angles at $\theta=0$ so that the initial magnetization
is $M_0=1$ (this distribution is unsteady). That was the first initial condition considered for the HMF model, the one that revealed the
existence of QSSs \cite{ar,latora,lrt}. These conditions have been extensively studied, with the purpose to
determine the scaling with $N$ of the QSS lifetime and of its magnetization. It is known that there are large fluctuations
from run to run, so it is necessary to perform an average on several runs to estimate the mentioned scaling.
In Ref. \cite{campa1}, it was shown that the fluctuations can be reduced by using the so-called isotropic
waterbag conditions, in which the velocities are not randomly extracted from the waterbag, but are taken
equally spaced. This is equivalent to using normal waterbag distributions and performing averages over many runs.
We adopt this strategy here, but now with the purpose to study the one particle distribution
functions in the QSS.

\begin{figure}
\begin{center}
\includegraphics[clip,scale=0.3]{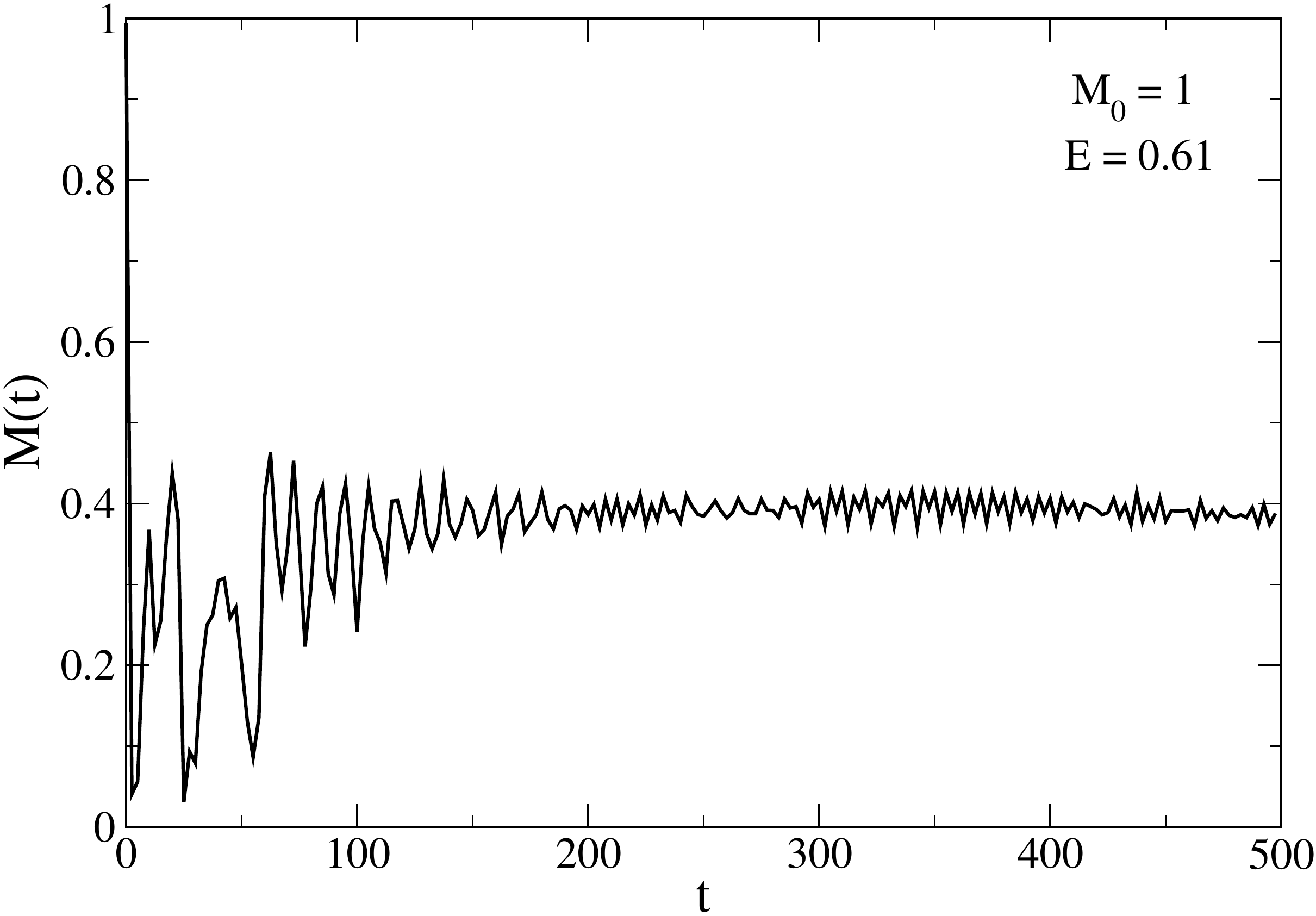}
\caption{Magnetization vs time for waterbag initial conditions with $M(0)=1$, at $E=0.61$.}
\label{M1E0p61magnetisation}
\end{center}
\end{figure}

\begin{figure}
\begin{center}
\includegraphics[clip,scale=0.3]{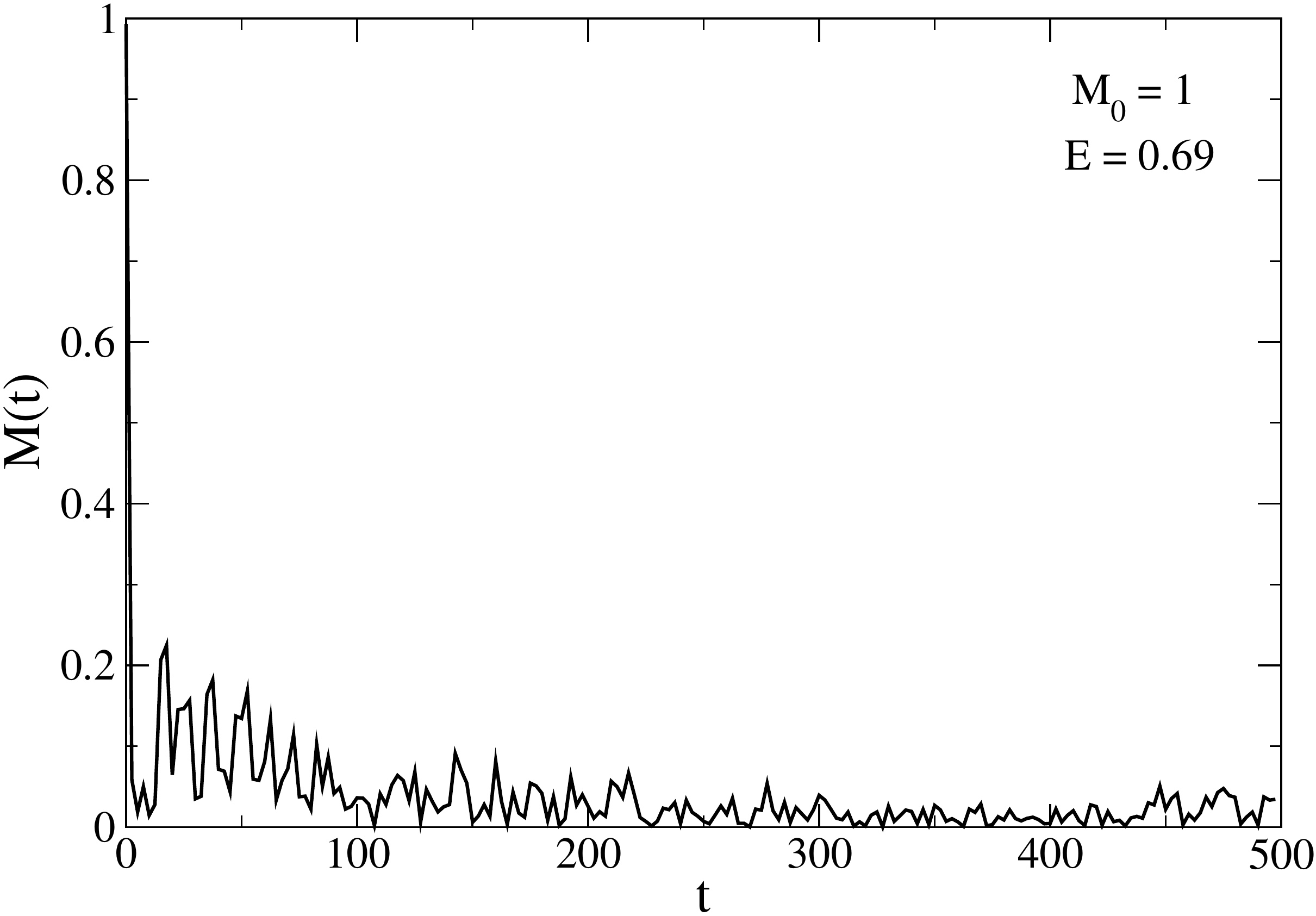}
\caption{Magnetization vs time for waterbag initial conditions with $M(0)=1$, at $E=0.69$.}
\label{M1E0p69magnetisation}
\end{center}
\end{figure}

We have performed runs at the energies $E=0.55$, $E=0.57$, $E=0.61$,
$E=0.64$, $E=0.65$, $E=0.66$, $E=0.665$, $E=0.67$ and $E=0.69$. As in Section
\ref{sec_simul}, we simulate a system with $2^{18}$ particles.  The
last energy is the one mostly studied in the literature, where it is
shown that the magnetization of the QSS vanishes in the large $N$
limit. We confirm this result here. On the other hand, we found that
at all the smaller energies the magnetization of the QSS does not
vanish. In Figures \ref{M1E0p61magnetisation} and
\ref{M1E0p69magnetisation} we show the time evolution of the
magnetization for the two energies $E=0.61$ and $E=0.69$,
respectively.

\begin{figure}
\begin{center}
\includegraphics[clip,scale=0.4]{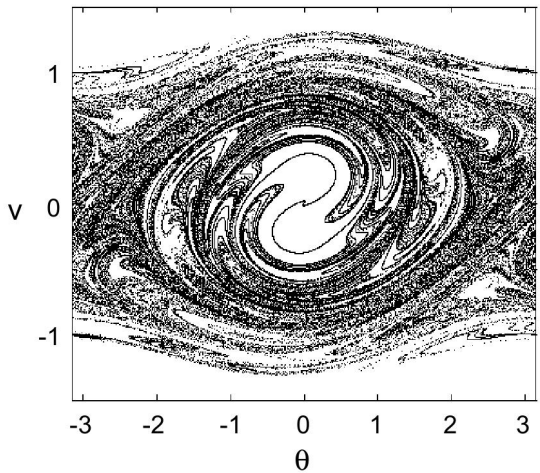}
\caption{Occupation of the one particle phase space in the early stage ($t=100$) of the QSS at $E=0. 61$, for
the run with waterbag initial conditions for the velocities, and with $M_0 = 1$. The plot shows a hole in
the center.}
\label{hole061early}
\end{center}
\end{figure}

\begin{figure}
\begin{center}
\includegraphics[clip,scale=0.35]{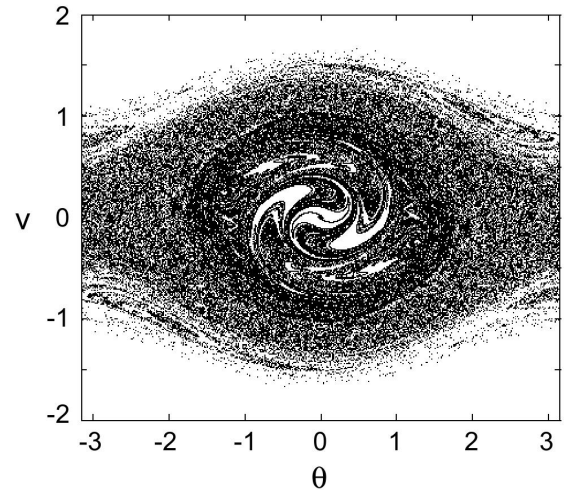}
\caption{Occupation of the one particle phase space at later times ($t=500$) of the QSS at $E=0.61$. The hole
at the center has shrunk considerably.}
\label{hole061late}
\end{center}
\end{figure}

In the early stages of the QSS, the occupation of the one particle phase space presents holes that tend
to disappear. This peculiarity is shown in some figures where we plot the location of all the particles
in the one particle phase space. The plots correspond to the two energies $E=0.61$ and $E=0.69$, and to two
different times for each energy. Figure \ref{hole061early} represents the phase space occupancy at $E=0.61$
at the time $t=100$, which is one fifth of the duration time of the run, as it can be seen from
Figure \ref{M1E0p61magnetisation}. The phase space occupancy at the latest time, $t=500$, is plotted in Figure
\ref{hole061late}.

\begin{figure}
\begin{center}
\includegraphics[clip,scale=0.3]{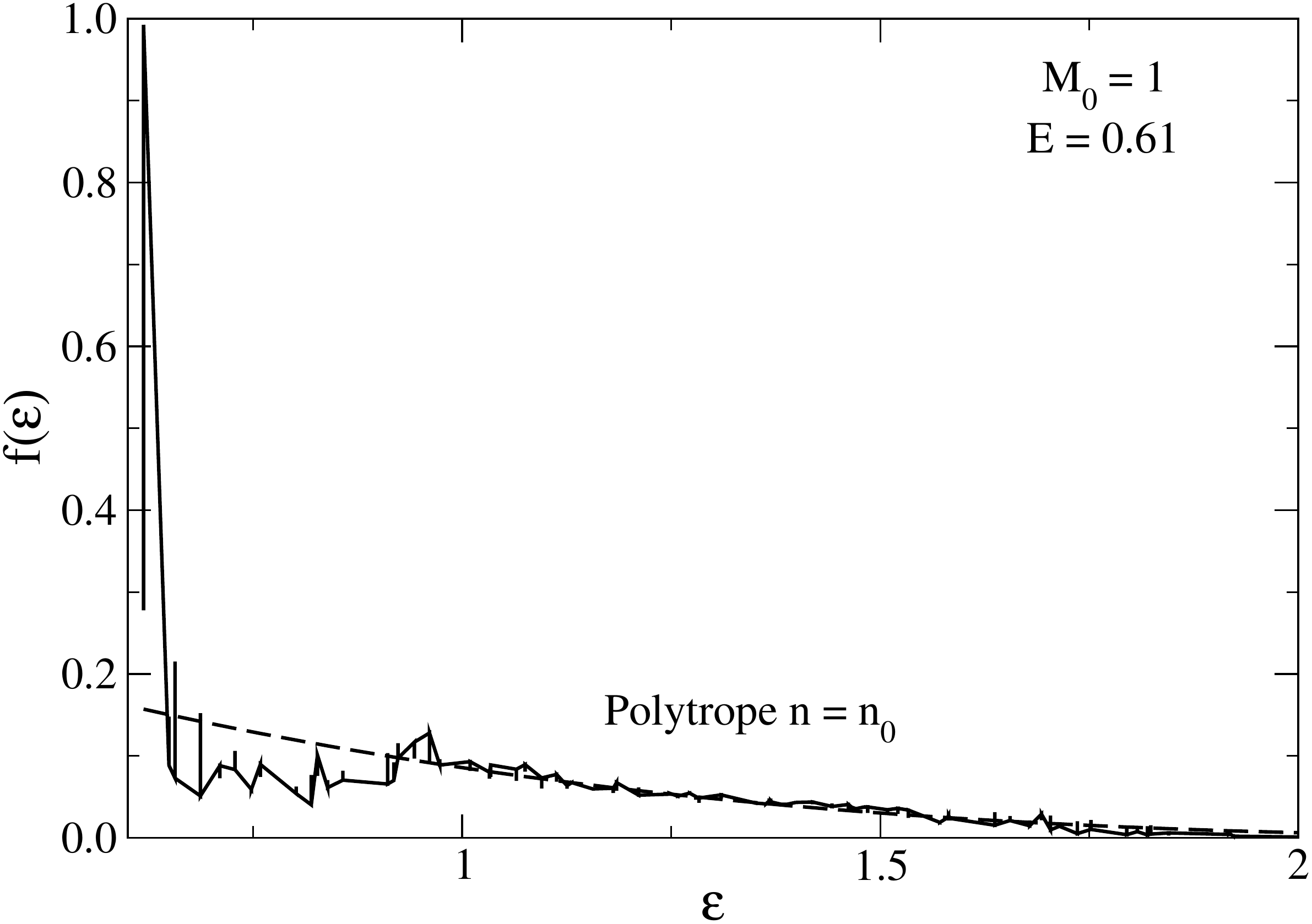}
\caption{Distribution function $f(\epsilon)$ in the QSS for $E=0.61$. It has been fitted, neglecting the high
peak at small energies, by a   $n_0\simeq 3.56$ polytrope.  }
\label{m1E0p61}
\end{center}
\end{figure}

Although the hole at the center of the phase space has shrunk
considerably at $t=500$, it is still very clearly visible. This
appears to be in apparent contrast with the numerical distribution
function $f(\epsilon)$ at the same time $t=500$, that is shown in
Figure \ref{m1E0p61}. As in the previous cases, the disposition along
a line of the numerical data confirms that the state is a
QSS. However, the interesting point to stress is the presence of a
pronounced peak at the smallest energies $\epsilon$. From the
expression of the individual energy $\epsilon$, we know that the
smallest energies correspond to the particles located near the center
of the one particle phase space. Then, we could argue that there is a
contradiction between this high peak in $f(\epsilon)$ and the hole in
the center of the phase space.

To clarify this point we have enlarged the region near the center of
the phase space (we do not report here these plots), and this has
revealed that the hole contains in fact many particles arranged along
thick filaments, that in the two-dimensional plots embracing the whole
phase space are not visible. These filaments could be explained by the
following argument.

We know that, as long as the evolution of
the one particle distribution function $f(\theta,v,t)$ is governed by the Vlasov equation, the mass of each of its
phase levels is conserved. However, the evolution causes a complex intertwinement of the various phase levels; this is
what allows to consider coarse-grained distributions, and justifies the fact that the numerical distributions, that
inevitably have an implicit averaging (at small scales as it may be) have to be compared with the theoretical
coarse-grained distributions. For the case under study in this Section, the fine-grained distribution function
has only two levels, zero and infinite (i.e., it is a $\delta$ function), since all the particles are initially
at $\theta = 0$. It may then be argued that for a while there remains very high values also for the coarse-grained
distribution function, that are gradually decreased by the mixing. Unless the coarse-grained distribution is analyzed
at a very small scale (but this would be possible only with a very large number of particles), high values can be
obtained only with particles arranged approximately along a one-dimensional region, i.e., a line.

\begin{figure}
\begin{center}
\includegraphics[clip,scale=0.3]{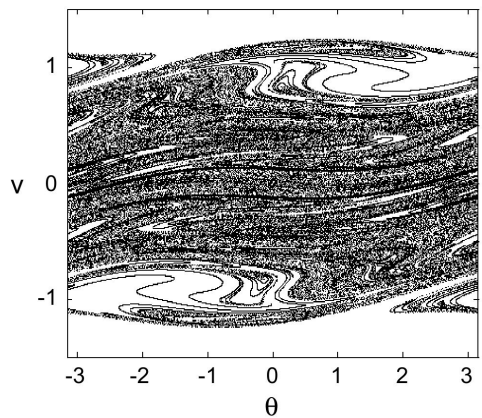}
\caption{Occupation of the one particle phase space in the early stage ($t=100$) of the QSS at $E=0. 69$, for
the run with waterbag initial conditions for the velocities, and with $M_0 = 1$. The plot shows two holes.}
\label{hole069early}
\end{center}
\end{figure}

\begin{figure}
\begin{center}
\includegraphics[clip,scale=0.35]{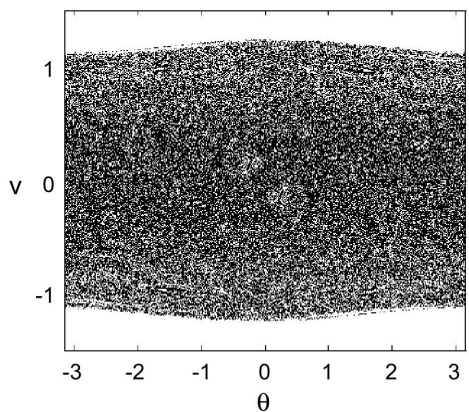}
\caption{Occupation of the one particle phase space at later times ($t=500$) of the QSS at $E=0.69$. The holes
have disappeared.}
\label{hole069late}
\end{center}
\end{figure}

In Figures \ref{hole069early} and \ref{hole069late} we show the
occupancy of the one particle phase space for the energy $E=0.69$, at
the times $t=100$ and $t=500$ respectively. Again, at the earlier time
the plot presents holes, although they are not at the center. We
remind, however, that now the QSS is not magnetized, as shown in
Figure \ref{M1E0p69magnetisation}. At the later time the holes have
disappeared.

\begin{figure}
\begin{center}
\includegraphics[clip,scale=0.3]{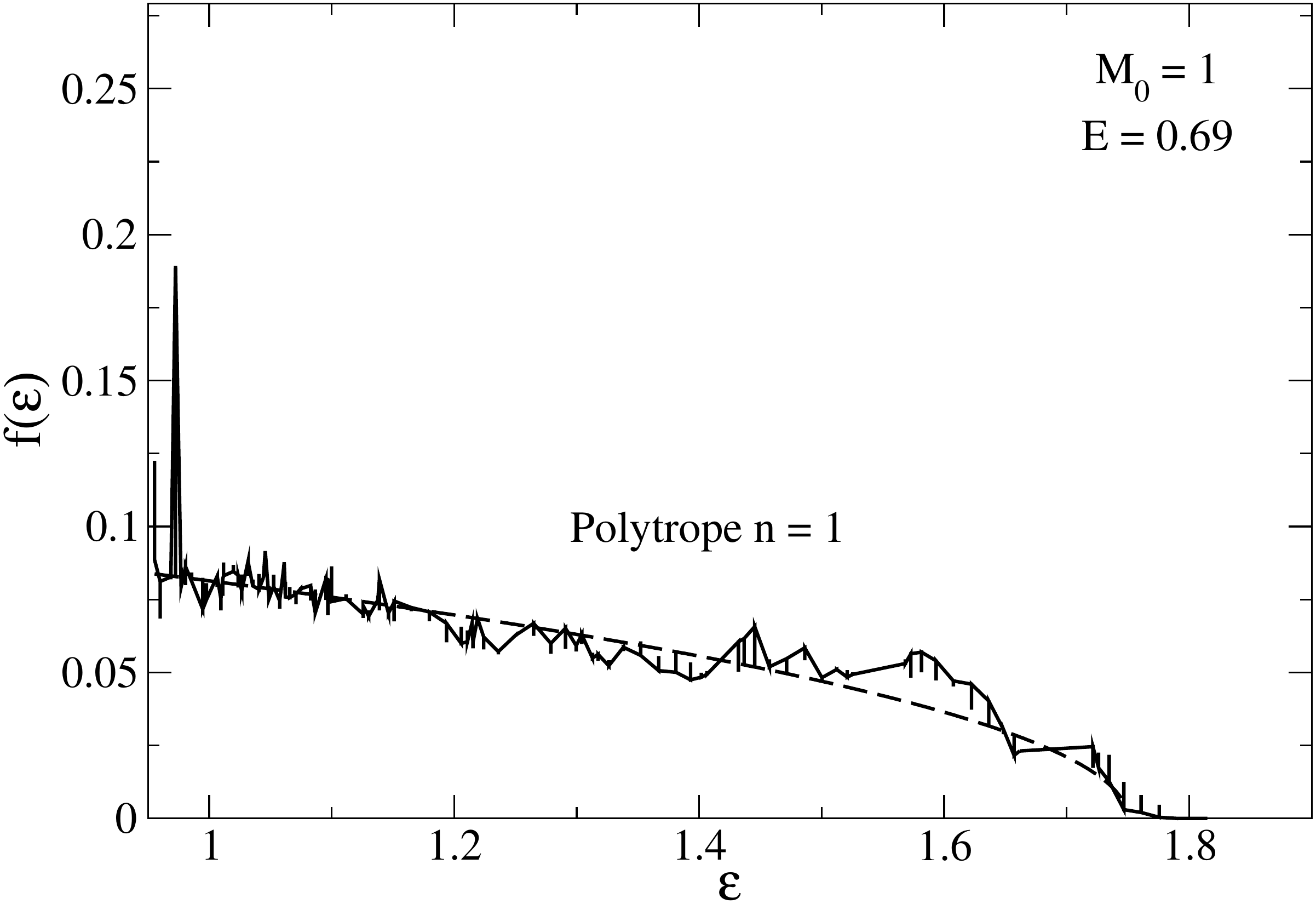}
\caption{Distribution function in the QSS for $E=0.69$. It has been fitted, neglecting the high peak
at small energy, by a $n=1$ polytrope.}
\label{M1E0p69}
\end{center}
\end{figure}

Also in this case the one particle distribution function at $t=500$
has a pronounced peak at the smaller energy $\epsilon$, as it can be
seen in Figure \ref{M1E0p69}. As a matter of fact, the peak is present
for all energies $E$ that have been simulated. However, neglecting the
high peak at small energy, we see that the distribution function at
$E=0.69$ can be fitted with a polytrope with $n=1$. In a previous work
\cite{campa1}, with numerical simulations performed with much less
particles, and therefore with stronger finite size effects and faster
QSS evolution, the peak probably had disappeared very soon, and no
evidence of it was found (the velocity distribution $f(v)$ could be
nicely fitted by a pure $n=1$ polytrope).

\begin{figure}
\begin{center}
\includegraphics[clip,scale=0.3]{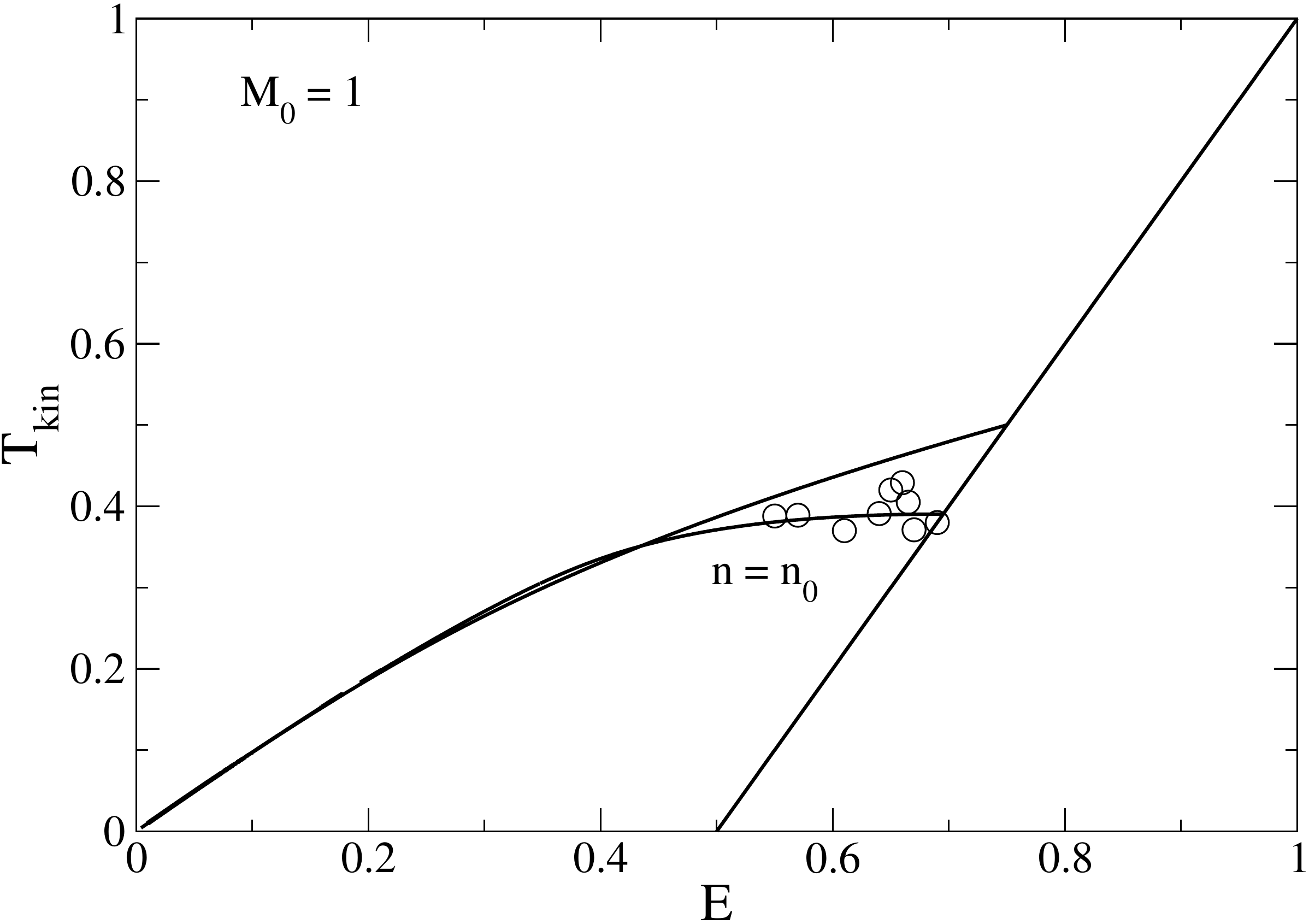}
\caption{Numerical caloric curve for the QSS reached from the waterbag $M(0)=1$ initial conditions.
It has been fitted by a polytrope with index $n_0\simeq 3.56$ for which $C_{kin}\rightarrow \infty$ close to
the bifurcation point. The fit is relatively good except for three points that reveal a sort
of ``hump''. This hump seems to be real (i.e. it is not a numerical artifact). It presents 
a region of negative kinetic specific heat.}
\label{m1}
\end{center}
\end{figure}

It is interesting to plot the numerical kinetic caloric curve obtained
from our simulations. This is done in Figure \ref{m1}. Except for the
curious points around the energy $E=0.66$ (more precisely, the points at
$E=0.65$, $E=0.66$ and $E=0.665$) that reveal a sort of
``hump'', it seems that $T_{kin}$ is almost constant with $E$. This
``plateau'' corresponds to a region with infinite kinetic specific
heat $C_{kin}=\infty$. Actually, the numerical caloric curve is
well-fitted by the physical caloric curve of $n_0\simeq 3.56$
polytropes that presents a region of infinite specific heat close to
the bifurcation point (see Section \ref{sec_phy}). The distribution
function of the QSS at $E=0.61$ is also well-fitted by a $n_0$
polytrope, except again for small energies (see Figure
\ref{m1E0p61}). These new results do not support the ``guess'' made in
our previous paper \cite{cc} that the numerical caloric cuve can be
fitted by $n=1$ polytropes. They are also in conflict with the early
claims of \cite{ar,latora} that the numerical caloric curve presents a
region of negative kinetic specific heat (this region is not clearly
apparent in our simulations, except in the hump). However, we know
that a system prepared in the waterbag distribution has a strange
dynamics, so that more averages may be needed to obtain reliable
results.

\section{Conclusion}

In spite of the large amount of numerical studies performed on the HMF
model, there are two main problems that are still poorly
understood. The first one is the anomalous scaling law with $N$ of the
homogeneous QSS\footnote{Some interesting results 
in this direction have been obtained recently \cite{firpo}.}; the second, treated in this work, is the
dependence of the QSS reached by the system from a generic initial
condition. The first problem should be restricted to the homogeneous
QSSs of one-dimensional systems, where the collisional effects vanish
at order ${1}/{N}$. The second problem is expected to be relevant to
long-range systems in general.

The Lynden-Bell theory determines, at least in principle, the distribution function of the QSS, although a technical
difficulty is to compute it in the generic case of multi-levels initial distributions. On the other hand, various
numerical simulations have shown that the theory works only in some cases, failing in others. Since the theory is
based on ergodicity or efficient mixing, the reason of the failure has to be traced to an incomplete
mixing \cite{incomplete}. If the system does not mix well, we should not expect any ``universality'': The QSS will
depend on the degree of mixing. This rises the level of difficulty of the problem since we do not have indications
on the further constraints that are imposed by the lack of full mixing.

In this work we have not offered a solution of this problem, in the sense that we have not proposed a way to predict
the QSS as a function of the initial condition. Rather, we have proposed to analyze the correspondence between
a class of stationary states of the Vlasov equation and the numerical QSS reached by the system starting from
various kinds of initial conditions. We have found numerically that, in many cases, the QSS is well-fitted by a
polytropic distribution with a compact support. The fact that the distribution has a compact support is consistent
with the phenomenology of incomplete relaxation according to which some parts of the phase space (usually those
with high energy $\epsilon$) are not sampled by the system. In this respect, the value of the index $n$ may be a measure of mixing.

Polytropic distributions were introduced in different situations:

(i) In astrophysics, they were introduced as particular steady states of the Vlasov equation \cite{eddington}.
Furthermore, it was shown that they extremize a ``pseudo-entropy'' functional
$S=-\int f^{1+1/(n-3/2)}\, d{\bf r}d{\bf v}$ at fixed mass $M$ and energy $E$ \cite{ipser}, and that {\it maxima}
of such functionals are dynamically Vlasov stable \cite{ih,aaantonov}. This is the interpretation that
we have followed in this paper.

(ii) Polytropic distributions were also introduced by Tsallis \cite{tsallis} from an approach of generalized
thermodynamics. In that case, the functional $S$ is interpreted as a generalized entropy and its maximization at
fixed mass and energy determines the most probable state of a non-ergodic system for which some regions of phase
space are forbidden by the dynamics. Several classes of generalized entropies are possible but the Tsallis entropy
is the simplest deviation to the Boltzmann entropy. We leave open the debate about the relevance of this interpretation
and refer to \cite{cstsallis,assise} for more discussions about this issue.

In this work, we have performed numerical simulations of the HMF model initially prepared in a Vlasov unstable steady
state, and we have analyzed the properties of the QSS reached by the system after the rapid relaxation. There have
been previous studies dedicated to the description of the QSS in the HMF model, and the following ones have treated
the subject under different conditions. In Refs. \cite{prl2,staniscia1,precommun,bachelard2}, the authors have compared
the numerically obtained QSS phase diagram with the theoretical one arising from the application of the Lynden-Bell
theory to rectangular waterbag initial conditions. They found that the results agree in some regions of the phase diagram,
and disagree in others. Morita and Kaneko \cite{mk} have analyzed the magnetization of the QSS reached by particular
initial conditions, i.e. of the same
form as that attained in the BG equilibrium, but with out-of-equilibrium parameter values; the value and the possible
oscillations of the QSS magnetization have been studied as a function of the parameters of the initial distribution.
Finally, Pakter and Levin \cite{levin} have studied the same two levels initial distributions as in \cite{prl2},
with the purpose to propose an alternative QSS one particle distribution function, with respect to the Lynden-Bell
functions, for the cases where the latter does not work.

From the point of view of the initial conditions analyzed, our work is in a sense complementary to those just cited.
We have studied several classes of initial distribution functions, some of which with only two levels, as in
\cite{prl2,staniscia1,precommun,bachelard2} and \cite{levin}, and some with continuous values, as in \cite{mk}. In
addition we have analyzed the
properties of the QSS distribution functions, showing that they can be fitted in many cases with polytropic
functions\footnote{Similarly, some QSSs in 2D turbulence have also been fitted by polytropic (Tsallis)
distributions \cite{boghosian,at,abe}, although this fit cannot be universal \cite{brands}.}. In this
respect, the spirit of our work is closer to \cite{levin}, where the authors try to find the expression for
the distribution function. Our results for the QSS reached from waterbag homogeneous initial conditions
confirm the results of Ref. \cite{levin}: we found that the QSS distribution function has a core-halo structure;
the limits of the core, in the one particle phase space, are given by a constant energy line, and 
the core can be fitted by a function $f(\epsilon)$ with a waterbag structure, i.e., a $n={1}/{2}$ polytrope.
In addition, we found that the magnetization in the QSS presents oscillations, reminiscent of those found in \cite{mk}.
This could be due to the absence
of Landau damping, analogously to what holds for the homogeneous waterbag function, characterized by purely
real proper frequencies for the corresponding linearized Vlasov equation.

We found that polytropic functions can fit the QSS distributions also in the other cases, with an index
depending on the class of initial condition, although the fit
is not very good for all the energies considered.  In particular, for energies slightly smaller than the instability
threshold the fit appears less satisfying, and probably those cases could be considered with a quasilinear
approach \cite{quasilinear}.
The fit with polytropic functions has allowed us to compare the numerical caloric curves with the kinetic
caloric curves computed on the basis on the theory of polytropes \cite{cc}. In particular, using this theory, we have
been able to explain the negative kinetic specific heat region observed in the numerical simulations for intermediate
energies. This is one of the main results of the paper. For lower energies, the system is not a pure polytrope but it
takes a core-halo structure. The core can be fitted by a polytrope, which may differ from the waterbag distribution.
The halo may also be fitted by a polytrope so that the core-halo state may be viewed as a ``mixture'' 
of two polytropes\footnote{For example, in Fig. \ref{feps055waterbag}, the distribution function 
in the halo is approximately 
constant so that the core-halo state may be fitted by two polytropes $n=1/2$ with different values of $f$.}.
Therefore, our results complement, and generalize, the findings in Ref. \cite{levin} for the waterbag initial condition.

An interesting problem is to understand {\it why} the QSS can be fitted by a core-halo state with a polytropic core.
Furthermore, the polytropic index of the QSS seems to be correlated to the initial distribution function. Indeed, for
the semi-ellipse and waterbag initial conditions, we find that the QSS can be fitted by a polytrope with index $n=1$
and $n=1/2$, respectively. This could be interpreted by saying that the system keeps memory of the initial condition.
For the waterbag case, the core does not mix well so it keeps its uniform distribution\footnote{We remark that, in all 
considered cases (see Figs. \ref{feps058gauss}, \ref{feps055gauss}, \ref{feps059ellipse}, \ref{feps055ellipse},  
\ref{feps0582waterbag}, and \ref{feps055waterbag}), the maximum value of 
the distribution function is conserved. This
is not a trivial result since we expect the coarse-grained distribution function $\overline{f}_{max}$ 
to {\it decrease} as the
system mixes (this is what the Lynden-Bell theory usually predicts). This shows that the 
core does not mix well. This property is also observed in self-gravitating 
systems \cite{jw2011} and in two-dimensional
turbulence \cite{chen}. The conservation of the maximum value of the distribution function can be explained qualitatively in terms of a kinetic 
theory (see Appendix A of \cite{scs,ktc}, Appendix B of \cite{staniscia2}, and \cite{hb4}).}. On the other hand, a halo is
formed probably due to parametric resonance \cite{levin}. As explained in this paper,
the  halo is necessary 
in order to conserve  the value of $f_0$ in the core while keeping the total energy unchanged. 
Similar explanations may hold for more general distributions.

We have also studied the approach to BG equilibrium of a system initially prepared in a homogeneous Vlasov stable
state. This approach is due to finite size effects. In Ref. \cite{campa2} we found that the slowly changing
homogeneous distribution function can be fitted by a sequence of polytropes with a time dependent index. Here we extended
that work, and we have found that this fit is good also in the successive time range, during which the system
progressively magnetizes. We found that the polytropic index increases with time, although the index in the early
stages of the magnetized phase appears to be slightly less than the critical index for the homogeneous polytrope. This
increase is to be expected since in the BG equilibrium state the index should approach infinity. The reason why polytropic
distributions fit well the collisional relaxation of the system remains a mystery. We recall, however, that a similar
observation has been made by Taruya \& Sakagami \cite{ts} for self-gravitating systems. Therefore, polytropic distributions
seem to be ubiquitous in long-range systems.

Finally, we have considered the QSS reached by a peculiar initial
state, in which all the rotators are at $\theta=0$, i.e., with the
initial magnetization equal to $1$. In this case, the dynamics is
complex due to the formation of phase space holes and thick filaments,
and the results have been more difficult to interpret. We found that,
except for a few energy values, the numerical caloric curve
presents a horizontal segment corresponding to an infinite 
kinetic specific heat.  However, we do not have an explanation for the
odd results for the three energies that do not comply with the others
and seem to reveal a hump.
We also found that the fit with polytropes is here more
problematic. For the only energy with a non magnetized QSS, $E=0.69$,
we found that the QSS distribution function can be fitted with a
semi-ellipse, i.e., an $n=1$ polytrope, only neglecting the pronounced
central peak at the smaller energies. We have shown that this peak
tends to disappear during the slow QSS evolution. For the others
energies with a magnetized QSS, the numerical caloric curve has been
compared to the kinetic caloric curve of $n_0\simeq 3.56$ polytropes
that presents an infinite kinetic specific heat $C_{kin}=+\infty$ (plateau) close
to the bifurcation point. The fit is fairly good, except for the three
anomalous points mentioned above. The distribution function at
$E=0.61$ has also been fitted by a $n_0$ polytrope, again excluding
the small energies.

Summarizing, we have shown that, for some classes of initial conditions, the QSS can be
characterized by polytropic distributions. This appears to be the norm rather than the exception. This does not solve the
problem of predicting the QSS structure as a
function of the initial conditions since there is no theory to predict the polytropic index $n$, and there exist cases
where the QSS is not even a polytrope (hence, polytropic distributions are {\it not} ``universal attractors''). The
problem of predicting the QSS would probably need a proper kinetic equation rather than
a theory based on pseudo-entropy functionals (see discussion in \cite{incomplete}). However, we hope that these results
can give some hint for the search
of a more quantitative explanation. We have also shown that for other classes 
of initial conditions the QSS can be described by a Lynden-Bell distribution (see Fig. \ref{distrlynden}). The distinction
between polytropic (Tsallis) and Lynden-Bell distributions may be related to the ergodic properties of the dynamics.
When the system ``mixes well'', the Lynden-Bell distribution is reached. Otherwise, the relaxation is 
incomplete and polytropic distributions may emerge. We conclude that both Lynden-Bell and Tsallis 
distributions may be relevant to describe QSSs depending on the efficiency of mixing as argued in \cite{epjb}.

\appendix

\end{document}